\documentclass[aps,prx,reprint,superscriptaddress,longbibliography]{revtex4-2}

\usepackage{graphicx}
\usepackage{amsmath,amssymb}
\usepackage{physics}
\usepackage{xcolor}
\usepackage[bookmarksnumbered]{hyperref}

\begin{document}

\title{Key observable for linear thermalization}
\author{Yuuya Chiba}
\email{chiba@as.c.u-tokyo.ac.jp}
\affiliation{Department of Basic Science, The University of Tokyo, 3-8-1 Komaba, Meguro, Tokyo 153-8902, Japan}

\author{Akira Shimizu}
\email{shmz@as.c.u-tokyo.ac.jp}
\affiliation{Institute for Photon Science and Technology, The University of Tokyo, 7-3-1 Hongo, Bunkyo-ku, Tokyo 113-0033, Japan}
\affiliation{RIKEN Center for Quantum Computing, 2-1 Hirosawa, Wako City, Saitama 351-0198, Japan}

\date{\today}

\begin{abstract}
For studies on thermalization of an isolated quantum many-body system,
the fundamental issue is 
to determine whether a given system thermalizes or not.
However, most studies tested 
only a small number of observables,
and it was unclear 
whether other observables thermalize.
Here, we study 
whether `linear thermalization' occurs 
for all additive observables:
We consider 
a quantum many-body system prepared in an equilibrium state
and its unitary time evolution
induced by a small change $\Delta f$ of 
a physical parameter $f$ of the Hamiltonian,
and examine 
whether \emph{all} additive observables relax to 
the equilibrium values 
in a manner fully consistent with thermodynamics
up to the linear order in $\Delta f$.
We find that 
the additive observable conjugate to $f$ is key for linear thermalization in that
its
linear thermalization guarantees,
under physically reasonable conditions,
linear thermalization of all additive observables.
Such a linear thermalization occurs
in the timescale of $\mathcal{O}(|\Delta f|^0)$,
and lasts at least for a period of $o(1/\sqrt{|\Delta f|})$.
We also consider 
linear thermalization 
against the change of other parameters,
and find that linear thermalization of 
the key observable against $\Delta f$
guarantees
its linear thermalization
against small changes of any other parameters.
Furthermore, we discuss the generalized susceptibilities 
for cross responses 
and their consistency 
between quantum mechanics and thermodynamics.
We demonstrate our main result 
by performing numerical calculations for spin models.
The present paper offers
an efficient way of judging linear thermalization
because it guarantees that 
examination of the single key observable is sufficient. 
\end{abstract}

\maketitle

\section{\label{sec:Introduction}Introduction}

Thermalization of isolated quantum many-body systems
has long been studied 
as the question of 
how 
the thermal equilibrium state arises from
the quantum unitary dynamics~\cite{
Neumann1929,
Bocchieri1957,Percival1961,
Deutsch1991,Srednicki1994,Tasaki1998,Srednicki1999,
Goldstein2006,Popescu2006,Sugita2006,
Rigol2008,Kim2014,Beugeling2014,Steinigeweg2014,Sirker2014,
Mondaini2016,Mondaini2017,Banuls2011,
Goldstein2015,Tasaki2016,Goldstein2017,
DePalma2015,Reimann2015,Shiraishi2017,Mori2017,Hamazaki2018,Lashkari2018,Anza2018,Brenes2020,Sugimoto2021,Dymarsky2022,Sugimoto2022,Wang2022a,
Biroli2010,Ikeda2013,Alba2015,Mori2016,Iyoda2017,
Kinoshita2006,Hofferberth2007,Gring2012,Trotzky2012,AduSmith2013,Langen2015,Kaufman2016,Bernien2017,Orioli2018,Tang2018,
Richerme2014,Clos2016,Smith2016,Hess2017,
Neill2016,Xu2018,Guo2021,Chen2021,
DarkwahOppong2022,Abuzarli2022,
DAlessio2016,Gogolin2016,Mori2018}.

As in classical systems, 
the relation of thermalization with 
chaos
has been attracting much attention.
It has brought a wealth of findings~\cite{Srednicki1999,Mondaini2016,Mondaini2017,DAlessio2016},
such as the relations to
the scrambling~\cite{Bohrdt2017,Schnaack2019,Zhu2022},
which was originally discussed in quantum information theory~\cite{Hayden2007,Sekino2008,Shenker2014},
and to the out-of-time-ordered correlations~\cite{Shenker2014a,Shenker2015,Roberts2015,Roberts2015a,Hosur2016},
which characterize the dynamical feature of quantum chaos~\cite{Hosur2016,Balachandran2021,Brenes2021}.

Many studies have also been devoted to 
the eigenstate thermalization hypothesis (ETH)~\cite{Neumann1929,Deutsch1991,Srednicki1994},
which states that every energy eigenstate represents an equilibrium state.
This hypothesis leads to  
thermalization,
and 
is expected to hold in quantum chaotic systems~\cite{
Rigol2008,Kim2014,Beugeling2014,Steinigeweg2014,Reimann2015,Mondaini2016,Mondaini2017}.
However, since there also exist many systems that do not satisfy the ETH,
how universally it holds
and when it fails 
are still the subjects of active research~\cite{
Shiraishi2017,Mori2017,Hamazaki2018,Lashkari2018,Anza2018,Brenes2020,Sugimoto2021,Dymarsky2022,Sugimoto2022,Wang2022a}.
Furthermore, 
while the ETH is a sufficient condition for thermalization,
whether it is also a necessary condition 
seems to be controversial because the answer 
depends on the choices of the initial state and 
observables that are employed to check thermalization~\cite{DePalma2015,Shiraishi2017,Mori2017}.

Researches in the opposite directions,
i.e., 
mechanisms for absence of thermalization
and related problems in nonthermalizing systems,
are also attracting much attention.
They include 
integrable systems~\cite{
Rigol2007,Fagotti2014,Vidmar2016,Pozsgay2014,Wouters2014,Mierzejewski2015,Ilievski2015,Ilievski2015a,Doyon2017,Mitsuhashi2022,Bertini2022,Yang2022,Cataldini2022},
many-body localization~\cite{Oganesyan2007,Pal2010,Imbrie2016,Roy2020,Danieli2020,McClarty2020,Kuno2020,Daumann2020,Vu2022,Balducci2022,Krajewski2022,Schreiber2015,Kondov2015,Choi2016,Smith2016,Wei2018,Nandkishore2015,Altman2015,Abanin2019},
many-body scars~\cite{Heller1984,Srednicki1994,Bernien2017,Turner2018,Turner2018a,Moudgalya2018,Moudgalya2018a,Alhambra2020a,Moudgalya2020,Bull2020,Windt2022,Desaules2022,Huang2022,Bull2022},
and Hilbert space fragmentation~\cite{DeTomasi2019,Moudgalya2021,Buca2022,Moudgalya2022,Bastianello2022,Yoshinaga2022,Kohlert2023},
which lead to 
absence of thermalization
and failure of the ETH.

In such nonthermalizing systems,
the entanglement entropy of energy eigenstate often behaves anomalously~\cite{
LeBlond2019,
Bauer2013,
Turner2018a,
DeTomasi2019,Moudgalya2021},
and hence is sometimes employed 
to discriminate between thermal and nonthermal
eigenstates~\cite{Turner2018a,Garrison2018,Ippoliti2022,Mueller2022,Comparin2022}.
However,
even if the entanglement entropy agrees with thermodynamic entropy,
it does not necessarily imply that 
the eigenstate 
represents the equilibrium state
because there exist many quantum 
states with the same entanglement entropy.
For this reason, 
many studies examined 
the expectation values of observables
to discriminate between thermal and
nonthermal states~\cite{Trotzky2012,Orioli2018,Kollath2007,Rigol2009,Rigol2009a,Banuls2011,Zill2015}.

When testing thermalization 
using 
observables,
most of previous studies examined 
only a small number of observables~\cite{Trotzky2012,Orioli2018,Kollath2007,Rigol2008,Rigol2009,Rigol2009a,Banuls2011,Kim2014,Beugeling2014,Steinigeweg2014,Zill2015,DePalma2015,Mondaini2016}.
However,
thermalization of such 
observables
does not necessarily guarantee thermalization
of other observables.
For example, 
the authors showed in Fig.~2(b) and (c) of Ref.~\cite{Chiba2020} that,
in the XXZ and the XY 
spin chains,
the magnetization of any nonzero wavenumber 
(such as the staggered magnetization) thermalizes
while the uniform (zero wavenumber) magnetization does not.
A simpler example is the case where 
thermalization trivially occurs by symmetry.
For instance,
when both the Hamiltonian and the initial state
are symmetric under spin rotation by $\pi$ around the $z$ axis,
the $x$ component of magnetization
is kept $0$ under the unitary time evolution
(because of the symmetry)
and hence thermalizes trivially.
Nevertheless, 
if this system is integrable
many other
observables
such as interactions between nearest neighbor spins
do not thermalize,
as in the case of Model~III
of this paper.
Another example, which seems nontrivial and most interesting, is the possibility that
\emph{all} one-body observables thermalize
whereas two- and more-body observables do not.
All these examples show that 
thermalization of \emph{some} observables
does not necessarily imply thermalization
of \emph{all} observables.

These issues have also been long-standing questions
even in the linear nonequilibrium regime.
For example, Kubo stated in his famous paper 
on the `Kubo formula'~\cite{Kubo1957} that 
if the system has an `ergodic property'
then his formula would give the isothermal response.
However, his discussion turned out wrong~\cite{Wilcox1968, Mazur1969, Suzuki1971}: 
Later studies proved that 
the Kubo formula gives the adiabatic response
under certain conditions~\cite{Wilcox1968, Mazur1969, Suzuki1971, Chiba2020},
and it can also give the isothermal response
under another condition 
by taking the limit of vanishing wavenumber~\cite{Chiba2020}.
However, 
the conditions given by these studies 
were for \emph{individual} observables,
which do not guarantee the conditions for \emph{all} observables.
In other words, 
the above-mentioned fundamental issue of thermalization has 
left unsolved 
even in the linear nonequilibrium regime.

In this paper, we study 
the unitary time evolution induced by a small change 
(quench)
$\Delta f$ of 
a physical parameter $f$,
and examine whether the quantum state relaxes to 
an equilibrium state 
that is fully consistent with 
thermodynamics 
up to $\mathcal{O}(\Delta f)$.
We call the relaxation in this sense \emph{linear thermalization},
which is obviously necessary for thermalization against an arbitrary 
magnitude of $\Delta f$.
We place a particular emphasis on the \emph{full}
consistency. 
That is, 
when comparing the quantum state with a thermal equilibrium state we examine \emph{all} additive observables 
because thermodynamics assumes that 
\emph{all} additive observables take macroscopically definite values in an equilibrium state~\cite{Landau1980,Oono2017,Shimizu2021},
although the state is specified by only \emph{a small number of} 
variables 
(such as temperature).
Furthermore, 
we consider the case where the initial state is an 
equilibrium state because thermodynamics 
basically 
treats transitions between equilibrium states.

We find that 
the additive observable $\hat{B}$ which is conjugate to $f$
is the key observable in linear thermalization.
We show rigorously that,
if its expectation value relaxes to 
the value 
predicted by thermodynamics,
so do the expectation values of \emph{all} additive observables.
Even when $\hat{B}$ is a simple one-body observable, 
its relaxation 
guarantees relaxation 
of all other additive observables including two- and more-body ones.
In addition to this theorem, we prove two propositions
which state that,
under reasonable conditions, 
the time fluctuations of the expectation values 
and the variances of all additive observables
are sufficiently small. 
These results mean that linear 
thermalization of the single observable $\hat{B}$
guarantees linear 
thermalization of all additive observables.
This linear thermalization occurs in timescale of 
$\mathcal{O}(|\Delta f|^0)$.
We prove that it lasts at least for a period of $o(1/\sqrt{|\Delta f|})$.
Furthermore, 
we show that linear 
thermalization of $\hat{B}$ against the quench of
$\Delta f$ 
implies linear 
thermalization of $\hat{B}$ against the quench of
any other parameters. 
Moreover, 
for the generalized 
susceptibilities (crossed susceptibilities) 
our theorem gives a necessary and sufficient condition for the
consistency between quantum mechanics and thermodynamics.
As demonstrations of the theorem,
we present numerical results for three models of spin systems.

These results dramatically reduce the costs
of experiments 
because 
a single quench experiment on the key observable in a timescale of 
$\mathcal{O}(|\Delta f|^0)$
gives rich information about 
all additive observables, a longer time scale $o(1/\sqrt{|\Delta f|})$,
and the quench of other parameters.

The paper is organized as follows.
Section~\ref{sec:Setup} explains the setup.
Section~\ref{sec:Criteria} defines linear thermalization
by introducing three criteria.
The theorem (main result) and two propositions 
are summarized in Sec.~\ref{sec:SummaryResults}.
We discuss the timescale of  linear thermalization
in Sec.~\ref{sec:Timescale},
and prove the third proposition about a longer timescale.
Generalized susceptibilities are analyzed in
Sec.~\ref{sec:GeneralizedSuscep},
where two corollaries are presented.
Numerical demonstrations are given in Sec.~\ref{sec:Numerical}.
We prove the theorem and propositions in Sec.~\ref{sec:Results}.
In Sec.~\ref{sec:Discussion}, 
we show that our results are also applicable to the case of continuous change of $f$, 
and explain a relation between our main result and the ETH.
Section~\ref{sec:Summary} summarizes the paper.

\section{\label{sec:Setup}Setup}

We study
an isolated quantum many-body system
that is defined on a lattice with $N$ sites and
described by a finite dimensional Hilbert space~\footnote{
For boson systems, the Hilbert space can be made finite dimensional
by fixing the total particle number
or
by introducing an appropriate cutoff.
}.
The system obeys the unitary time evolution
generated by the Hamiltonian $\hat{H}(f)$, 
which depends on a physical parameter $f$~\footnote{
We assume that all matrix elements of $\hat{H}(f)$ are twice differentiable.
}
such as
an external magnetic field.
We use units where the reduced Planck constant $\hbar$ and the Boltzmann constant $k_{\text{B}}$ are unity.

We investigate 
the system prepared in an equilibrium state
and its time evolution induced by a change of $f$.
We specifically consider a \emph{quench} process, 
in which $f$ is changed discontinuously.
This does not mean any loss of generality because 
the same results can also be obtained for 
continuous change of $f$, as shown in Sec.~\ref{sec:Nonquench}.

Since we are interested in the consistency with thermodynamics,
we consider the case 
where each equilibrium state of the system is uniquely specified macroscopically 
by an appropriate set of variables. 
Here, the number of the variables in the set is finite and independent of $N$.
(This is one of the basic assumptions of thermodynamics~\footnote{
If one lifted this basic assumption of thermodynamics~\cite{Callen1985,Ruelle1969},
as in the generalized Gibbs ensemble,
almost all 
systems including integrable systems would satisfy \emph{Criterion}~(i)
given in Sec.~\ref{sec:Criteria}~\cite{Sirker2014,Doyon2017}.
}.)
To be specific,
we here assume that 
$f$, $N$, and the inverse temperature $\beta$ is such a set of variables.
Then,
equilibrium states can be represented by
the canonical Gibbs state
\begin{align}
\hat{\rho}^{\text{can}}(\beta,f):=e^{-\beta \hat{H}(f)}/Z(\beta,f),
\end{align}
where $Z(\beta,f)$ is the partition function.

For time $t<0$, $f$ takes a constant 
value $f_{0}$,
which defines the initial Hamiltonian, 
\begin{align}
\hat{H}_{0}:=\hat{H}(f_{0}),
\end{align}
and the system 
is in an equilibrium state 
at a finite inverse temperature $\beta_{0}$,
represented by
\begin{align}
\hat{\rho}^{\text{eq}}_{0}:=\hat{\rho}^{\text{can}}(\beta_{0},f_{0}).
\label{eq:rho_0}
\end{align}

\begin{figure}
\includegraphics[width=\linewidth]{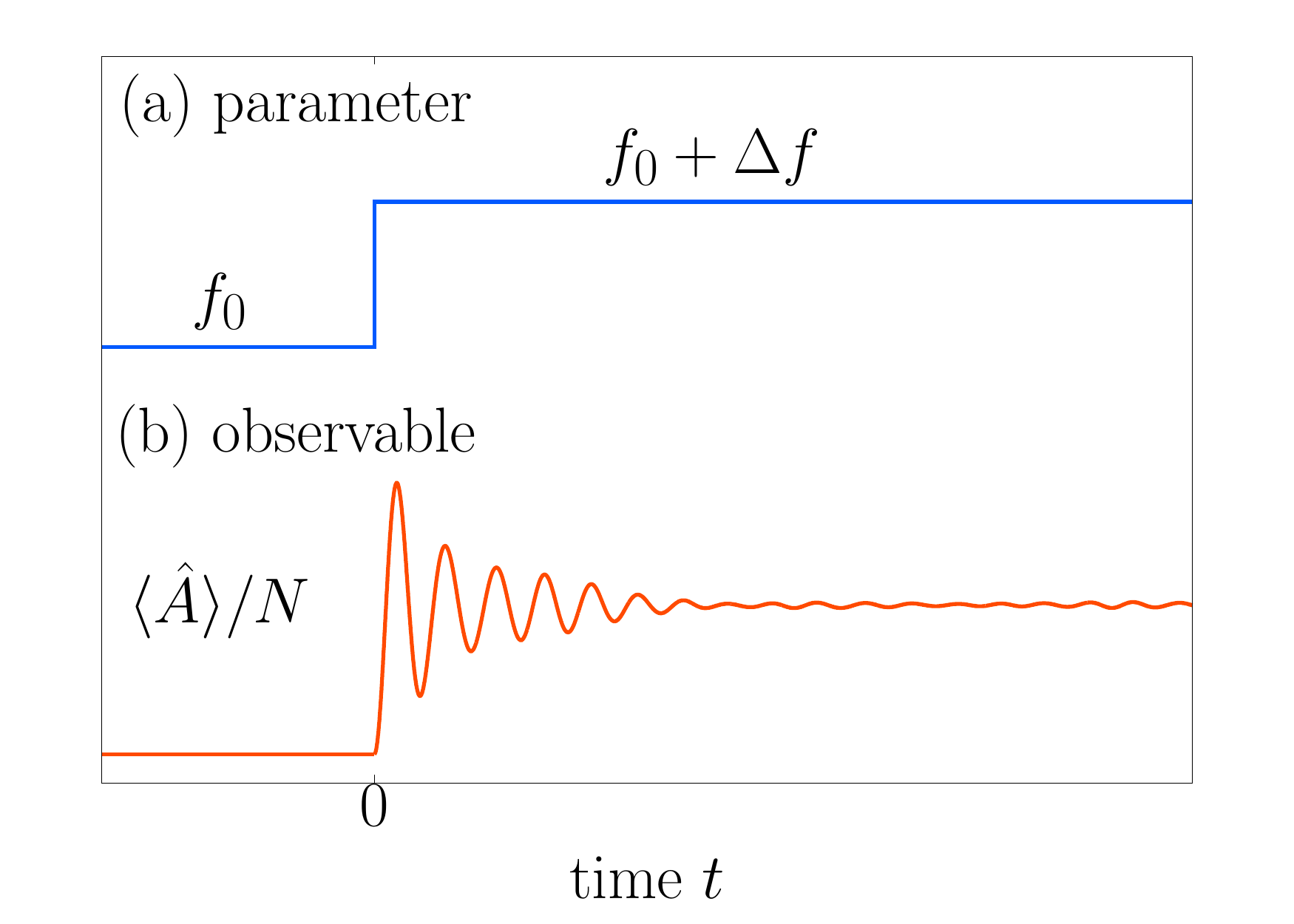}
\caption{\label{fig:Setup_TEvolution}%
(a) The quench process considered in this paper.
See Sec.~\ref{sec:Setup} for details.
(b) A schematic plot of 
typical time evolution of 
the expectation value of 
an additive observable per site
after the quench at $t=0$.
}
\end{figure}

At $t=0$,
$f$ is changed
from $f_{0}$ to another constant value $f_{0}+\Delta f$ discontinuously,
as shown in Fig.~\ref{fig:Setup_TEvolution}(a).
Due to this quench of $f$, the state for $t>0$ evolves according to
the Schr\"{o}dinger equation described by
the postquench Hamiltonian 
\begin{align}
\hat{H}_{\Delta f}:=\hat{H}(f_{0}+\Delta f).
\end{align}
Consequently the expectation values of additive observables
evolve in time,
as shown in Fig.~\ref{fig:Setup_TEvolution}(b).
Here,
we say an observable 
$\hat{A}$ is \emph{additive} when it is the sum of 
local observables $\hat{a}_{\vb*{r}}$
over the whole system, 
\begin{align}
\hat{A} = \sum_{\vb*{r}} \hat{a}_{\vb*{r}}, 
\label{eq:Additive}
\end{align}
where we say an observable $\hat{a}_{\vb*{r}}$ is \emph{local} 
when its support consists of sites 
within 
a distance of at most $\mathcal{O}(N^0)$ from the site $\vb*{r}$.

Note that a local observable $\hat{a}_{\vb*{r}}$
 is \emph{not} necessarily a one-body observable.
It can be, say, a two-spin observable 
such as the one given by Eq.~(\ref{eq:Mzz}) below.
Note also that 
the additive observable $\hat{A}$ can be noninvariant
under a spatial translation.
For example, 
for a magnetic field of a wavenumber $\vb*{k}$ with magnitude $f$,
\begin{align}
h({\vb*{r}})= f \sin({\vb*{k}} \cdot {\vb*{r}}),
\label{eq:h=fsin(kr)}
\end{align}
its interaction with spins,
\begin{align}
- \sum_{\vb*{r}} h({\vb*{r}}) \hat{\sigma}^z_{\vb*{r}}
=- f \sum_{\vb*{r}} \sin({\vb*{k}} \cdot {\vb*{r}})\hat{\sigma}^z_{\vb*{r}},
\label{eq:-h_sigma^z}
\end{align}
is an additive observable with 
$\hat{a}_{\vb*{r}}= -h({\vb*{r}}) \hat{\sigma}^z_{\vb*{r}}$.
This example also shows that
our theory is applicable to the case where a spatially-varying external field
is applied.
Since we allow such an $\vb*{r}$-dependent coefficient in $\hat{a}_{\vb*{r}}$,
we exclude strange cases where the operator norm $\| \hat{a}_{\vb*{r}} \|_{\infty}$ diverges 
as 
$|\vb*{r}|\to\infty$
by imposing 
\begin{align}
\| \hat{a}_{\vb*{r}} \|_{\infty} \le C_{A} \quad\text{for all }\vb*{r},
\label{eq:Bound_Norm}
\end{align}
where $C_{A}$ is a constant independent of $N$ and $\vb*{r}$.
From this restriction, any additive observable $\hat{A}$ satisfies 
\begin{align}
\|\hat{A}\|_{\infty}=\mathcal{O}(N).
\end{align}

We also make a natural assumption that $\hat{H}(f)$ and its derivative,
\begin{align}
\hat{B} :=
-\frac{\partial \hat{H}}{\partial f}(f_{0}),
\label{eq:def_B}
\end{align}
are additive observables.
Indeed this assumption is satisfied
for the above example of $\hat{H}(f)$
[see also Eq.~(\ref{eq:B_for_simple_example}) below].
We say that
the observable $\hat{B}$ is \emph{conjugate} to the parameter $f$,
and vice versa.

We examine whether the system 
relaxes to the equilibrium state
for small $|\Delta f|$.
For this purpose, 
we consider the case where 
an equilibrium state exists
not only for $f=f_0$ but also for $f=f_0 +\Delta f$.
That is, we exclude, for instance,
electric conductors in a uniform electric field.
Furthermore, 
to exclude uninteresting divergences, 
we also limit ourselves to the case where 
no phase transition occurs in the neighborhood of the initial equilibrium state.
This implies that
the initial equilibrium state is stable
against
the changes of parameters conjugate to 
any additive observables.
(This condition is precisely expressed
by Eq.~(\ref{eq:OutOfEquilibrium}) in Sec.~\ref{sec:Exp_Ave}.)

\section{\label{sec:Criteria}Criteria for linear thermalization}

Since we consider the case of small $|\Delta f|$, 
we examine the consistency with 
thermodynamics 
up to the linear order in $\Delta f$:
We say
\emph{linear thermalization} occurs for an additive observable $\hat{A}$
if the following three criteria are satisfied up to $\mathcal{O}(\Delta f)$.

\emph{Criterion}~(i)
(consistency of expectation value): 
The long time average of the expectation value of $\hat{A}$
after the quench
is sufficiently close to
the equilibrium value.
More precisely,
\begin{align}
\overline{\langle \hat{A}^{\Delta f}(t)\rangle^{\text{eq}}_{0}}^{\mathcal{T}}
-
\langle \hat{A}\rangle^{\text{eq}}_{\Delta f}
=o(N)
\label{eq:Question_Exp_Ave}
\end{align}
for sufficiently long time $\mathcal{T}$.
(See Sec.~\ref{sec:Timescale}
for detailed discussions on $\mathcal{T}$.)
Here
\begin{align}
\hat{A}^{\Delta f}(t):=e^{i\hat{H}_{\Delta f}t}\hat{A}e^{-i\hat{H}_{\Delta f}t}
\end{align}
is the Heisenberg operator of $\hat{A}$
evolved by $\hat{H}_{\Delta f}$,
and,
for any $t$-dependent quantity $g(t)$,
\begin{align}
\overline{g(t)}^{\mathcal{T}}
:=
\frac{1}{\mathcal{T}}\int_{0}^{\mathcal{T}}\dd{t}
g(t).
\label{eq:DEF_FiniteTave}
\end{align}
Furthermore,
$\langle\bullet\rangle^{\text{eq}}_{0}:=\Tr\bigl[\hat{\rho}^{\text{eq}}_{0}\bullet\bigr]$
is the expectation value
in the initial state
$\hat{\rho}^{\text{eq}}_{0}$,
while
$\langle\bullet\rangle^{\text{eq}}_{\Delta f}:=\Tr\bigl[\hat{\rho}^{\text{eq}}_{\Delta f}\bullet\bigr]$
is the expectation value
in
\begin{align}
\hat{\rho}^{\text{eq}}_{\Delta f}:=\hat{\rho}^{\text{can}}(\beta_{\Delta f},f_{0}+\Delta f).
\label{eq:rho_Df}
\end{align}
The latter
state $\hat{\rho}^{\text{eq}}_{\Delta f}$ represents the 
\emph{final} equilibrium state predicted by thermodynamics.
Its inverse temperature $\beta_{\Delta f}$
is determined from energy conservation
\begin{align}
\langle\hat{H}_{\Delta f}\rangle_{\Delta f}^{\text{eq}}
=\langle\hat{H}_{\Delta f}\rangle_{0}^{\text{eq}}.
\label{eq:EnergyConservation}
\end{align}
Note that 
$\beta_{\Delta f} \neq \beta_0$ 
even in $\mathcal{O}(\Delta f)$,
as explicitly given by Eq.~(\ref{eq:beta_Df}) 
in Appendix~\ref{sec:Derivation_Exp_Ave}.

\emph{Criterion}~(ii) (equilibration): 
At almost all $t>0$,
the expectation value of $\hat{A}$ 
is sufficiently close to its long time average.
That is, 
their difference is macroscopically negligible in the sense that
\begin{align}
\overline{\Bigl|
\langle \hat{A}^{\Delta f}(t)\rangle^{\text{eq}}_{0}
-\overline{\langle \hat{A}^{\Delta f}(t)\rangle^{\text{eq}}_{0}}^{\mathcal{T}}
\Bigr|^2}^{\mathcal{T}}
=o(N^2)
\label{eq:Question_Exp_Fluc}
\end{align}
for sufficiently long time $\mathcal{T}$.

\emph{Criterion}~(iii)
(smallness of variance): 
At almost all $t>0$,
the variance of $\hat{A}$
is macroscopically negligible such that
\begin{align}
\overline{\text{Var}_{0}[\hat{A}^{\Delta f}(t)]}^{\mathcal{T}}
=o(N^2)
\label{eq:Question_Var}
\end{align}
for sufficiently long time $\mathcal{T}$.
Here
$\text{Var}_{0}[\hat{\bullet}]
:=\langle(\hat{\bullet})^2\rangle^{\text{eq}}_{0}
-(\langle\hat{\bullet}\rangle^{\text{eq}}_{0})^2$.

Thermodynamics assumes that 
\emph{all} additive observables take macroscopically definite values in an equilibrium state~\cite{Landau1980,Oono2017,Shimizu2021}.
Since we are interested in full consistency with thermodynamics,
we examine whether linear thermalization occurs, 
i.e., whether these criteria are satisfied,
for \emph{all} additive observables.

\section{\label{sec:SummaryResults}Summary of results}

Among three criteria of the previous section,
\emph{Criterion}~(i) 
has been studied most intensively
because 
in many cases
it discriminates thermalizing systems from nonthermalizing ones.
For instance, 
it was observed in many integrable systems
that relaxation to some steady states occurs,
indicating that \emph{Criteria}~(ii) and (iii) are satisfied,
whereas 
they are nonthermal states,
which do not satisfy \emph{Criterion}~(i)~\cite{Rigol2007,Pozsgay2014,Fagotti2014,Vidmar2016,Gring2012,AduSmith2013,Langen2015}.

Therefore
we place a theorem about \emph{Criterion}~(i) as our main result.
We also obtain additional results which show that 
\emph{Criteria}~(ii) and (iii) are easily satisfied
in our setting, 
under conditions
weaker than those of previous studies~\cite{Short2012,Srednicki1996,Srednicki1999}.
These results are summarized in this section.
We will extend them slightly in Sec.~\ref{sec:Timescale},
and thereby discuss the timescale of the linear thermalization.

\subsection{\label{sec:Summary_Theorem}Theorem for \emph{Criterion}~(i)}

Our main result is that 
if the additive observable $\hat{B}$ conjugate to $f$, 
given by Eq.~(\ref{eq:def_B}), 
satisfies \emph{Criterion}~(i) of Sec.~\ref{sec:Setup}
up to $\mathcal{O}(\Delta f)$, then so do \emph{all} additive observables.
To be precise, we obtain
\\
\emph{Theorem} (main result): 
\begin{align}
\lim_{\mathcal{T} \to \infty}
\lim_{\Delta f\to 0}
\frac{
\overline{\langle \hat{A}^{\Delta f}(t)\rangle^{\text{eq}}_{0}}^{\mathcal{T}}
-\langle \hat{A}\rangle^{\text{eq}}_{\Delta f}}{\Delta f}
=o(N)
\label{eq:Main_Exp_Ave}
\end{align}
holds for every additive observable $\hat{A}$
if and only if
it holds for $\hat{A}=\hat{B}$,
\begin{align}
\lim_{\mathcal{T} \to \infty}
\lim_{\Delta f\to 0}
\frac{
\overline{\langle \hat{B}^{\Delta f}(t)\rangle^{\text{eq}}_{0}}^{\mathcal{T}}
-\langle \hat{B}\rangle^{\text{eq}}_{\Delta f}}{\Delta f}
=o(N).
\label{eq:Cond_Exp_Ave}
\end{align}

This theorem identifies $\hat{B}$,
which is conjugate to $f$, 
 as the key observable for linear thermalization of all additive observables.
That is, 
one can judge
whether
the long time average of
the expectation value of
\emph{every} additive observable $\hat{A}$ is consistent with
thermodynamics
by examining the \emph{single} observable $\hat{B}$.
This striking result has the following 
significant features.

Firstly, it 
is in sharp contrast with the existing approaches to 
\emph{Criterion}~(i)~\cite{Rigol2008,Kim2014,Beugeling2014,Steinigeweg2014,Khodja2015,Sirker2014},
such as the ETH.
In fact, previous studies clarified that the ETH for 
an 
observable implies \emph{Criterion}~(i) for that observable.
However,
it did not guarantee
either the ETH or \emph{Criterion}~(i) for other observables.
By contrast,
our \emph{Theorem} does guarantee that
\emph{Criterion}~(i) is satisfied by all additive 
observables
up to $\mathcal{O}(\Delta f)$
if it is satisfied by 
$\hat{B}$.
(We will demonstrate this point numerically in Sec.~\ref{sec:Numerical}.)
This result may be most surprising in the case
where $\hat{B}$ is a one-body observable:
If 
Eq.~(\ref{eq:Cond_Exp_Ave})
holds for that one-body observable,
then it also holds for all other additive observables
including two- and more-body ($\mathcal{O}(N^0)$-body) ones.

Secondly, 
the ETH for $\hat{B}$ implies condition~(\ref{eq:Cond_Exp_Ave})
but the converse is not necessarily true,
as will be discussed in Sec.~\ref{sec:ETH_imlies_Cond_Exp_Ave}.

Thirdly, 
our \emph{Theorem} has significant meanings about the generalized susceptibilities for cross responses, as will be discussed in Sec.~\ref{sec:GeneralizedSuscep}.

Fourthly,
we can extend the result
such that
$\hat{A}$ and $\hat{B}$ in Eqs.~(\ref{eq:Main_Exp_Ave}) and (\ref{eq:Cond_Exp_Ave}) are not restricted to additive observables,
as will be discussed in 
Sec.~\ref{sec:ExtensionNonadditive}.

\subsection{\label{sec:Summary_Proposition1}Proposition for \emph{Criterion}~(ii)}

Under the reasonable condition
that $\hat{H}_{0}$ does not have exponentially 
many resonances,
all additive observables satisfy
\emph{Criterion}~(ii)
up to $\mathcal{O}(\Delta f)$.
That is, we obtain 
\\
\emph{Proposition}~$1$: 
If the maximum number of resonances $D_{\text{res}}$ 
[defined by Eq.~(\ref{def:Dres}) in Sec.~\ref{sec:Exp_Fluc}]
satisfies 
\begin{align}
D_{\text{res}}
\Tr \bigl[(\hat{\rho}^{\text{eq}}_{0})^2\bigr]
= o(1/N^2),
\label{eq:Assumption_Resonances}
\end{align}
then
\begin{align}
\lim_{\mathcal{T} \to \infty}
\lim_{\Delta f\to 0}
\overline{\biggl|\frac{
\langle \hat{A}^{\Delta f}(t)\rangle^{\text{eq}}_{0}
-\overline{\langle \hat{A}^{\Delta f}(t)\rangle^{\text{eq}}_{0}}^{\mathcal{T}}
}{\Delta f}\biggr|^2}^{\mathcal{T}}
=o(N^2)
\label{eq:Main_Exp_Fluc}
\end{align}
for every additive observable $\hat{A}$.
[A detailed expression of the right-hand side (r.h.s) will be given 
in Sec.~\ref{sec:Exp_Fluc}.]

This proposition is obtained by adapting
the argument of Short and Farrelly~\cite{Short2012}
to our setting.
While their result contains the effective dimension of the initial state,
our condition~(\ref{eq:Assumption_Resonances}) instead 
contains its purity $\Tr \bigl[(\hat{\rho}^{\text{eq}}_{0})^2\bigr]$.
Since the postquench Hamiltonian $\hat{H}_{\Delta f}$ is different from the initial Hamiltonian $\hat{H}_{0}$,
evaluation of the effective dimension of the initial state $\hat{\rho}^{\text{eq}}_{0}$
in terms of $\hat{H}_{\Delta f}$ is not an easy task. 
By contrast,
its purity 
can be evaluated easily as shown in Appendix~\ref{sec:Derivation_Exp_Fluc}.

Note that 
our condition
is weaker than 
the ``nonresonance condition,"
$D_{\text{res}}=1$,
of the previous studies~\cite{Tasaki1998,Reimann2008,Linden2009}.
The nonresonance condition often fails even in nonintegrable systems,
e.g., when energy eigenvalues have degeneracies due to symmetries.
By contrast, 
condition~(\ref{eq:Assumption_Resonances}) is expected to hold
not only in such systems but also in wider classes of systems,
including 
interacting integrable systems
(such as Model III of Sec.~\ref{sec:Numerical}),
because 
\begin{align}
\Tr \bigl[(\hat{\rho}^{\text{eq}}_{0})^2\bigr]
=e^{-\Theta(N)}
\label{eq:Exp_Fluc_3}
\end{align}
at any nonzero temperature $1/\beta_0$
as shown in Appendix~\ref{sec:Derivation_Exp_Fluc}.

\subsection{\label{sec:Summary_Proposition2}Proposition for \emph{Criterion}~(iii)}

Under the reasonable condition that 
the fluctuation of every additive observable in the initial equilibrium state
is sufficiently small,
the variances of all additive observables after the quench remain
small enough
such that
\emph{Criterion}~(iii) is satisfied 
up to $\mathcal{O}(\Delta f)$.
To be precise, we find
\\
\emph{Proposition}~$2$: 
If 
the fourth order central moment
of every additive observable $\hat{A}$
in the initial state satisfies
\begin{align}
\langle (\hat{A} - \langle \hat{A} \rangle^{\text{eq}}_{0} )^4 \rangle^{\text{eq}}_{0}
=\mathcal{O}(N^2),
\label{eq:Assumption_4thMoment}
\end{align}
then 
\begin{align}
\lim_{\mathcal{T}\to\infty}\lim_{\Delta f\to 0}
\left|
\frac{
\overline{\text{Var}_{0}[\hat{A}^{\Delta f}(t)]}^{\mathcal{T}}
-\text{Var}_{0}[\hat{A}]
}{\Delta f}
\right|
=\mathcal{O}\bigl(N^{\frac{3}{2}}\bigr)
\label{eq:Main_Var_Ave}
\end{align}
for every additive observable $\hat{A}$.
[A detailed expression of the r.h.s will be given as 
Eq.~(\ref{eq:Main_Var_Ave_Explicit}) in Sec.~\ref{sec:Var}.]

Note that condition~(\ref{eq:Assumption_4thMoment}) also bounds 
the variance 
in the initial state as 
\begin{align}
\text{Var}_{0}[\hat{A}]
\le\sqrt{\langle (\hat{A} - \langle \hat{A} \rangle^{\text{eq}}_{0} )^4 \rangle^{\text{eq}}_{0}}
=\mathcal{O}(N),
\label{eq:Var=O(N)}
\end{align}
which 
follows from the Cauchy-Schwarz inequality.
By inserting this into 
Eq.~(\ref{eq:Main_Var_Ave}), 
we have~\footnote{
The order of three limits (regarding $N$, $\mathcal{T}$ and $\Delta f$)
in Eq.~(\ref{eq:Main_Var_2})
should be understood as in Eq.~(\ref{eq:Main_Var_Ave}).
}
\begin{align}
\overline{\text{Var}_{0}[\hat{A}^{\Delta f}(t)]}^{\mathcal{T}}
=
\mathcal{O}(N)
+
\Delta f \times \mathcal{O}\bigl(N^{\frac{3}{2}}\bigr)
+ o(\Delta f),
\label{eq:Main_Var_2}
\end{align}
which shows that \emph{Criterion}~(iii) is satisfied
up to $\mathcal{O}(\Delta f)$.

Since we exclude phase transition points,
our condition~(\ref{eq:Assumption_4thMoment})
seems plausible.
By contrast, 
previous results on \emph{Criterion}~(iii)~\cite{Srednicki1996,DAlessio2016}
required a condition that is stronger than the ETH.
The above proposition shows that 
the criterion is satisfied
under a much weaker condition in our setting.

To sum up 
these theorem and propositions,
linear thermalization of the single key observable $\hat{B}$ guarantees
linear thermalization of all additive observables,
under physically reasonable conditions.

\section{\label{sec:Timescale}Timescale of linear thermalization}

When studying thermalization theoretically,
it is customary to take the limit $\mathcal{T} \to \infty$~\cite{DAlessio2016,Mori2018}, 
as we did above.
However, more detailed information about the timescale is 
necessary because in experiments thermalization occurs in 
reasonably short timescales~\cite{Trotzky2012,Orioli2018}.
The timescale is particularly nontrivial when 
``prethermalization''~\cite{Berges2004,Moeckel2008,Moeckel2009,Moeckel2010,Kollar2011,Mori2018,Mallayya2019} occurs, i.e., when  
the system first relaxes to a nonthermal quasi-steady state,
and at some time stage after that, 
it relaxes to a true thermal equilibrium state.
In 
nearly integrable systems, the
timescale for the relaxation to a true thermal equilibrium state
crucially depends on the magnitude of $\Delta f$ 
and is typically of $\Theta(1/|\Delta f|^2)$~\cite{Pauli1928,VanHove1955,VanHove1956,VanVliet1977}.
Thus 
the $\Delta f$ dependence of the timescale 
is very important.
In this section, 
we investigate it for
linear thermalization.

\subsection{\label{sec:TimescaleOurResults}Timescale of linear thermalization in \emph{Theorem} and \emph{Propositions} $1$ and $2$}

\begin{figure}
\includegraphics[width=\linewidth]{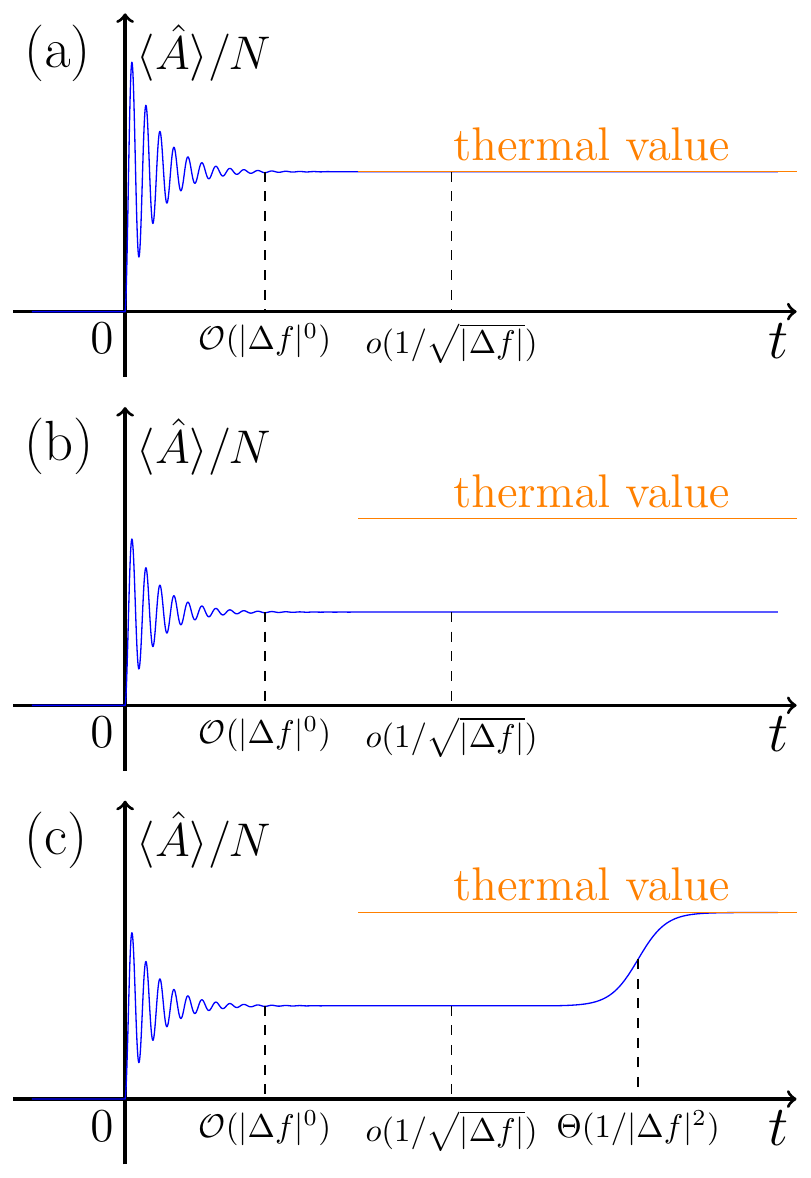}
\caption{\label{fig:Timescale}%
Schematic plots of the time evolution 
of $\langle\hat{A}^{\Delta f}(t)\rangle^{\text{eq}}_{0}/N$
in three typical cases.
(a)~Linear thermalization occurs, 
corresponding to cases ($1$) of Sec.~\ref{sec:TimescaleOurResults} 
and ($1^{\prime}$) of \ref{sec:LongerTimescale}.
(b)~Linear thermalization does not occur, 
corresponding to cases ($2$) of Sec.~\ref{sec:TimescaleOurResults} 
and ($2^{\prime}$) of \ref{sec:LongerTimescale}. 
(c)~Prethermalization occurs,
which is discussed in Sec.~\ref{sec:Prethermalization}.
}
\end{figure}

Our results of Sec.~\ref{sec:SummaryResults}, 
namely
\emph{Theorem} and \emph{Propositions}~$1$ and $2$, 
take the two limits
$\mathcal{T}\to\infty$ and $\Delta f\to 0$
in the following order~\footnote{
Note that $N\to\infty$ is taken after $\mathcal{T}\to\infty$ throughout this paper.
Such an order of limits is standard 
and is taken in most studies on thermalization~\cite{Sirker2014,DAlessio2016,Mori2018}.
}:
\begin{align}
\lim_{\mathcal{T}\to\infty}\lim_{\Delta f\to 0}
[\mbox{function of $\mathcal{T}$ and $\Delta f$]}.
\label{eq:Lim_T_Lim_Df}
\end{align}
In this order of limits, 
$\mathcal{T}$ is smaller than
any timescale that grows as 
$\Delta f \to 0$~\footnote{If the two limits are taken in the reverse order,
then $\mathcal{T}$ is much larger than
any timescale that depends on $\Delta f$.}.
That is, it 
extracts 
the behavior 
of the system 
in the time interval $[0, \mathcal{T}]$ 
such that $\mathcal{T}=\Theta(|\Delta f|^0)$.
This leads to 
the following observations:\\
\emph{Timescale in Theorem and Propositions $1$ and $2$: }
($1$) If linear thermalization occurs 
in the limit~(\ref{eq:Lim_T_Lim_Df}),
as in \emph{Theorem} and \emph{Propositions}~$1$ and $2$,
then 
it occurs
in some $t = \mathcal{O}(|\Delta f|^0)$,
as illustrated in Fig.~\ref{fig:Timescale}(a).
($2$) On the other hand, 
if linear thermalization does not occur in the limit~(\ref{eq:Lim_T_Lim_Df}),
then 
it does not occur
in any timescale of $\mathcal{O}(|\Delta f|^0)$,
as in Fig.~\ref{fig:Timescale}(b).

That is, the timescale of thermalization is independent of the magnitude of $\Delta f$.
Behaviors in longer timescales will be 
discussed in the following two subsections.

\subsection{\label{sec:LongerTimescale}Extension to a longer timescale}

We now extend our results 
to a longer timescale
$o(1/\sqrt{|\Delta f|})$,
and explain its implications for
linear thermalization.

Let $\mathcal{T}_{\Delta f}$ be a timescale that grows as $\Delta f\to 0$
satisfying
$\mathcal{T}_{\Delta f}=o(1/\sqrt{|\Delta f|})$.
As will be shown in Sec.~\ref{sec:Proof_Timescale},
we find that the values of the left-hand sides
of Eqs.~(\ref{eq:Main_Exp_Ave}), (\ref{eq:Main_Exp_Fluc})
and (\ref{eq:Main_Var_Ave})
do not change 
when 
the limit~(\ref{eq:Lim_T_Lim_Df})
is replaced with
\begin{align}
\lim_{\Delta f\to 0}
[\mbox{function of $\mathcal{T}$ and $\Delta f$ with $\mathcal{T}$ 
replaced by $\mathcal{T}_{\Delta f}$]}.
\label{eq:Lim_Df_LongerT}
\end{align}
That is, we obtain the following proposition~\footnote{
We here take $\mathcal{T}_{\Delta f}=o(1/\sqrt{|\Delta f|})$
in order to keep the right-hand sides of inequalities~(\ref{eq:finite_Df_T_Exp_Ave}), (\ref{eq:finite_Df_T_Exp_Fluc}) and (\ref{eq:finite_Df_T_Var}) 
sufficiently small.
Note that these inequalities 
are rigorous but not so tight.
We expect that tighter estimations will allow to
investigate a longer timescale, 
which will be a subject of future studies.
}.\\
\emph{Proposition}~3: 
For any $\mathcal{T}_{\Delta f}=o(1/\sqrt{|\Delta f|})$
and for any additive observable $\hat{A}$,
\begin{align}
&\lim_{\Delta f\to 0}
\frac{
\overline{\langle \hat{A}^{\Delta f}(t)\rangle^{\text{eq}}_{0}}^{\mathcal{T}_{\Delta f}}
-\langle \hat{A}\rangle^{\text{eq}}_{\Delta f}}{\Delta f}
\notag\\
&=\lim_{\mathcal{T}\to \infty}\lim_{\Delta f\to 0}
\frac{
\overline{\langle \hat{A}^{\Delta f}(t)\rangle^{\text{eq}}_{0}}^{\mathcal{T}}
-\langle \hat{A}\rangle^{\text{eq}}_{\Delta f}}{\Delta f},
\label{eq:Main_Exp_Ave_longer}
\end{align}
\begin{align}
&\lim_{\Delta f\to 0}
\overline{\biggl|\frac{
\langle \hat{A}^{\Delta f}(t)\rangle^{\text{eq}}_{0}
-\overline{\langle \hat{A}^{\Delta f}(t)\rangle^{\text{eq}}_{0}}^{\mathcal{T}_{\Delta f}}
}{\Delta f}\biggr|^2}^{\mathcal{T}_{\Delta f}}
\notag\\
&=\lim_{\mathcal{T} \to \infty}
\lim_{\Delta f\to 0}
\overline{\biggl|\frac{
\langle \hat{A}^{\Delta f}(t)\rangle^{\text{eq}}_{0}
-\overline{\langle \hat{A}^{\Delta f}(t)\rangle^{\text{eq}}_{0}}^{\mathcal{T}}
}{\Delta f}\biggr|^2}^{\mathcal{T}},
\label{eq:Main_Exp_Fluc_longer}
\end{align}
\begin{align}
&\lim_{\Delta f\to 0}
\frac{
\overline{\text{Var}_{0}[\hat{A}^{\Delta f}(t)]}^{\mathcal{T}_{\Delta f}}
-\text{Var}_{0}[\hat{A}]
}{\Delta f}
\notag\\
&=\lim_{\mathcal{T}\to\infty}\lim_{\Delta f\to 0}
\frac{
\overline{\text{Var}_{0}[\hat{A}^{\Delta f}(t)]}^{\mathcal{T}}
-\text{Var}_{0}[\hat{A}]
}{\Delta f}.
\label{eq:Main_Var_Ave_longer}
\end{align}

This proposition shows
that 
\emph{Theorem}, \emph{Propositions}~1 and 2
hold
even when 
$\mathcal{T}$ is replaced with a longer timescale $\mathcal{T}_{\Delta f}$,
and leads to the following observations:\\
\emph{Timescale of linear thermalization: }
($1^{\prime}$) If linear thermalization occurs 
in some $t = \mathcal{O}(|\Delta f|^0)$,
it lasts at least for a period of $o(1/\sqrt{|\Delta f|})$.
Conversely, 
if linear thermalization occurs 
at some $o(1/\sqrt{|\Delta f|})$, 
it already occurs at some
$t = \mathcal{O}(|\Delta f|^0)$.
See Fig.~\ref{fig:Timescale}(a).
($2^{\prime}$) On the other hand, 
absence of linear thermalization in a shorter timescale of $\mathcal{O}(|\Delta f|^0)$
guarantees
its absence in a longer timescale of $o(1/\sqrt{|\Delta f|})$, and vice versa.
See Fig.~\ref{fig:Timescale}(b).

This result, together with the results of Secs.~\ref{sec:SummaryResults} and \ref{sec:TimescaleOurResults},
implies, in particular, that
linear thermalization of the single key observable $\hat{B}$ 
in the timescale of $\mathcal{O}(|\Delta f|^0)$
guarantees linear thermalization of all additive observables
not only in the same timescale but also in a longer timescale of $o(1/\sqrt{|\Delta f|})$.

\subsection{\label{sec:Prethermalization} Stationarity and implications for prethermalization}

We here discuss 
stationarity of the system 
and 
its implication for prethermalization.

From Eqs.~(\ref{eq:Main_Exp_Ave_longer})--(\ref{eq:Main_Var_Ave_longer}),
we 
obtain the following observations:
\\
\emph{Stationarity throughout a certain time region}:
Suppose that
conditions~(\ref{eq:Assumption_Resonances}) and (\ref{eq:Assumption_4thMoment}) for \emph{Propositions}~1 and 2 are fulfilled.
Then,
throughout a time region from $\mathcal{O}(|\Delta f|^0)$ to $o(1/\sqrt{|\Delta f|})$,
all 
additive observables take macroscopically 
definite and stationary values,
up to $\mathcal{O}(\Delta f)$.
In other words,
the system relaxes to a macroscopic state
and stays in the same macroscopic state
throughout this time region,
up to $\mathcal{O}(\Delta f)$.

Note that this stationary 
state can be either thermal or nonthermal,
depending on whether condition~(\ref{eq:Cond_Exp_Ave}) of
our \emph{Theorem} is satisfied or not.
If 
the state is thermal as illustrated
in Fig.~\ref{fig:Timescale}(a),
it means that linear thermalization occurs.
On the other hand,
if the state is nonthermal 
[i.e., if condition~(\ref{eq:Cond_Exp_Ave}) is not satisfied]
as in Fig.~\ref{fig:Timescale}(b), 
it is a nonthermal stationary state.

The latter case 
includes
systems which exhibit 
prethermalization~\cite{Berges2004,Moeckel2008,Moeckel2009,Moeckel2010,Kollar2011,Mori2018,Mallayya2019}.
The prethermalization 
often occurs 
when $\Delta f$ switches the system from integrable to nonintegrable.
(Shiraishi proved the existence of a system in which 
such switching is possible 
by an arbitrary nonvanishing value of $\Delta f$~\cite{Shiraishi2019}.)
In a typical case of such prethermalization,
the system first relaxes to a nonthermal stationary state
in some timescale of $\mathcal{O}(|\Delta f|^0)$,
and then relaxes to the true thermal equilibrium state
at some timescale of $\Theta(1/|\Delta f|^2)$~\cite{Mori2018},
as illustrated in Fig.~\ref{fig:Timescale}(c).
For such a system, 
our results 
detect
the nonthermal stationary 
state
in the time region 
from $\mathcal{O}(|\Delta f|^0)$ to $o(1/\sqrt{|\Delta f|})$
and the absence of linear thermalization 
in this time region.

\section{\label{sec:GeneralizedSuscep}Generalized Susceptibilities for Cross Responses}

Our \emph{Theorem} has
significant meanings about 
the generalized susceptibilities for cross responses.

In response to 
the change of a parameter (such as an external field), 
not only its conjugate observable $\hat{B}$ 
but also other observables 
often change their values.
The magnetoelectric effect and 
the piezoelectric effect are well-known examples.
Such responses of observables $\hat{A}$ that are not conjugate to 
the changed parameter
are called \emph{cross responses}, 
and have been attracting much attention~\cite{Onsager1931,Qi2011,Ominato2019,Bender2019,Fischer2022}.
They are
characterized by the \emph{generalized susceptibilities}
(crossed susceptibilities),
which we denote by $\chi(A|B)$.

For example, when 
a magnetic field of an arbitrary wavenumber $\vb*{k}$, 
Eq.~(\ref{eq:h=fsin(kr)}), 
is applied to a spin system, 
Eq.~(\ref{eq:-h_sigma^z}) yields
the additive observable conjugate to $f$ as
\begin{align}
\hat{B} 
=
\sum_{\vb*{r}} \sin({\vb*{k}} \cdot {\vb*{r}})\hat{\sigma}^z_{\vb*{r}},
\label{eq:B_for_simple_example}
\end{align}
which is the total magnetization $\hat{M}_{\vb*{k}}$ of wavenumber $\vb*{k}$.
Then, if we take $\hat{A}$ to be 
the total electric polarization
$\hat{P}_{\vb*{k}}$ of wavenumber $\vb*{k}$ 
the generalized susceptibility $\chi(A|B)$ is the magnetoelectric susceptibility at wavenumber $\vb*{k}$~\footnote{
When the system is an electrical conductor, 
one should keep $\vb*{k} \neq \vb*{0}$ 
(and, if necessary, take the $\vb*{k} \to \vb*{0}$ limit afterwards)
to obtain the response of electric polarization.
Otherwise, one would obtain an electric current rather than the polarization,
as explained in Ref.~\cite{Mahan2000}.
},
whereas if we take $\hat{A}=\hat{B}=\hat{M}_{\vb*{k}}$ the corresponding 
susceptibility $\chi(B|B)$ is just 
the ordinary magnetic susceptibility at wavenumber $\vb*{k}$.

We here compare two types of generalized susceptibilities.
One is $\chi(A|B)$ obtained in quantum mechanics,
\begin{align}
\chi^{\text{QM}}_{N}(A|B)
&:=\lim_{\mathcal{T}\to\infty}\lim_{\Delta f\to 0}
\frac{
\overline{\langle \hat{A}^{\Delta f}(t)\rangle^{\text{eq}}_{0}}^{\mathcal{T}}
-\langle \hat{A}\rangle^{\text{eq}}_{0}
}{\Delta f N},
\label{eq:DEF_chi^QM}
\end{align}
which is defined by 
the Schr\"{o}dinger dynamics.
The other is $\chi(A|B)$ predicted by thermodynamics~\footnote{
Although thermodynamic susceptibility is often defined
by a second derivative of the thermodynamic limit of a certain thermodynamic function,
we expect that it will coincide with the $N\to\infty$ limit of Eq.~(\ref{eq:DEF_chi^TD}),
as is usually expected in equilibrium statistical mechanics.
In other words,
the order of $N\to\infty$ and $\Delta f\to 0$ 
will not matter in 
the definition of $\chi^{\text{TD}}$, Eq.~(\ref{eq:DEF_chi^TD}).},
\begin{align}
\chi^{\text{TD}}_{N}(A|B)
&:=\lim_{\Delta f\to 0}
\frac{
\langle \hat{A}\rangle^{\text{eq}}_{\Delta f}
-\langle \hat{A}\rangle^{\text{eq}}_{0}
}{\Delta f N},
\label{eq:DEF_chi^TD}
\end{align}
where the final state $\hat{\rho}^{\text{eq}}_{\Delta f}$ is determined by thermodynamics and equilibrium statistical mechanics
using the energy conservation, 
Eq.~(\ref{eq:EnergyConservation}).

From these definitions,
Eqs.~(\ref{eq:Main_Exp_Ave}) and (\ref{eq:Cond_Exp_Ave}) are equivalent to
\begin{align}
\lim_{N \to \infty} \chi^{\text{QM}}_{N}(A|B)
&=
\lim_{N \to \infty} \chi^{\text{TD}}_{N}(A|B),
\label{eq:chi^QM(A|B)-chi^TD(A|B)}
\\
\lim_{N \to \infty} \chi^{\text{QM}}_{N}(B|B)
&=
\lim_{N \to \infty} \chi^{\text{TD}}_{N}(B|B),
\label{eq:chi^QM(B|B)-chi^TD(B|B)}
\end{align}
respectively.
Therefore, our \emph{Theorem} can be rephrased as follows.\\
\emph{Consistency of cross responses}: 
Equation~(\ref{eq:chi^QM(A|B)-chi^TD(A|B)}) holds for every additive observable $\hat{A}$ 
if it holds for $\hat{B}$.

Furthermore, the following symmetries follow from Eqs.~(\ref{eq:chi^QM_BogoProd}) 
and (\ref{eq:chi^TD_BogoProd})
in Appendix~\ref{sec:Derivation_Exp_Fluc}:
\begin{align}
\chi^{\text{QM}}_{N}(A|B) &= \chi^{\text{QM}}_{N}(B|A),
\\
\chi^{\text{TD}}_{N}(A|B) &= \chi^{\text{TD}}_{N}(B|A).
\end{align}
(The latter is just a Maxwell relation of thermodynamics.) 
From these relations, Eq.~(\ref{eq:chi^QM(A|B)-chi^TD(A|B)}) 
can be rewritten as
\begin{align}
\lim_{N \to \infty} \chi^{\text{QM}}_{N}(B|A)
&=
\lim_{N \to \infty} \chi^{\text{TD}}_{N}(B|A).
\label{eq:chi^QM(B|A)-chi^TD(B|A)}
\end{align}
Therefore, we obtain the following result.\\
\emph{Corollary}~1
(\emph{consistency of responses to other parameters}):
Equation~(\ref{eq:chi^QM(B|A)-chi^TD(B|A)}) holds for every additive observable $\hat{A}$ if 
Eq.~(\ref{eq:chi^QM(B|B)-chi^TD(B|B)}) holds for $\hat{B}$.

That is,
the quantum-mechanical response of an additive observable $\hat{B}$ to 
the parameter 
$f_A$ conjugate to an \emph{arbitrary} additive observable $\hat{A}$
is consistent with 
thermodynamics if 
the quantum-mechanical response of $\hat{B}$
to its own conjugate parameter 
$f$ is consistent with thermodynamics.

Moreover, this corollary has the following implication for linear thermalization.\\
\emph{Corollary}~2:
Under the conditions~(\ref{eq:Assumption_Resonances}) and (\ref{eq:Assumption_4thMoment}) for \emph{Propositions}~1 and 2,
linear thermalization of $\hat{B}$
against the quench of its conjugate parameter $f$
implies
linear thermalization of the same observable $\hat{B}$
against the quench of any other parameter 
$f_A$ that is conjugate to an \emph{arbitrary} additive observable $\hat{A}$.

These corollaries dramatically reduce
the costs of experiments and 
theoretical calculations 
of linear thermalization and cross responses.
For example, suppose that one wants to examine 
the cross response of
the magnetization $\hat{M}$ of a spin system 
against the quench of an interaction parameter $J$,
whose quench is, however, technically difficult. 
In such a case, 
one can perform an alternative experiment
in which an external magnetic field 
that is conjugate to $\hat{M}$ is quenched. 
If
the quantum-mechanical response of $\hat{M}$ is consistent with thermodynamical one
in the latter experiment,
then \emph{Corollary}~$1$ guarantees their consistency 
in the former experiment.

\section{\label{sec:Numerical}Examples}

In this section, we demonstrate our \emph{Theorem}
using one-dimensional spin systems. 
For a nonintegrable model,
we first 
show linear thermalization of $\hat{B}$,
which implies, according to our \emph{Theorem}, 
linear thermalization of all other $\hat{A}$'s.
We demonstrate it for typical $\hat{A}$'s.
We also present integrable models
in which 
linear thermalization does not occur neither for $\hat{B}$
nor for typical $\hat{A}$'s.

\subsection{\label{sec:Models}Models}

If a system had 
many symmetries,
one would have to 
investigate thermalization separately
in individual symmetry sectors
not to overlook the degeneracy of energy eigenvalues.
To avoid such 
complicated procedures,
we construct our model Hamiltonian by adding two extra terms 
to the 
one-dimensional
XYZ model in a magnetic field~\cite{Shiraishi2019} as 
\begin{align}
\hat{H} &=-\sum_{r=1}^{N}
\Bigl(J_{xx}\hat{\sigma}_{r}^x\hat{\sigma}_{r+1}^x
 +J_{yy}\hat{\sigma}_{r}^y\hat{\sigma}_{r+1}^y
 +J_{zz}\hat{\sigma}_{r}^z\hat{\sigma}_{r+1}^z
 +h_{z}\hat{\sigma}^{z}_{r}
\notag\\
&\hspace{48pt} +J_{yz}\hat{\sigma}_{r}^y\hat{\sigma}_{r+1}^z
 +h_{x}\hat{\sigma}_{r}^x
\Bigr),
\label{eq:Model}
\end{align}
where 
$\hat{\sigma}_{N+1}^{\mu}=\hat{\sigma}_{1}^{\mu}$ ($\mu=x,y,z$).
The last two terms are the extra terms
that break all known symmetries,
except for the translation symmetry, 
of the conventional XYZ model in a magnetic field.
Indeed, 
the $J_{yz}$ term breaks the lattice inversion symmetry and the spin $\pi$ rotation symmetry around $z$ axis.
Furthermore, 
the $J_{yz}$ and the $h_{x}$ terms break the complex conjugate symmetry.

We believe this model,
when all parameters are nonzero and $J_{xx}\neq J_{yy}$,
is nonintegrable in the sense that it 
has no local conserved quantities other than the Hamiltonian,
because so is 
the conventional model
with $J_{yz} = h_{x} =0$~\cite{Shiraishi2019}.
As a support of this belief, we have confirmed 
in Appendix~\ref{sec:LevelSpacing}
that the energy level statistics
in each momentum sector is described 
by the Gaussian unitary ensemble (GUE)
of the
random matrix theory~\cite{Mehta2004,Oganesyan2007,Atas2013}.

We tabulate the values of the parameters used in the numerical calculations
as Model I in Table~\ref{tbl:Models}.
The values are taken
in such a way that the ratio of every two of them is irrational
in order to avoid a possible accidental symmetry.

\begin{table}
\caption{\label{tbl:Models}%
Values of the parameters
in the Hamiltonian, Eq.~(\ref{eq:Model}), for 
the three models.} 
\begin{ruledtabular}
\begin{tabular}{lcccccc}
Model & $J_{xx}$ & $J_{yy}$ & $J_{zz} \ (=f)$ & $h_z$ & $J_{yz}$ & $h_x$
\\
 & fixed    & fixed    & initial value $f_0$ & fixed & fixed & fixed
\\ \hline
I (nonintegrable) & $\cos 1$ & $1$ & $e$ & $\ln 5$ & $\ln 3$ & $\pi$
\\ 
II (integrable) & $0$ & $1$ & $e$ & $0$ & $\ln 3$ & $\pi$
\\ 
III (integrable) & $1$ & $1$ & $e$ & $\ln 5$ & $0$ & $0$
\end{tabular}
\end{ruledtabular}
\end{table}%

For comparison,
we also study Model~II in which $J_{xx} = h_z=0$ (see Table \ref{tbl:Models}).
Although this model is slightly different from known models~\cite{Lieb1961},
we find it integrable in the sense that it 
can be mapped to a noninteracting fermionic system
by the Jordan-Wigner transformation.
We give its analytic solutions in Appendix~\ref{sec:AnalyticNoninteracting}.

In addition,
Model~III in which $J_{xx}=J_{yy}$ and $J_{yz}=h_{x}=0$ (see Table~\ref{tbl:Models})
is studied.
It is just 
the XXZ model,
whose energy eigenstates and additive conserved quantities
are constructed
by using the Bethe ansatz~\cite{Bethe1931,Yang1966,Yang1966a,Yang1966b,Sklyanin1979,Takhtadzhan1979,Baxter1982,Korepin1993,Takahashi1999}.

These models
cover three 
typical types of systems:
the nonintegrable systems,
the ``noninteracting integrable systems,''
and the ``interacting integrable systems'' that are
solvable by the Bethe ansatz.

In all these models, 
we choose $J_{zz}$ as the quench parameter $f$
in order to demonstrate that our additive observables are 
not restricted to one-body observables.
In fact,
the additive observable conjugate to $f=J_{zz}$ is the 
two-spin operator, 
\begin{align}
\hat{B}
=\sum_{r=1}^{N}\hat{\sigma}_{r}^z\hat{\sigma}_{r+1}^z
=:\hat{M}^{zz}.
\label{eq:Mzz}
\end{align}
We write 
$\hat{H}$ of Eq.~(\ref{eq:Model}) as $\hat{H}(f)$,
and $\hat{H}_0 = \hat{H}(f_0)$, where the initial value $f_0$ of
$f=J_{zz}$ is given in Table~\ref{tbl:Models}.
Taking the initial state as the canonical Gibbs state $\hat{\rho}^{\text{eq}}_{0}$ of $\hat{H}_0$ with the inverse temperature $\beta_{0}=0.15$, 
we study the quench process in which 
$f=J_{zz}$ is changed suddenly.

We calculate  $\chi^{\text{QM}}_{N}(A|B)$ and $\chi^{\text{TD}}_{N}(A|B)$
for several choices of $\hat{A}$ including the case of $\hat{A}=\hat{B}$.
For this purpose, we express the susceptibilities as 
Eqs.~(\ref{eq:chi^QM_BogoProd}) and (\ref{eq:chi^TD_BogoProd})
of Appendix~\ref{sec:Derivation_Exp_Ave},
and calculate them by performing the exact diagonalization of $\hat{H}_{0}$
from $N=6$ to $19$.

We here present the results for 
Models I and II, whereas those for Model III are presented 
in Appendix~\ref{sec:NumericalInteracting}.

\subsection{\label{sec:Numerical_chi_B}Susceptibilities of \texorpdfstring{$\hat{B}$}{B}}

First we calculate the two susceptibilities of $\hat{B}$,
$\chi^{\text{QM}}_{N}(B|B)$ and $\chi^{\text{TD}}_{N}(B|B)$,
and examine whether
condition~(\ref{eq:chi^QM(B|B)-chi^TD(B|B)}),
which is equivalent to condition~(\ref{eq:Cond_Exp_Ave}),
is satisfied.

The two susceptibilities of Model~I 
are plotted against the system size $N$
in Fig.~\ref{fig:chi(zz)_XYZ,XY}(a).
They approach the same value as $N$ is increased.
The inset of Fig.~\ref{fig:chi(zz)_XYZ,XY}(a) shows
the $N$ dependence of
their difference, 
$\chi^{\text{TD}}_{N}(B|B)-\chi^{\text{QM}}_{N}(B|B)$,
in a log-log plot,
indicating a power-law decay.
From these results we conclude 
that Model~I satisfies condition~(\ref{eq:chi^QM(B|B)-chi^TD(B|B)})
and, equivalently, condition (\ref{eq:Cond_Exp_Ave}).
\begin{figure}
\includegraphics[width=\linewidth]{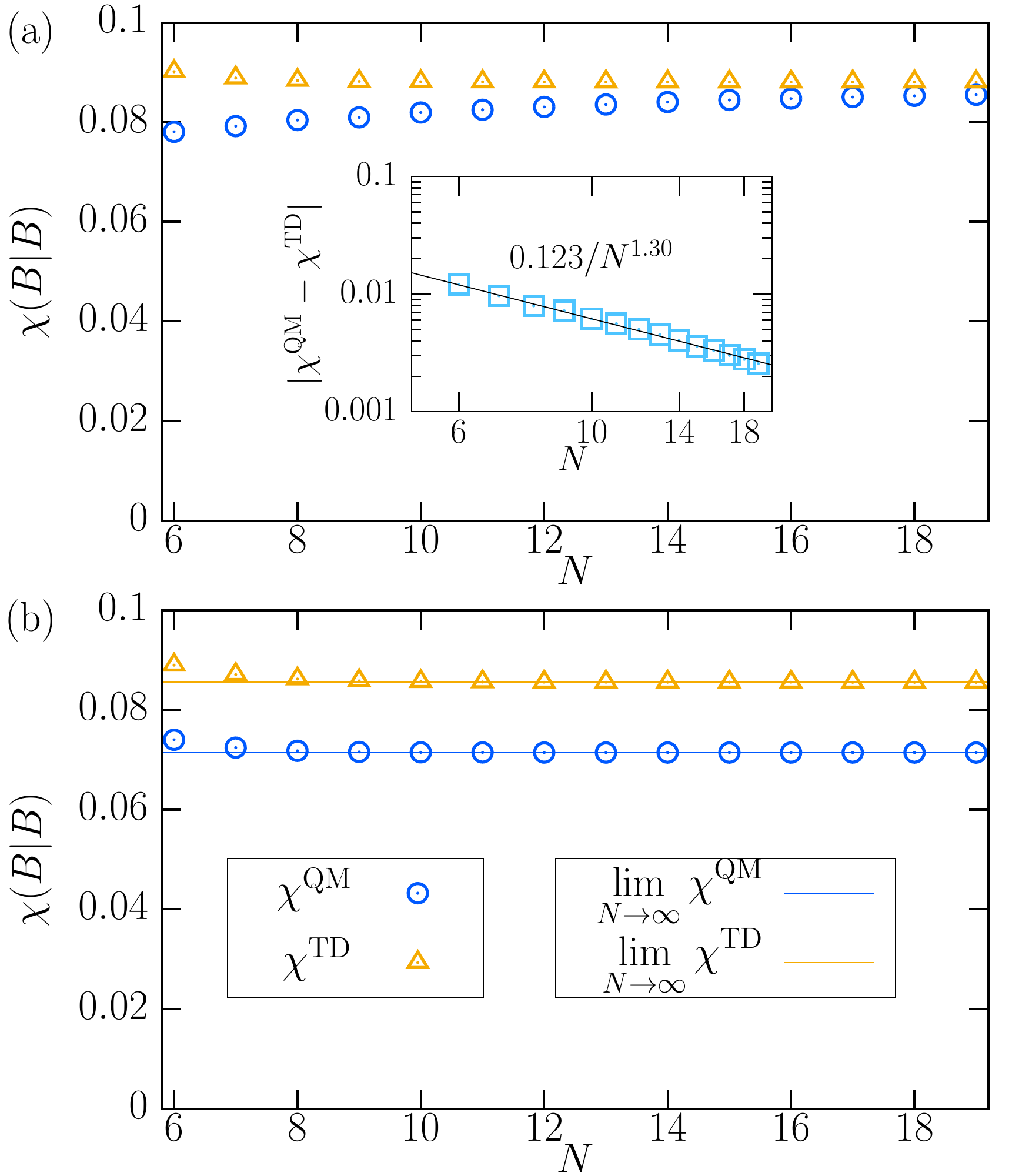}
\caption{\label{fig:chi(zz)_XYZ,XY}%
$\chi^{\text{QM}}_{N}(B|B)$ (blue circle) and $\chi^{\text{TD}}_{N}(B|B)$ (orange triangle)
against the system size $N$
in (a) Model~I (nonintegrable) and in (b) Model~II (integrable).
The solid lines in (b) show the thermodynamic 
limits of these susceptibilities.
Inset of (a): A log-log plot of
$N$ dependence of $|\chi^{\text{QM}}_{N}(B|B)-\chi^{\text{TD}}_{N}(B|B)|$ in Model~I, 
which can be fitted by a function $a/N^b$ with constants $a=0.123(5)$, $b=1.30(2)$, and
the solid line shows the fitting function $0.123/N^{1.30}$.
}
\end{figure}%

For comparison,
we plot the two susceptibilities of Model~II 
against $N$ in Fig.~\ref{fig:chi(zz)_XYZ,XY}(b).
The solid lines depict
their thermodynamic limits, 
which are calculated from the analytic solutions~(\ref{eq:Analytic_chi^QM}) and (\ref{eq:Analytic_chi^TD}) 
given in Appendix~\ref{sec:AnalyticNoninteracting}.
These results clearly show 
that Model~II violates condition~(\ref{eq:chi^QM(B|B)-chi^TD(B|B)})
and hence condition (\ref{eq:Cond_Exp_Ave}).

\subsection{\label{sec:Numerical_chi_A}Susceptibilities of \texorpdfstring{$\hat{A}$}{A}}

Next we calculate the susceptibilities of additive observables $\hat{A}$
that are \emph{not} conjugate to the quench parameter 
$f=J_{zz}$
in order to demonstrate our 
\emph{Theorem} 
(in the form rephrased in Sec.~\ref{sec:GeneralizedSuscep}).
That is, we demonstrate that
Eq.~(\ref{eq:chi^QM(A|B)-chi^TD(A|B)}) is satisfied for 
such $\hat{A}$'s in Model~I
while it is violated in Model~II.

As typical $\hat{A}$'s, 
we choose the following observables for the demonstration.
The first one is the sum of single-site observables,
\begin{align}
\hat{M}^{x}
&:=\sum_{r=1}^{N}\hat{\sigma}_{r}^x.
\end{align}
The second one is the sum of two-spin observables, 
\begin{align}
\hat{M}^{xx}
&:=\sum_{r=1}^{N}\hat{\sigma}_{r}^x\hat{\sigma}_{r+1}^x.
\end{align}
The third one is also the sum of two-spin observables, but 
the two spins are the next nearest to each other,
\begin{align}
\hat{M}^{z1z}
&:=\sum_{r=1}^{N}\hat{\sigma}_{r}^z\hat{\sigma}_{r+2}^z,
\end{align}
where $\hat{\sigma}_{N+2}^z=\hat{\sigma}_{2}^z$.

The susceptibilities of  
$\hat{A}=\hat{M}^{x}$
of Model~I are plotted against the system size $N$
in Fig.~\ref{fig:chi(x)_XYZ,XY}(a),
and those of $\hat{A}=\hat{M}^{xx}$ and $\hat{M}^{z1z}$
are plotted in Figs.~\ref{fig:chi(xx)_XYZ,XY}(a) and \ref{fig:chi(z1z)_XYZ,XY}(a)
of Appendix~\ref{sec:AdditionalNumerical}, respectively.
In each figure,
$\chi^{\text{QM}}_{N}(A|B)$ and $\chi^{\text{TD}}_{N}(A|B)$ 
approach the same value
with increasing $N$.
The insets of these figures show
the difference of the susceptibilities
in a log-log plot.
They indicate power law decays,
as in the case of Fig.~\ref{fig:chi(zz)_XYZ,XY}(a).
From these results, we confirm that 
Eq.~(\ref{eq:chi^QM(A|B)-chi^TD(A|B)}) is satisfied for
all the three additive observables. 
\begin{figure}
\includegraphics[width=\linewidth]{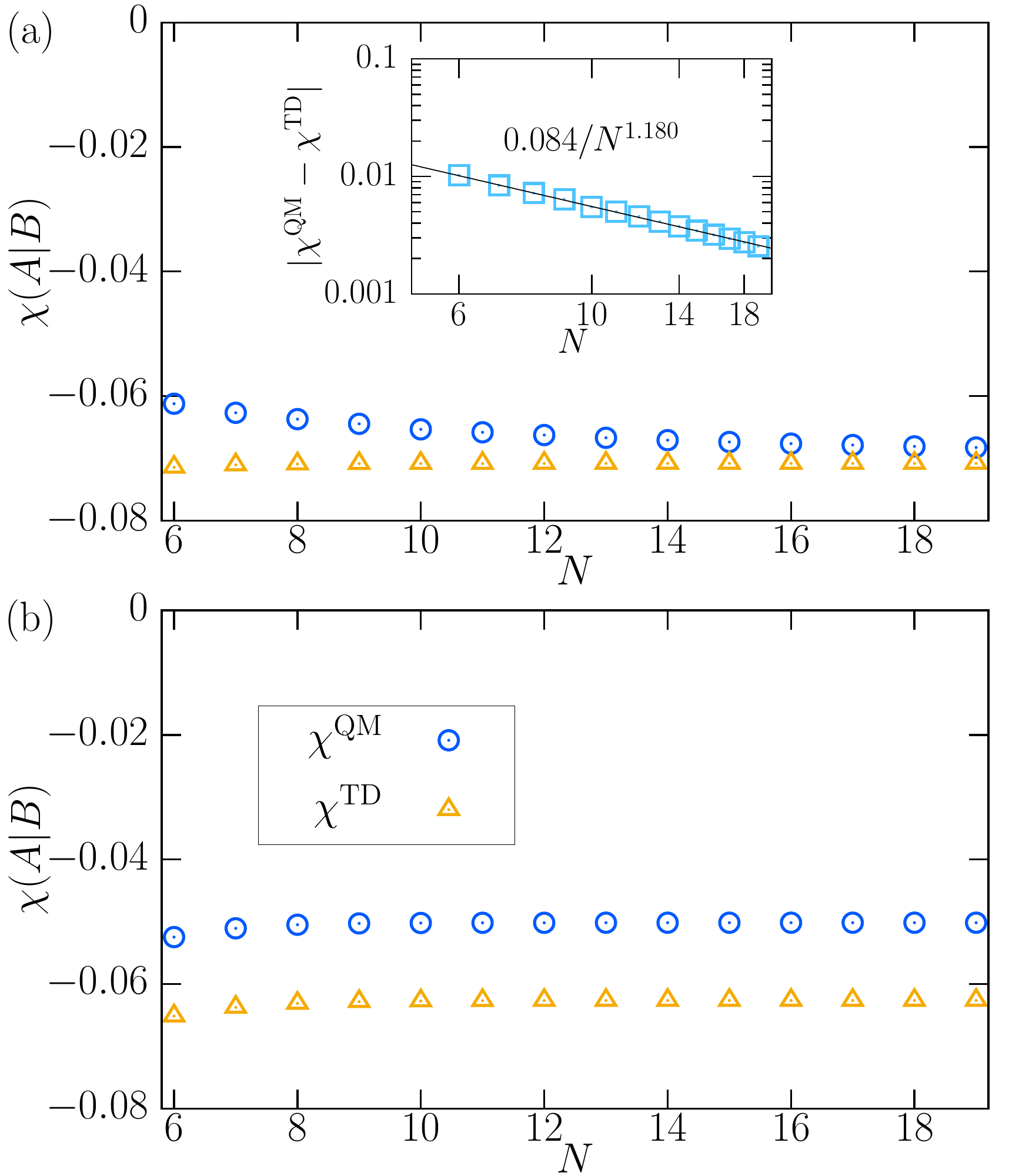}
\caption{\label{fig:chi(x)_XYZ,XY}%
$\chi^{\text{QM}}_{N}(A|B)$ (blue circle) and $\chi^{\text{TD}}_{N}(A|B)$ (orange triangle)
for $\hat{A}=\hat{M}^{x}$
against the system size $N$
in (a) Model~I (nonintegrable) and in (b) Model~II (integrable).
Inset of (a): A log-log plot of
$N$ dependence of $|\chi^{\text{QM}}_{N}(A|B)-\chi^{\text{TD}}_{N}(A|B)|$ in Model~I, which  
can be fitted by a function $a/N^b$ with constants $a=0.084(2)$, $b=1.180(9)$,
and the solid line shows the fitting function $0.084/N^{1.180}$.
}
\end{figure}%

For comparison, 
the susceptibilities of  
$\hat{A}=\hat{M}^{x}$, $\hat{M}^{xx}$, and $\hat{M}^{z1z}$
of Model~II are plotted 
in Figs.~\ref{fig:chi(x)_XYZ,XY}(b), \ref{fig:chi(xx)_XYZ,XY}(b), and \ref{fig:chi(z1z)_XYZ,XY}(b), respectively.
In all these figures,
$\chi^{\text{QM}}_{N}(A|B)$ and $\chi^{\text{TD}}_{N}(A|B)$ 
deviate from each other.
Therefore
we conclude that, in Model II,
Eq.~(\ref{eq:chi^QM(A|B)-chi^TD(A|B)}) is not satisfied for
the three additive observables~\footnote{
Although one can construct an observable that satisfies Eq.~(\ref{eq:chi^QM(A|B)-chi^TD(A|B)}) by taking a fine-tuned linear combination of observables 
that violate it, this does not conflict with our \emph{Theorem} at all. 
}. 
This is consistent with our \emph{Theorem} because 
this integrable model violates condition~(\ref{eq:chi^QM(B|B)-chi^TD(B|B)}).

\section{\label{sec:Results}Outlines of the proofs}

In this section, 
we describe outlines of 
proofs of 
\emph{Theorem} and \emph{Propositions}
of Secs.~\ref{sec:SummaryResults} and \ref{sec:Timescale}.
A combination of these outlines and the detailed discussions 
in Appendix \ref{sec:Derivation} gives the complete proofs.

\subsection{\label{sec:Exp_Ave}Outline of the proof of \emph{Theorem}}

Since 
our \emph{Theorem} in Sec.~\ref{sec:SummaryResults} can be rephrased as the
\emph{Consistency of cross responses} of Sec.~\ref{sec:GeneralizedSuscep}, 
we prove the latter.

It is known that 
the generalized susceptibilities can be expressed in terms of 
the canonical correlation~\cite{Kubo1957,KTH,Petz1993},
which is defined 
for arbitrary operators $\hat{X}$ and $\hat{Y}$
by
\begin{align}
\langle\hat{X};\hat{Y}\rangle^{\text{eq}}_{0}
:=\int_{0}^{1}\dd{\lambda}
\Tr\bigl[(\hat{\rho}^{\text{eq}}_{0})^{1-\lambda}\hat{X}^{\dagger}(\hat{\rho}^{\text{eq}}_{0})^{\lambda}\hat{Y}\bigr].
\label{eq:BogoProd}
\end{align}
We start from such expressions.
We then introduce an projection superoperator.
We finally make use of the fact that 
the canonical correlation defines an inner product,
i.e., it satisfies all of the axioms of an inner product
(hence is called the Kubo-Mori-Bogoliubov inner product).

In Appendix \ref{sec:Derivation_Exp_Ave}, 
we have expressed 
$\chi_{N}^{\text{QM}}(A|B)$ 
and
$\chi^{\text{TD}}_{N}(A|B)$
using the canonical correlations
as Eqs.~(\ref{eq:chi^QM_BogoProd}) and ~(\ref{eq:chi^TD_BogoProd}),
respectively.
These expressions yield
\begin{align}
&\chi^{\text{TD}}_{N}(A|B)-\chi^{\text{QM}}_{N}(A|B)\notag\\
&=\frac{1}{N}\langle\beta_{0}\delta\hat{B};\delta\overline{\hat{A}^{0}}\rangle^{\text{eq}}_{0}
-\frac{\langle\beta_{0}\delta\hat{B}\delta\hat{H}_{0}\rangle^{\text{eq}}_{0}\langle\delta\hat{H}_{0}\delta\hat{A}\rangle^{\text{eq}}_{0}}{N\langle(\delta\hat{H}_{0})^2\rangle^{\text{eq}}_{0}},
\label{eq:chi^TD-chi^QM_Cor}
\end{align}
where
$
\delta\hat{\bullet}:=\hat{\bullet}-\langle\hat{\bullet}\rangle^{\text{eq}}_{0}
$ is the deviation from the initial equilibrium value,
and  we have introduced 
the Heisenberg operator,
\begin{align}
\hat{X}^{0}(t):=e^{i\hat{H}_{0}t}\hat{X}e^{-i\hat{H}_{0}t},
\label{eq:X^0(t)}
\end{align}
and its long time average,
\begin{align}
\overline{\hat{X}^{0}}
:=\lim_{\mathcal{T}\to\infty}
\overline{\hat{X}^{0}(t)}^{\mathcal{T}}.
\label{eq:LongTimeAveragedOperator}
\end{align}
Here, we have put the superscript $0$ because $\hat{X}^{0}(t)$ 
evolves by the initial Hamiltonian $\hat{H}_{0}$.
Since $[\overline{\hat{X}^{0}},\hat{H}_{0}]=0$ holds
for any operator $\hat{X}$, 
we can rewrite the first term of Eq.~(\ref{eq:chi^TD-chi^QM_Cor}) as
\begin{align}
&\langle\delta\hat{B};\delta\overline{\hat{A}^{0}}\rangle^{\text{eq}}_{0}
=\langle\delta\hat{B}\delta\overline{\hat{A}^{0}}\rangle^{\text{eq}}_{0}
=
\langle\delta\hat{B}^{0}(t)\delta\overline{\hat{A}^{0}}\rangle^{\text{eq}}_{0}
\nonumber\\
&\quad
=\langle\delta\overline{\hat{B}^{0}}\delta\overline{\hat{A}^{0}}\rangle^{\text{eq}}_{0}.
\label{eq:BogoProd_commute_Tave}
\end{align}

Now we introduce the following projection superoperator,
\begin{align}
\mathcal{P}[\hat{X}]:=
\delta\overline{\hat{X}^{0}}
-\frac{\langle\delta\hat{H}_{0}\delta\overline{\hat{X}^{0}}\rangle^{\text{eq}}_{0}}{\langle(\delta\hat{H}_{0})^2\rangle^{\text{eq}}_{0}}
\delta\hat{H}_{0}.
\end{align}
It 
projects an operator $\hat{X}$
onto the operator subspace whose elements (operators)
commute with $\hat{H}_{0}$
and are orthogonal to $\hat{1}$ and $\hat{H}_{0}$
(under the Kubo-Mori-Bogoliubov inner product).
By using this superoperator, 
Eq.~(\ref{eq:chi^TD-chi^QM_Cor})
can be written as
\begin{align}
\chi^{\text{TD}}_{N}(A|B)-\chi^{\text{QM}}_{N}(A|B)
&=\frac{\beta_{0}}{N}
\langle\mathcal{P}[\hat{B}];\mathcal{P}[\hat{A}]\rangle^{\text{eq}}_{0}.
\label{eq:chi^TD-chi^QM_P}
\end{align}
Since the r.h.s. is an inner product, 
we apply the Cauchy-Schwarz inequality, and obtain
\begin{align}
&|\chi^{\text{TD}}_{N}(A|B)-\chi^{\text{QM}}_{N}(A|B)|^2\notag\\
&\le
\frac{(\beta_{0})^2}{N^2}
\langle\mathcal{P}[\hat{A}];\mathcal{P}[\hat{A}]\rangle^{\text{eq}}_{0}
\langle\mathcal{P}[\hat{B}];\mathcal{P}[\hat{B}]\rangle^{\text{eq}}_{0}
\nonumber\\
&=
\Bigl(\chi^{\text{TD}}_{N}(A|A)-\chi^{\text{QM}}_{N}(A|A)\Bigr)
\Bigl(\chi^{\text{TD}}_{N}(B|B)-\chi^{\text{QM}}_{N}(B|B)\Bigr)
\notag\\
&\le\chi^{\text{TD}}_{N}(A|A)\Bigl(\chi^{\text{TD}}_{N}(B|B)-\chi^{\text{QM}}_{N}(B|B)\Bigr),
\label{eq:chi^TD-chi^QM_Schwarz}
\end{align}
where we have used 
$\chi^{\text{QM}}_{N}(A|A) \geq 0$,
which follows from Eq.~(\ref{eq:chi^QM_BogoProd}),
and
$\chi^{\text{TD}}_{N}(B|B) \geq \chi^{\text{QM}}_{N}(B|B)$~\cite{Wilcox1968,Suzuki1971}, 
which is obvious from Eq.~(\ref{eq:chi^TD-chi^QM_P}).
Since we exclude phase transition points as stated in Sec.~\ref{sec:Setup},
it is required that
\begin{align}
\chi^{\text{TD}}_{N}(A|A)
=\mathcal{O}(N^0)
\qquad \text{for any }\hat{A}
\label{eq:OutOfEquilibrium}
\end{align}
from thermodynamics and equilibrium statistical mechanics~\footnote{
We can prove Eq.~(\ref{eq:OutOfEquilibrium})
for any additive observable $\hat{A}$ defined by Eq.~(\ref{eq:Additive})
when correlation functions in $\hat{\rho}^{\text{eq}}_{0}$ decay sufficiently fast.
In fact, 
Eqs.~(\ref{eq:chi^TD_BogoProd}) and (\ref{eq:BogoProd_Symmetrized}) show that
$\chi^{\text{TD}}_{N}(A|A)$ is bounded from above 
by $\beta_{0}\langle(\delta\hat{A})^2\rangle^{\text{eq}}_{0}/N$.
}.
Therefore, 
Eq.~(\ref{eq:chi^TD-chi^QM_Schwarz}) yields
the \emph{Consistency of cross responses} of Sec.~\ref{sec:GeneralizedSuscep}, 
and hence our \emph{Theorem} in Sec.~\ref{sec:SummaryResults}.

\subsection{\label{sec:Exp_Fluc}Outline of the proof of \emph{Proposition}~$1$}

Let $|\nu\rangle$ be an eigenstate of $\hat{H}_{0}$
with an energy eigenvalue $E_{\nu}$.
We take these eigenstates such that
they form
an orthonormal basis
even if the eigenvalues are degenerate. 
The maximum number of resonances
in condition (\ref{eq:Assumption_Resonances})
is defined by
\begin{align}
D_{\text{res}}
:=
\max_{\nu_1, \nu_2 \ (E_{\nu_1} \neq E_{\nu_2})}\, 
\sum_{\nu_3, \nu_4 \ (E_{\nu_3} \neq E_{\nu_4})}
\delta_{E_{\nu_{1}}-E_{\nu_{2}},E_{\nu_{3}}-E_{\nu_{4}}}.
\label{def:Dres}
\end{align}

As shown in Appendix~\ref{sec:Derivation_Exp_Fluc},
the left hand side of Eq.~(\ref{eq:Main_Exp_Fluc})
can be rewritten as
\begin{align}
&\lim_{\mathcal{T} \to \infty}
\lim_{\Delta f\to 0}
\overline{\biggl|\frac{
\langle \hat{A}^{\Delta f}(t)\rangle^{\text{eq}}_{0}
-\overline{\langle \hat{A}^{\Delta f}(t)\rangle^{\text{eq}}_{0}}^{\mathcal{T}}
}{\Delta f N}\biggr|^2}^{\mathcal{T}}\notag\\
&\quad =
\lim_{\mathcal{T}\to\infty}
\overline{
\bigl(\chi_{N}^{\text{QM}}(A|B;t)
-\chi^{\text{QM}}_{N}(A|B)\bigr)^2
}^{\mathcal{T}}.
\label{eq:Exp_Fluc_1}
\end{align}
Here $\chi_{N}^{\text{QM}}(A|B;t)$ is the time-dependent susceptibility,
\begin{align}
\chi_{N}^{\text{QM}}(A|B;t)
&:=
\lim_{\Delta f \to 0}
\frac{
\langle \hat{A}^{\Delta f}(t)\rangle^{\text{eq}}_{0}
-\langle\hat{A}\rangle^{\text{eq}}_{0}
}{N\Delta f},
\label{eq:DEF_chi^QM_t}
\end{align}
which is related to $\chi^{\text{QM}}_{N}(A|B)$ via
\begin{align}
\chi^{\text{QM}}_{N}(A|B)
=\lim_{\mathcal{T}\to\infty}
\overline{\chi_{N}^{\text{QM}}(A|B;t)}^{\mathcal{T}}.
\label{eq:Details_chi^QM_interchange}
\end{align}
The r.h.s. of Eq.~(\ref{eq:Exp_Fluc_1}) 
is bounded from above by
\begin{align}
&\lim_{\mathcal{T}\to\infty}
\overline{
\bigl(\chi_{N}^{\text{QM}}(A|B;t)-\chi^{\text{QM}}_{N}(A|B)\bigr)^2
}^{\mathcal{T}}
\notag\\
&\quad \le
\frac{(\beta_{0})^2\|\hat{A}\|_{\infty}^2\|\hat{B}\|_{\infty}^2}{N^2}
D_{\text{res}}\Tr \bigl[(\hat{\rho}^{\text{eq}}_{0})^2\bigr].
\label{eq:Exp_Fluc_2}
\end{align}
The proof of this inequality,
shown in Appendix~\ref{sec:Derivation_Exp_Fluc},
is similar to
that by 
Short and Farrelly~\cite{Short2012}.
Combining 
Eqs.~(\ref{eq:Exp_Fluc_1}) and (\ref{eq:Exp_Fluc_2})
with the condition~(\ref{eq:Assumption_Resonances}),
we prove Eq.~(\ref{eq:Main_Exp_Fluc}) 
for every additive observable $\hat{A}$.

We remark that Eq.~(\ref{eq:Exp_Fluc_2}) can be extended to
finite $\mathcal{T}$,
as in Ref.~\cite{Short2012}.

\subsection{\label{sec:Var}Outline of the proof of \emph{Proposition}~2}

As explained in 
Appendix~\ref{sec:Derivation_Var},
we can show that
\begin{align}
&\lim_{\Delta f\to 0}
\frac{
\text{Var}_{0}[\hat{A}^{\Delta f}(t)]
-\text{Var}_{0}[\hat{A}]
}{\Delta f}\notag\\
&\quad =
\langle\beta_{0}\delta\hat{B};(\delta \hat{A})^2\rangle^{\text{eq}}_{0}
-\langle\beta_{0}\delta\hat{B};\bigl(\delta \hat{A}^{0}(t)\bigr)^2\rangle^{\text{eq}}_{0},
\label{eq:VarChangeRate_can}
\end{align}
where $\delta\hat{A}^{0}(t)=\hat{A}^{0}(t)-\langle\hat{A}^{0}(t)\rangle^{\text{eq}}_{0}=\hat{A}^{0}(t)-\langle\hat{A}\rangle^{\text{eq}}_{0}$.
An upper bound of the last term 
is obtained by the Cauchy-Schwarz inequality as
\begin{align}
&\bigl|\langle\delta\hat{B};\bigl(\delta \hat{A}^{0}(t)\bigr)^2\rangle^{\text{eq}}_{0}\bigr|^2\notag\\
&\quad \le
\langle\delta\hat{B};\delta\hat{B}\rangle^{\text{eq}}_{0}
\langle\bigl(\delta \hat{A}^{0}(t)\bigr)^2;\bigl(\delta \hat{A}^{0}(t)\bigr)^2\rangle^{\text{eq}}_{0}\notag\\
&\quad =
\langle\delta\hat{B};\delta\hat{B}\rangle^{\text{eq}}_{0}
\langle(\delta \hat{A})^2;(\delta \hat{A})^2\rangle^{\text{eq}}_{0}
\notag\\
&\quad \le 
\langle(\delta\hat{B})^2\rangle^{\text{eq}}_{0}
\langle(\delta \hat{A})^4\rangle^{\text{eq}}_{0}.
\label{eq:VarChangeRate_CauchySchwarz}
\end{align}%
Here the last line follows from Eq.~(\ref{eq:BogoProd_Symmetrized}) of Appendix~\ref{sec:Derivation_Var}.
By using this inequality, we have
\begin{align}
\lim_{\Delta f\to 0}
&\left|
\frac{
\text{Var}_{0}[\hat{A}^{\Delta f}(t)]
-\text{Var}_{0}[\hat{A}]
}{\Delta f}
\right|
\notag\\
&\leq
2\beta_{0}\sqrt{\langle(\delta\hat{B})^2\rangle^{\text{eq}}_{0}\langle(\delta\hat{A})^4\rangle^{\text{eq}}_{0}}.
\label{eq:Result_VarChangeRate}
\end{align}
Since the time average of the left-hand side (l.h.s.) of this inequality
can also be bounded from above by the same quantity,
we have
\begin{align}
&\lim_{\mathcal{T}\to\infty}\lim_{\Delta f\to 0}
\left|
\frac{
\overline{\text{Var}_{0}[\hat{A}^{\Delta f}(t)]}^{\mathcal{T}}
-\text{Var}_{0}[\hat{A}]
}{\Delta f}
\right|
\notag\\
&\qquad\le
2\beta_{0}\sqrt{\langle(\delta\hat{B})^2\rangle^{\text{eq}}_{0}\langle(\delta\hat{A})^4\rangle^{\text{eq}}_{0}}.
\label{eq:Main_Var_Ave_Explicit}
\end{align}
Here, we have used interchangeability of
the time integration and the limit $\Delta f\to 0$,
which is shown in Eqs.~(\ref{eq:FormulasInterchange})--(\ref{eq:FormulasInterchange_3}) of Appendix~\ref{sec:Derivation_Timescale}.
Combining this with the condition (\ref{eq:Assumption_4thMoment})
and its consequence (\ref{eq:Var=O(N)}), 
we obtain Eq.~(\ref{eq:Main_Var_Ave}).

It should be remarked that 
Eq.~(\ref{eq:Result_VarChangeRate})
holds 
at an arbitrary time $t>0$
without taking time average.
That is, the variance remains small at \emph{all} $t>0$,
although \emph{Criterion}~(iii) 
requires it only for \emph{almost all} 
$t>0$.

\subsection{\label{sec:Proof_Timescale}Outline of the proof of \emph{Proposition}~3}

We use the following inequalities
that are proved in Appendix~\ref{sec:Derivation_Timescale},
\begin{align}
&\Bigl|\frac{
\overline{\langle \hat{A}^{\Delta f}(t)\rangle^{\text{eq}}_{0}}^{\mathcal{T}}
-\langle \hat{A}\rangle^{\text{eq}}_{0}
}{\Delta f N}
-\overline{\chi_{N}^{\text{QM}}(A|B;t)}^{\mathcal{T}}\Bigr|
\notag\\
&\le
\bigl(D_{1}\mathcal{T}+D_{2}\mathcal{T}^2\bigr)|\Delta f|,
\label{eq:finite_Df_T_Exp_Ave}
\\
&\Bigl|\overline{\Bigl(\frac{
\langle \hat{A}^{\Delta f}(t)\rangle^{\text{eq}}_{0}
-\langle \hat{A}\rangle^{\text{eq}}_{0}
}{\Delta f N}\Bigr)^2}^{\mathcal{T}}
-\overline{\bigl(\chi_{N}^{\text{QM}}(A|B;t)\bigr)^2}^{\mathcal{T}}\Bigr|
\notag\\
&\le
\bigl(D_{3}\mathcal{T}+D_{4}\mathcal{T}^2\bigr)
|\Delta f|
\notag\\
&\qquad
+\bigl(D_{5}\mathcal{T}^2
+D_{6}\mathcal{T}^3
+D_{7}\mathcal{T}^4\bigr)
|\Delta f|^2,
\label{eq:finite_Df_T_Exp_Fluc}
\\
&\Bigl|\frac{
\overline{\langle \bigl(\hat{A}^{\Delta f}(t)\bigr)^2\rangle^{\text{eq}}_{0}}^{\mathcal{T}}
-\langle \hat{A}^2\rangle^{\text{eq}}_{0}
}{\Delta f N^2}
-\overline{
\tilde{\chi}_{N}(A^2;t)
}^{\mathcal{T}}\Bigr|
\notag\\
&\le
\bigl(D_{8}\mathcal{T}+D_{9}\mathcal{T}^2\bigr)
|\Delta f|.
\label{eq:finite_Df_T_Var}
\end{align}
Here 
\begin{align}
\tilde{\chi}_{N}(A^2;t):=
\lim_{\Delta f\to 0}\frac{
\langle \bigl(\hat{A}^{\Delta f}(t)\bigr)^2\rangle^{\text{eq}}_{0}
-\langle \hat{A}^2\rangle^{\text{eq}}_{0}
}{\Delta f N^2},
\label{eq:DEF_tilde_chi_t}
\end{align}
and
$D_1, \cdots, D_9$ are nonnegative constants 
of $\mathcal{O}(|\Delta f|^0)$ that are independent of $\mathcal{T}$. 
By noting that the 
right-hand sides of these equations
vanish in the limit~(\ref{eq:Lim_Df_LongerT}),
we can prove 
Eqs.~(\ref{eq:Main_Exp_Ave_longer})--(\ref{eq:Main_Var_Ave_longer}) as follows.

Firstly, combining 
Eq.~(\ref{eq:finite_Df_T_Exp_Ave}) with Eq.~(\ref{eq:Details_chi^QM_interchange}),
we have
\begin{align}
\lim_{\Delta f\to 0}
\frac{
\overline{\langle \hat{A}^{\Delta f}(t)\rangle^{\text{eq}}_{0}}^{\mathcal{T}_{\Delta f}}
-\langle \hat{A}\rangle^{\text{eq}}_{0}
}{\Delta f N}
=\chi_{N}^{\text{QM}}(A|B).
\label{eq:chi^QM_longer}
\end{align}
Using Eqs.~(\ref{eq:DEF_chi^QM}), (\ref{eq:DEF_chi^TD}) and (\ref{eq:chi^QM_longer}),
we obtain
Eq.~(\ref{eq:Main_Exp_Ave_longer}).

Next we evaluate
\begin{align}
&\overline{\biggl|\frac{
\langle \hat{A}^{\Delta f}(t)\rangle^{\text{eq}}_{0}
-\overline{\langle \hat{A}^{\Delta f}(t)\rangle^{\text{eq}}_{0}}^{\mathcal{T}}
}{\Delta f N}\biggr|^2}^{\mathcal{T}}
\notag\\
&=
\overline{\Bigl(\frac{
\langle \hat{A}^{\Delta f}(t)\rangle^{\text{eq}}_{0}
-\langle \hat{A}\rangle^{\text{eq}}_{0}
}{\Delta f N}\Bigr)^2}^{\mathcal{T}}
-\Bigl(\frac{
\overline{\langle \hat{A}^{\Delta f}(t)\rangle^{\text{eq}}_{0}}^{\mathcal{T}}
-\langle \hat{A}\rangle^{\text{eq}}_{0}
}{\Delta f N}\Bigr)^2.
\label{eq:Proof_Timescale_Exp_Fluc}
\end{align}
By taking the limit~(\ref{eq:Lim_Df_LongerT}),
we can evaluate the first and second terms
from 
Eqs.~(\ref{eq:finite_Df_T_Exp_Fluc}) and (\ref{eq:chi^QM_longer}),
respectively.
Then we have
\begin{align}
&\lim_{\Delta f\to 0}
\overline{\biggl|\frac{
\langle \hat{A}^{\Delta f}(t)\rangle^{\text{eq}}_{0}
-\overline{\langle \hat{A}^{\Delta f}(t)\rangle^{\text{eq}}_{0}}^{\mathcal{T}_{\Delta f}}
}{\Delta f N}\biggr|^2}^{\mathcal{T}_{\Delta f}}
\notag\\
&=\lim_{\mathcal{T}\to\infty}
\overline{
\bigl(\chi_{N}^{\text{QM}}(A|B;t)\bigr)^2
}^{\mathcal{T}}
-\bigl(\chi^{\text{QM}}_{N}(A|B)\bigr)^2.
\end{align}
Combining this with Eq.~(\ref{eq:Exp_Fluc_1}),
we obtain Eq.~(\ref{eq:Main_Exp_Fluc_longer}).

Finally we evaluate
\begin{align}
&
\frac{
\overline{\text{Var}_{0}[\hat{A}^{\Delta f}(t)]}^{\mathcal{T}}
-\text{Var}_{0}[\hat{A}]
}{\Delta f N^2}
\notag\\
&=\frac{
\overline{\langle \bigl(\hat{A}^{\Delta f}(t)\bigr)^2\rangle^{\text{eq}}_{0}}^{\mathcal{T}}
-\langle \hat{A}^2\rangle^{\text{eq}}_{0}
}{\Delta f N^2}
-\overline{\Bigl(\frac{
\langle \hat{A}^{\Delta f}(t)\rangle^{\text{eq}}_{0}
-\langle \hat{A}\rangle^{\text{eq}}_{0}
}{\Delta f N}\Bigr)^2\Delta f}^{\mathcal{T}}
\notag\\
&\qquad
-2\frac{\langle \hat{A}\rangle^{\text{eq}}_{0}}{N}
\frac{
\overline{\langle \hat{A}^{\Delta f}(t)\rangle^{\text{eq}}_{0}}^{\mathcal{T}}
-\langle \hat{A}\rangle^{\text{eq}}_{0}
}{\Delta f N}
\label{eq:Proof_Timescale_Var}
\end{align}
By taking the limit~(\ref{eq:Lim_Df_LongerT}),
we can evaluate the first and third terms
from 
Eqs.~(\ref{eq:finite_Df_T_Var}) and (\ref{eq:chi^QM_longer}),
respectively.
Note that the second term vanishes in this limit
because of Eq.~(\ref{eq:finite_Df_T_Exp_Fluc}).
As a result,
we have
\begin{align}
&\lim_{\Delta f\to 0}
\frac{
\overline{\text{Var}_{0}[\hat{A}^{\Delta f}(t)]}^{\mathcal{T}_{\Delta f}}
-\text{Var}_{0}[\hat{A}]
}{\Delta f N^2}
\notag\\
&=
\lim_{\mathcal{T}\to\infty}\overline{
\tilde{\chi}_{N}(A^2;t)
}^{\mathcal{T}}
-2\frac{\langle \hat{A}\rangle^{\text{eq}}_{0}}{N}
\chi^{\text{QM}}_{N}(A|B)
\notag\\
&=\lim_{\mathcal{T}\to\infty}\overline{
\Bigl(\lim_{\Delta f\to 0}\frac{
\text{Var}_{0}[\hat{A}^{\Delta f}(t)]
-\text{Var}_{0}[\hat{A}]
}{\Delta f N^2}
\Bigr)}^{\mathcal{T}}.
\label{eq:Var_ChangeRate_longer}
\end{align}
In the last line,
we have used 
\begin{align}
&\lim_{\Delta f\to 0}\frac{
\text{Var}_{0}[\hat{A}^{\Delta f}(t)]
-\text{Var}_{0}[\hat{A}]
}{\Delta f N^2}
\notag\\
&=
\tilde{\chi}_{N}(A^2;t)
-2\frac{\langle \hat{A}\rangle^{\text{eq}}_{0}}{N}
\chi^{\text{QM}}_{N}(A|B;t),
\end{align}
which follows 
from Eqs.~(\ref{eq:Var_Change}) and (\ref{eq:Leibniz_chi^QM}) 
of Appendix~\ref{sec:Derivation_Var}.
From
Eqs.~(\ref{eq:FormulasInterchange})--(\ref{eq:FormulasInterchange_3})
of Appendix~\ref{sec:Derivation_Timescale},
the time integration and the limit $\Delta f\to 0$ can be interchanged,
and hence
Eq.~(\ref{eq:Var_ChangeRate_longer}) 
yields  
Eq.~(\ref{eq:Main_Var_Ave_longer}).

\section{\label{sec:Discussion}Discussions}

\subsection{\label{sec:ExtensionNonadditive}Extension to nonadditive observables}

In the above proof,
additivity of $\hat{A}$ and $\hat{B}$ is not crucial.
In fact, 
we can extend our \emph{Theorem} as follows.
\\
\emph{Extension to nonadditive observables}:
Suppose that the observable $\hat{B}$ defined by Eq.~(\ref{eq:def_B}) is 
not necessarily additive.
If Eq.~(\ref{eq:Main_Exp_Ave}) holds for $\hat{B}$
then it holds for any observable $\hat{A}$ that 
is not necessarily additive but satisfies
\begin{align}
\chi^{\text{TD}}_{N}(A|A)-\chi^{\text{QM}}_{N}(A|A)=\mathcal{O}(N^0).
\end{align}

This result will be particularly 
useful when discussing nonlocal properties of the system, such as the entanglement.
Nevertheless, 
we have focused on 
additive observables 
because we are interested in consistency with thermodynamics 
in this paper.

\subsection{\label{sec:Nonquench}Case of continuous change of $f$}

We have studied the case of a quench process, 
in which $f$ is jumped discontinuously.
Our \emph{Theorem} holds also for 
processes in which $f$ is changed continuously as
\begin{align}
f(t)=f_{0}+\lambda(t)\Delta f.
\end{align}
Here, 
$\lambda(t)$ is a continuously differentiable function such that 
$\lambda(t)=0$ for $t \leq 0$ and 
$\lambda(t)=1$ for $t^* \leq t$,
where $t^*$ is a constant independent of $\Delta f$ and $\mathcal{T}$.

The validity of our \emph{Theorem} for this process is shown as follows.
$\chi^{\text{QM}}_{N}(A|B)$
of this process agrees with that of the quench process because
$\chi^{\text{QM}}_{N}(A|B)$ is the zero-frequency component 
of a linear-response coefficient and hence
independent of the time profile of $f$,
as will be shown in Sec.~\ref{sec:LinearResponse}.
$\chi^{\text{TD}}_{N}(A|B)$ is also 
independent of the time profile of $f$
because it
agrees with the adiabatic susceptibility
regardless of the time profile of $f$.
In fact, the entropy does not change 
in $\mathcal{O}(\Delta f)$ because 
if the entropy increased by $\mathcal{O}(\Delta f)$ then 
it would decrease when the sign of $\Delta f$ is inverted,
in contradiction to the second law of thermodynamics.
Therefore, our \emph{Theorem} holds independently of details of the process.

\subsection{\label{sec:ETH_imlies_Cond_Exp_Ave}ETH for \texorpdfstring{$\hat{B}$}{B} implies condition~(\ref{eq:Cond_Exp_Ave}) 
but they are inequivalent}

In this subsection, we show that
the ETH for $\hat{B}$ implies condition~(\ref{eq:Cond_Exp_Ave})
but the converse is not necessarily true.

In Ref.~\cite{Chiba2020}, we have shown that
if the ETH (referred to as the ``strong ETH'' there)
is satisfied for the uniform magnetization
then 
condition~(8) of Ref.~\cite{Chiba2020} holds.
This condition is equivalent to
Eq.~(4) of Ref.~\cite{Chiba2020},
which states that
the $\vb*{k}=\vb*{0}$ components of two types of magnetic susceptibilities,
denoted by $\chi^{\text{qch}}_{N}(\vb*{0})$ and $\chi^{S}_{N}(\vb*{0})$
there,
coincide.
We can easily show that the proof is applicable
when 
$\chi^{\text{qch}}_{N}(\vb*{0})$ and $\chi^{S}_{N}(\vb*{0})$ 
are replaced with the generalized susceptibilities
$\chi^{\text{QM}}_{N}(\hat{B}|\hat{B})$ and $\chi^{\text{TD}}_{N}(\hat{B}|\hat{B})$  introduced in Sec.~\ref{sec:GeneralizedSuscep}, respectively, and
the uniform magnetization with $\hat{B}$.
Therefore,
the ETH for $\hat{B}$ implies
Eq.~(\ref{eq:chi^QM(B|B)-chi^TD(B|B)}) of the present paper,
which is equivalent to condition~(\ref{eq:Cond_Exp_Ave})
as explained in Sec.~\ref{sec:GeneralizedSuscep}.

Note that
this result is not so obvious
because, 
as shown in Sec.~\ref{sec:Timescale},
the timescale of linear thermalization in condition~(\ref{eq:Cond_Exp_Ave})
is 
shorter than
the timescale where the ETH is usually applied.
In other words, 
when applying the ETH,
one usually take $\mathcal{T}\to\infty$ before $\Delta f \to 0$,
which differs from the limit 
(\ref{eq:Lim_T_Lim_Df}) employed
in condition~(\ref{eq:Cond_Exp_Ave}).

On the other hand, 
condition~(\ref{eq:Cond_Exp_Ave}) 
for $\hat{B}$
is weaker than its ETH.
In fact, Shiraishi and Mori~\cite{Shiraishi2017,Mori2017} constructed systems
in which a certain observable violates the ETH but satisfies 
\emph{Criterion}~(i)
when the initial state is 
an arbitrary equilibrium state at a nonzero temperature.
Hence, when $f$ is chosen as the parameter 
conjugate to that observable,
these systems satisfy condition~(\ref{eq:Cond_Exp_Ave}) but violate the ETH for $\hat{B}$.

\subsection{How to test our results experimentally}

In this section,
we discuss a way of testing  our results experimentally. 
For concreteness, we discuss how to measure
$\chi_{N}^{\text{QM}}(A|B)$ and $\chi_{N}^{\text{TD}}(A|B)$.
Other quantities such as fluctuation can also be measured in 
a similar manner.

To measure $\chi_{N}^{\text{QM}}(A|B)$ experimentally, 
prepare the system in an equilibrium state~\footnote{
Even if the state 
is not exactly equal to the canonical Gibbs state
$\hat{\rho}^{\text{eq}}_{0}$, it only makes  
irrelevant difference of $o(N^0)$ in the susceptibilities.}.
This can be achieved, for example, 
by making the system be in thermal contact with 
a heat bath of inverse temperature $\beta_0$.
Obtain $\langle \hat{A}\rangle^{\text{eq}}_{0}$ by measuring 
$\hat{A}$ in this state.
After that, detach the heat bath, so that 
the system undergoes the unitary time evolution.
Then, 
$\langle \hat{A}^{\Delta f}(t)\rangle^{\text{eq}}_{0}$
evolves as schematically shown in Fig.~\ref{fig:Setup_TEvolution}.
By measuring this evolution,
one can judge whether the system relaxes to a (quasi)-steady state or not.
Note that the (quasi)-steady state can be a nonthermal state
because 
the relaxation occurs much more easily than 
thermalization,
as pointed out at the beginning of
Sec.~\ref{sec:SummaryResults}.
When the relaxation occurs, 
one obtains
$\overline{\langle \hat{A}^{\Delta f}(t)\rangle^{\text{eq}}_{0}}^{\mathcal{T}}$
and the relaxation time.
By taking $|\Delta f|$ small enough and $\mathcal{T}$ long enough~\footnote{
For the relative values of $|\Delta f|$ and $\mathcal{T}$,
it is sufficient to take $\mathcal{T}^2|\Delta f|$ small, 
according to Eq.~(\ref{eq:finite_Df_T_Exp_Ave}).
This means focusing on the timescale of $o(1/\sqrt{|\Delta f|})$.
}, 
and extrapolating the data to $\Delta f \to 0$ and $\mathcal{T} \to \infty$,
one obtains $\chi_{N}^{\text{QM}}(A|B)$
as Eq.~(\ref{eq:DEF_chi^QM}).
One also obtains the $\Delta f$ dependence of the relaxation time
from the $\Delta f$ dependence of the data.

For $\chi_{N}^{\text{TD}}(A|B)$, 
its measurement might look difficult when the system 
does not thermalize under the unitary evolution.
Fortunately, one can fully utilize a heat bath to 
realize equilibrium states 
because $\chi_{N}^{\text{TD}}(A|B)$
is a state function and therefore 
independent 
[apart from a finite-size-effect term of $o(N^0)$]
of how the equilibrium state is prepared.
Firstly, set $f=f_{0}$ and $\beta = \beta_{0}$,
and obtain $\langle\hat{A}\rangle^{\text{eq}}_{0}$
by measuring $\hat{A}$.
Then, supply a small amount of energy $d U$ 
to the total system (composed of the system and the bath),
and measure the resultant decrease $d \beta$ of the inverse temperature 
and the change $dB$ of the equilibrium value of $\hat{B}$.
Since the thermal capacity of the bath is known, 
one obtains the specific heat $c_{0}$ of the system from $dU$ and $d \beta$.
Furthermore, 
$dB/d \beta$ gives
\begin{align}
\eval{\Bigl(\pdv{\beta}\Tr\Bigl[\hat{\rho}^{\text{can}}(\beta,f_{0})\frac{\hat{B}}{N}\Bigr]\Bigr)}_{\beta=\beta_{0}}
=-\langle\delta\hat{H}_{0}\delta\hat{B}\rangle^{\text{eq}}_{0}/N.
\label{eq:dB_df}
\end{align}
Then, using this and Eq.~(\ref{eq:c_0}),
one can determine the inverse temperature $\beta_{\Delta f}$,
given by Eq.~(\ref{eq:beta_Df}),
of the final equilibrium state that 
is predicted by thermodynamics after the quench of $\Delta f$~\footnote{
Although we have derived 
Eqs~(\ref{eq:dB_df}), (\ref{eq:beta_Df}) and (\ref{eq:c_0})
using statistical mechanics, we can also derive them using 
thermodynamics.
}.
Finally, 
set $f=f_{0} + \Delta f$ and $\beta = \beta_{\Delta f}$,
and obtain 
$\langle\hat{A}\rangle^{\text{eq}}_{\Delta f}$
by measuring $\hat{A}$.
By substituting the measured values for 
$\langle\hat{A}\rangle^{\text{eq}}_{0}$
 and
$\langle\hat{A}\rangle^{\text{eq}}_{\Delta f}$
of Eq.~(\ref{eq:DEF_chi^TD}),
one obtains $\chi_{N}^{\text{TD}}(A|B)$.

\subsection{\label{sec:LinearResponse}Relation to linear response theory}

The quantum-mechanical susceptibility,
$\chi_{N}^{\text{QM}}(A|B)$,
is closely related to that given by the linear response theory.
Hence, the consistency of $\chi_{N}^{\text{QM}}(A|B)$
with $\chi_{N}^{\text{TD}}(A|B)$,
discussed in Sec.~\ref{sec:GeneralizedSuscep},
is also related to the validity of 
the linear response theory.
We finally discuss these points.

Following the pioneering studies by Callen, Welton, Greene, Takahashi,
and Nakano~\cite{Callen1951,Callen1952,Greene1952,Takahasi1952,Nakano1956}, 
Kubo established the linear response theory 
for quantum systems ~\cite{Kubo1957}.
He assumed that the system, which is initially in an equilibrium state,
is {\em isolated from environments} 
while an external force is applied adiabatically,
which means, in our notation, that
\begin{align}
f(t)=f_{0}+\Delta f e^{\varepsilon t}\cos\omega t.
\label{eq:f(t)_omega}
\end{align}
Here $\varepsilon$ is a small positive number
and $\omega$ is the frequency.
Then, the Kubo formula for 
the response of 
$\hat{A}/N$
reads
\begin{align}
\chi_{N}^{\text{Kubo}}(A|B)[\omega+i\varepsilon]
=\int_{0}^{\infty}\dd{t}e^{(i\omega-\varepsilon)t}
\frac{i}{N}\langle[\hat{A}^{0}(t),\hat{B}]\rangle^{\text{eq}}_{0},
\label{eq:Kubo_formula}
\end{align}
where $\hat{A}^{0}(t)$ is defined by Eq.~(\ref{eq:X^0(t)}).

Note that the validity of this formula is never obvious
because it depends on the degree of complexity of the dynamics.
The conditions for the validity have been 
discussed 
by Kubo himself~\cite{Kubo1957}
and by many authors~\cite{KTH,Wilcox1968,Suzuki1971,Chiba2020}.
For technical reasons, 
these discussions 
have been made for finite $N$, 
although it is sometimes stressed that 
the thermodynamic limit should be taken {\em before} taking other 
limits such as $\varepsilon \to +0$~\cite{Pines1966,Giuliani2005,Zubarev1974,Zubarev1996,Zubarev1997}.

If we keep $N$ finite following these studies,
we can show that~\cite{Chiba2020}
\begin{align}
\lim_{\varepsilon\to +0}\chi_{N}^{\text{Kubo}}(A|B)[0+i\varepsilon]
=\chi_{N}^{\text{QM}}(A|B).
\label{eq:chi^Kubo=chi^QM}
\end{align}
Therefore, the validity of the Kubo formula 
can be judged from 
the validity of $\chi_{N}^{\text{QM}}(A|B)$,
which was discussed in Sec.~\ref{sec:GeneralizedSuscep}.
That is, 
the validity for the response of  
the key observable against $\Delta f$ guarantees
the validity for those of other observables
and for the responses against any other parameters.
By contrast, 
the previous conditions for the validity~\cite{Kubo1957,KTH,Wilcox1968,Suzuki1971,Chiba2020}
required investigations of individual observables and responses.

It is worth mentioning that the above validity means 
the agreement of the {\em response}
between quantum mechanics and thermodynamics.
On the other hand, 
the classical limit of Eq.~(\ref{eq:Kubo_formula}) 
yields the ``fluctuation-dissipation theorem'' (FDT).
It states that
\begin{align}
[\mbox{response at $\omega$}]
=
\beta \times [\mbox{correlation spectrum at $\omega$}],
\label{eq:FDT_classical}
\end{align}
where the r.h.s.\ is not a formal one but the {\em observed} spectrum. 
In classical systems the FDT, despite its name, holds 
even for non-dissipative responses~\cite{Takahasi1952}.
Many authors discussed its quantum corrections~\cite{Nyquist1928,Callen1951,Nakano1956,Kubo1957,KTH,Koch1982,Deblock2003,Billangeon2006,Zakka-Bajjani2007,Basset2010,Parmentier2012,Altimiras2014,Parlavecchio2015}.
Although their results partially disagree with each other, 
it can be interpreted as due to different assumptions 
on the ways of measurements.
However, it is recently shown rigorously
that the observed fluctuation in macroscopic systems
is independent of details of the ways of measurements
when the measurement is performed 
as ideally as possible~\cite{Fujikura2016,Shimizu2017}.
This universal result clarified
when the FDT [in the form of Eq.~(\ref{eq:FDT_classical})]
holds and when 
it is violated~\cite{Fujikura2016,Shimizu2017,Kubo2018, Kubo2022}.

These studies on the FDT also demonstrate
that the nonequilibrium statistical mechanics is never trivial
even in the linear response regime.

\section{\label{sec:Summary}Summary}

We have studied how quantum mechanics is consistent with thermodynamics
with respect to 
infinitesimal transitions between equilibrium states.
We suppose that an isolated quantum many-body system
is prepared in an equilibrium state,
and then a parameter $f$ of the Hamiltonian is changed
by a small amount $\Delta f$,
which induces the unitary time evolution.
By inspecting the expectation values and the variances of
\emph{all} additive observables, 
we have investigated whether linear thermalization occurs, i.e., 
whether the system relaxes 
to the equilibrium state that is fully consistent with 
thermodynamics up to the linear order in $\Delta f$.

By comparing the long time average of the expectation value
of an arbitrary additive observable $\hat{A}$
with
its equilibrium value predicted by thermodynamics,
we have obtained \emph{Theorem} described in Sec.~\ref{sec:Summary_Theorem}.
It states roughly that 
the two values coincide for \emph{every} $\hat{A}$ 
if and only if they coincide for a \emph{single} additive observable.
This key observable is identified
as the additive observable $\hat{B}$ that is conjugate to $f$.
We have also pointed out that 
this condition for $\hat{B}$ 
is weaker than the ETH for $\hat{B}$.

To reinforce \emph{Theorem}, we have then proved two propositions.
\emph{Proposition}~1, described in Sec.~\ref{sec:Summary_Proposition1}, 
shows that 
the time fluctuation of the expectation value  (after changing $f$)
is macroscopically negligible for every $\hat{A}$
as long as 
the number of resonating pairs of energy eigenvalues (before changing $f$)
is not exponentially 
large.
\emph{Proposition}~2, described in Sec.~\ref{sec:Summary_Proposition2}, 
shows that 
the variances of all $\hat{A}$'s 
remain macroscopically negligible
if 
fluctuations of $\hat{A}$'s
in the initial equilibrium state
have reasonable magnitudes as Eq.~(\ref{eq:Assumption_4thMoment}).
Reinforced with 
these propositions, \emph{Theorem} ensures
that, under the reasonable conditions, 
the linear thermalization of the key observable $\hat{B}$
implies 
linear thermalization of all additive observables,
which means 
full consistency with thermodynamics.

We have also proved \emph{Proposition}~3 in Sec.~\ref{sec:Timescale},
which extends the above theorem and propositions to a longer timescale.
It show that when 
linear thermalization occurs,
it occurs in a 
timescale that is 
independent of the magnitude of $\Delta f$,
$t = \mathcal{O}(|\Delta f|^0)$,
and it lasts at least 
for a period not shorter than $o(1/\sqrt{|\Delta f|})$.
On the other hand, 
when linear thermalization does not occur
by $t = \mathcal{O}(|\Delta f|^0)$,
it does not occur 
at least in a period of $o(1/\sqrt{|\Delta f|})$.

Furthermore, we have shown that 
\emph{Theorem} has significant meanings about 
the generalized susceptibilities for cross responses,
which have been
attracting much attention in condensed matter physics.
\emph{Theorem} guarantees that 
the quantum-mechanical susceptibility
of \emph{every} additive observable $\hat{A}$
coincides with the thermodynamical one
if
those of the key 
observable $\hat{B}$ coincide. 
We have also obtained two corollaries, 
described in Sec.~\ref{sec:GeneralizedSuscep},
about response
of $\hat{B}$ to 
another 
parameter 
$f_A$ that is conjugate to an \emph{arbitrary} additive observable $\hat{A}$.
\emph{Corollary}~1 states that
the quantum-mechanical susceptibility of $\hat{B}$ to $f_A$
and the thermodynamical one
coincide 
if 
those of $\hat{B}$
to its own conjugate parameter 
$f$ coincide.
This is rephrased in terms of linear thermalization as \emph{Corollary}~2:
Linear thermalization of $\hat{B}$ against $\Delta f$ 
implies 
linear thermalization of $\hat{B}$ against 
any other $\Delta f_A$. 

We have demonstrated \emph{Theorem} by numerically
calculating the generalized susceptibilities in three models.
In a nonintegrable model,
we have first shown linear thermalization of $\hat{B}$,
and then confirmed that of 
other $\hat{A}$'s 
predicted by \emph{Theorem}.
We have also studied two integrable models; 
one can be mapped to noninteracting fermions, 
the other is solvable by the Bethe ansatz.
We have found that 
linear thermalization does not occur
either for $\hat{B}$ or for other $\hat{A}$'s
in these models.
This is again consistent with \emph{Theorem}.

Our results will dramatically reduce
the costs of experiments and 
theoretical calculations 
of linear thermalization and cross responses
because testing them for a single key observable
against the change of its conjugate parameter 
in a timescale of $\mathcal{O}(|\Delta f|^0)$
gives much information about those for all additive observables,
about those against the changes of any other parameters,
and about a longer timescale.

\begin{acknowledgments}
We thank N. Shiraishi, R. Hamazaki and T. Mori for 
helpful 
comments on 
the manuscript
and Y. Yoneta for fruitful discussions.
Y.C. is supported by 
Japan Society for the Promotion of Science KAKENHI Grant No. JP21J14313.
A.S. is supported by
Japan Society for the Promotion of Science KAKENHI Grant No. JP22H01142.
\end{acknowledgments}

\appendix

\section{\label{sec:Derivation}Derivation of relations used in proofs}

\subsection{\label{sec:Derivation_Exp_Ave}Relations used in proof of \emph{Theorem}}

From the linear response theory~\cite{Kubo1957,KTH,Wilcox1968,Suzuki1971},
the time-dependent quantum-mechanical susceptibility
defined by Eq.~(\ref{eq:DEF_chi^QM_t})
can be written as
\begin{align}
&\chi_{N}^{\text{QM}}(A|B;t)\notag\\
&=\frac{1}{N}\langle\beta_{0}\delta\hat{B};\delta\hat{A}\rangle^{\text{eq}}_{0}
-
\frac{1}{N}\langle\beta_{0}\delta\hat{B};\delta\hat{A}^{0}(t)\rangle^{\text{eq}}_{0},
\label{eq:chi^QM_t_BogoProd}
\end{align}
which results in 
\begin{align}
&\chi_{N}^{\text{QM}}(A|B)
\notag\\
&=
\frac{1}{N}\langle\beta_{0}\delta\hat{B};\delta\hat{A}\rangle^{\text{eq}}_{0}
-
\frac{1}{N}\langle\beta_{0}\delta\hat{B};\delta\overline{\hat{A}^{0}}\rangle^{\text{eq}}_{0}.
\label{eq:chi^QM_BogoProd}
\end{align}

For $\chi^{\text{TD}}_{N}(A|B)$, 
we obtain $\beta_{\Delta f}$
from energy conservation, Eq.~(\ref{eq:EnergyConservation}),
as
\begin{align}
\beta_{\Delta f}-\beta_{0}
=-\frac{
\langle\beta_{0}\delta\hat{B}\delta\hat{H}_{0}\rangle^{\text{eq}}_{0}
}{\langle(\delta\hat{H}_{0})^2\rangle^{\text{eq}}_{0}}\Delta f
+o(\Delta f).
\label{eq:beta_Df}
\end{align}
Here,
since $(\beta_{0},f_{0})$ is not at a phase transition point,
the specific heat
\begin{align}
c_{0}:=
\beta_{0}^2\frac{\langle(\delta\hat{H}_{0})^2\rangle^{\text{eq}}_{0}}{N}
\label{eq:c_0}
\end{align}
takes a positive finite value for sufficiently large $N$,
$c_{0}=\Theta(N^0)$.
By substituting Eq.~(\ref{eq:beta_Df}) into Eq.~(\ref{eq:DEF_chi^TD}),
we have 
\begin{align}
&\chi_{N}^{\text{TD}}(A|B)
\notag\\
&=
\frac{1}{N}\langle\beta_{0}\delta\hat{B};\delta\hat{A}\rangle^{\text{eq}}_{0}
-
\frac{
\langle\beta_{0}\delta\hat{B}\delta\hat{H}_{0}\rangle^{\text{eq}}_{0}
\langle\delta\hat{H}_{0}\delta\hat{A}\rangle^{\text{eq}}_{0}
}{N\langle(\delta\hat{H}_{0})^2\rangle^{\text{eq}}_{0}}.
\label{eq:chi^TD_BogoProd}
\end{align}

\subsection{\label{sec:Derivation_Exp_Fluc}Relations used in proof of \emph{Proposition}~$1$}

We derive Eqs.~(\ref{eq:Exp_Fluc_1}), (\ref{eq:Exp_Fluc_2}) and
(\ref{eq:Exp_Fluc_3}).

First, we derive Eq.~(\ref{eq:Exp_Fluc_1}).
The temporal fluctuation of $\langle \hat{A}^{\Delta f}(t)\rangle^{\text{eq}}_{0}$ can be divided into two terms,
\begin{align}
&\overline{\biggl|\frac{
\langle \hat{A}^{\Delta f}(t)\rangle^{\text{eq}}_{0}
-\overline{\langle \hat{A}^{\Delta f}(t)\rangle^{\text{eq}}_{0}}^{\mathcal{T}}
}{\Delta f N}\biggr|^2}^{\mathcal{T}}\notag\\
&=
\overline{\biggl|
\frac{
\langle \hat{A}^{\Delta f}(t)\rangle^{\text{eq}}_{0}
-\langle \hat{A}\rangle^{\text{eq}}_{0}
}{\Delta f N}
\biggr|^2}^{\mathcal{T}}
-\biggl|
\frac{
\overline{\langle \hat{A}^{\Delta f}(t)\rangle^{\text{eq}}_{0}}^{\mathcal{T}}
-\langle \hat{A}\rangle^{\text{eq}}_{0}
}{\Delta f N}
\biggr|^2.
\label{eq:Exp_Fluc_TwoParts}
\end{align}
We evaluate these two terms in the limit $\Delta f\to 0$.
The first term will be evaluated in Appendix~\ref{sec:Derivation_Timescale} as Eq.~(\ref{eq:FormulasInterchange_2}).
From Eq.~(\ref{eq:FormulasInterchange}), which will also be given in Appendix~\ref{sec:Derivation_Timescale},
the second term is evaluated as
\begin{align}
\lim_{\Delta f\to 0}
\biggl|
\frac{
\overline{\langle \hat{A}^{\Delta f}(t)\rangle^{\text{eq}}_{0}}^{\mathcal{T}}
-\langle \hat{A}\rangle^{\text{eq}}_{0}
}{\Delta f N}
\biggr|^2
=
\Bigl(\overline{\chi_{N}^{\text{QM}}(A|B;t)}^{\mathcal{T}}\Bigr)^2.
\label{eq:Exp_Fluc_chi^2}
\end{align}
Therefore,
Eq.~(\ref{eq:Exp_Fluc_TwoParts}) is evaluated,
in the limit $\Delta f\to 0$, as
\begin{align}
&\lim_{\Delta f\to 0}
\overline{\biggl|\frac{
\langle \hat{A}^{\Delta f}(t)\rangle^{\text{eq}}_{0}
-\overline{\langle \hat{A}^{\Delta f}(t)\rangle^{\text{eq}}_{0}}^{\mathcal{T}}
}{\Delta f N}\biggr|^2}^{\mathcal{T}}
\notag\\
&=
\overline{
\Bigl(
\chi_{N}^{\text{QM}}(A|B;t)
-\overline{\chi_{N}^{\text{QM}}(A|B;t)}^{\mathcal{T}}
\Bigr)^2
}^{\mathcal{T}}
\notag\\
&=
\overline{
\bigl(\chi_{N}^{\text{QM}}(A|B;t)-\chi^{\text{QM}}_{N}(A|B)\bigr)^2
}^{\mathcal{T}}
\notag\\
&\quad -
\Bigl(
\overline{\chi_{N}^{\text{QM}}(A|B;t)}^{\mathcal{T}}-\chi^{\text{QM}}_{N}(A|B)
\Bigr)^2.
\end{align}
By taking the limit $\mathcal{T}\to\infty$,
we obtain Eq.~(\ref{eq:Exp_Fluc_1}).

Next, we derive Eq.~(\ref{eq:Exp_Fluc_2}).
Using 
Eqs.~(\ref{eq:BogoProd}), (\ref{eq:chi^QM_t_BogoProd}), (\ref{eq:chi^QM_BogoProd}) and
the eigenstates of $\hat{H}_{0}$, we have
\begin{align}
&\chi_{N}^{\text{QM}}(A|B;t)-\chi^{\text{QM}}_{N}(A|B)\notag\\
&=\frac{1}{N}\langle\beta_{0}\delta\hat{B};\delta\overline{\hat{A}^{0}}\rangle^{\text{eq}}_{0}
-
\frac{1}{N}\langle\beta_{0}\delta\hat{B};\delta\hat{A}^{0}(t)\rangle^{\text{eq}}_{0}
\notag\\
&=
-\sum_{\nu_{1},\nu_{2}(E_{\nu_{1}}\neq E_{\nu_{2}})}\frac{\beta_{0}}{N}
\int_{0}^{1}\dd{\lambda}
\rho_{\nu_{1}}^{1-\lambda}
\langle \nu_{1}|\hat{B}|\nu_{2}\rangle \rho_{\nu_{2}}^{\lambda}
\mel{\nu_{2}}{\hat{A}}{\nu_{1}}
\notag\\
&\hspace{110pt}\times e^{-i(E_{\nu_{1}}-E_{\nu_{2}})t},
\end{align}
where 
\begin{align}
\rho_{\nu}:=e^{-\beta_{0}E_{\nu}}/Z(\beta_{0},f_{0})
\label{eq:rho_n_can}
\end{align}
and we have used
\begin{align}
\overline{\hat{A}^{0}}
=\sum_{\nu_{1},\nu_{2} \ (E_{\nu_{1}}=E_{\nu_{2}})}
\ket{\nu_{1}}\mel{\nu_{1}}{\hat{A}}{\nu_{2}}\bra{\nu_{2}}.
\end{align}
Hence we have
\begin{align}
&\chi_{N}^{\text{QM}}(A|B;t)-\chi^{\text{QM}}_{N}(A|B)\notag\\
&=
-\sum_{\nu_{1},\nu_{2} \ (E_{\nu_{1}}\neq E_{\nu_{2}})}
\frac{1}{N}
\frac{\rho_{\nu_{2}}-\rho_{\nu_{1}}}{E_{\nu_{1}}-E_{\nu_{2}}}
\langle \nu_{1}|\hat{B}|\nu_{2}\rangle
\langle \nu_{2}|\hat{A}|\nu_{1}\rangle\notag\\
&\hspace{110pt}\times e^{-i(E_{\nu_{1}}-E_{\nu_{2}})t}
\notag\\
&=
-\sum_{\nu_{1},\nu_{2} \ (E_{\nu_{1}}\neq E_{\nu_{2}})}
v_{(\nu_{1},\nu_{2})}
e^{-i(E_{\nu_{1}}-E_{\nu_{2}})t},
\end{align}
where, for $E_{\nu_{1}}\neq E_{\nu_{2}}$,
we have introduced
\begin{align}
v_{(\nu_{1},\nu_{2})}
&:=
\frac{1}{N}\frac{\rho_{\nu_{2}}-\rho_{\nu_{1}}}{E_{\nu_{1}}-E_{\nu_{2}}}
\langle \nu_{1}|\hat{B}|\nu_{2}\rangle
\langle \nu_{2}|\hat{A}|\nu_{1}\rangle\\
&=v_{(\nu_{2},\nu_{1})}^*.
\end{align}
Using this expression, 
the time fluctuation of 
$\chi_{N}^{\text{QM}}(A|B;t)$
is evaluated as
\begin{align}
&\lim_{\mathcal{T}\to\infty}
\overline{
\bigl(\chi_{N}^{\text{QM}}(A|B;t)-\chi^{\text{QM}}_{N}(A|B)\bigr)^2
}^{\mathcal{T}}\notag\\
&=
\sum_{(\nu_{1},\nu_{2})\in\mathcal{G}}
\sum_{(\nu_{3},\nu_{4})\in\mathcal{G}}
v_{(\nu_{3},\nu_{4})}^* v_{(\nu_{1},\nu_{2})}
\notag\\
&\hspace{80pt}\times\lim_{\mathcal{T}\to\infty}\overline{e^{i(E_{\nu_{3}}-E_{\nu_{4}}-E_{\nu_{1}}+E_{\nu_{2}})t}}^{\mathcal{T}}
\label{eq:SusFluc_FiniteTime}
\notag\\
&=
\sum_{(\nu_{1},\nu_{2})\in\mathcal{G}}
\sum_{(\nu_{3},\nu_{4})\in\mathcal{G}}
v_{(\nu_{3},\nu_{4})}^* v_{(\nu_{1},\nu_{2})}
\delta_{E_{\nu_{1}}-E_{\nu_{2}},E_{\nu_{3}}-E_{\nu_{4}}}.
\end{align}
By using $|v_{(\nu_{3},\nu_{4})}^* v_{(\nu_{1},\nu_{2})}|\le (|v_{(\nu_{3},\nu_{4})}|^2+|v_{(\nu_{1},\nu_{2})}|^2)/2$,
we have
\begin{align}
&\lim_{\mathcal{T}\to\infty}
\overline{
\bigl(\chi_{N}^{\text{QM}}(A|B;t)-\chi^{\text{QM}}_{N}(A|B)\bigr)^2
}^{\mathcal{T}}\notag\\
&\le
\sum_{(\nu_{1},\nu_{2})\in\mathcal{G}}
|v_{(\nu_{1},\nu_{2})}|^2
\sum_{(\nu_{3},\nu_{4})\in\mathcal{G}}
\delta_{E_{\nu_{1}}-E_{\nu_{2}},E_{\nu_{3}}-E_{\nu_{4}}
\notag}\\
&\le
D_{\text{res}}
\sum_{(\nu_{1},\nu_{2})\in\mathcal{G}}
|v_{(\nu_{1},\nu_{2})}|^2.
\end{align}
The term $|v_{(\nu_{1},\nu_{2})}|^2$ can be bounded from above
by using
\begin{align}
\frac{1}{\beta_{0}}\frac{\rho_{\nu_{2}}-\rho_{\nu_{1}}}{E_{\nu_{1}}-E_{\nu_{2}}}
\le 
\frac{\rho_{\nu_{2}}+\rho_{\nu_{1}}}{2},
\label{eq:sinh_cosh_can}
\end{align}
which
follows from 
the inequality,
$\sinh x/x\le \cosh x$.
Combining this inequality with
$|(\rho_{\nu_{2}}+\rho_{\nu_{1}})/2|^2\le(\rho_{\nu_{2}}^2+\rho_{\nu_{1}}^2)/2$,
we have
\begin{align}
&\sum_{(\nu_{1},\nu_{2})\in\mathcal{G}}
|v_{(\nu_{1},\nu_{2})}|^2\notag\\
&\le \frac{(\beta_{0})^2}{N^2}
\sum_{(\nu_{1},\nu_{2})\in\mathcal{G}}
\frac{\rho_{\nu_{2}}^2+\rho_{\nu_{1}}^2}{2}
|\langle \nu_{1}|\hat{B}|\nu_{2}\rangle|^2
|\langle \nu_{2}|\hat{A}|\nu_{1}\rangle|^2
\notag\\
&\le \frac{(\beta_{0})^2}{N^2}\|\hat{B}\|_{\infty}^2
\sum_{(\nu_{1},\nu_{2})\in\mathcal{G}}
\rho_{\nu_{1}}^2
|\langle \nu_{2}|\hat{A}|\nu_{1}\rangle|^2
\notag\\
&\le
\frac{(\beta_{0})^2\|\hat{A}\|_{\infty}^2\|\hat{B}\|_{\infty}^2}{N^2}
\Tr \bigl[(\hat{\rho}^{\text{eq}}_{0})^2\bigr].
\end{align}
From these expressions, we obtain Eq.~(\ref{eq:Exp_Fluc_2}).

Finally, we show that
the purity of $\hat{\rho}^{\text{eq}}_{0}$ is exponentially small with respect to $N$,
i.e., Eq.~(\ref{eq:Exp_Fluc_3}).
The purity of canonical ensemble satisfies
\begin{align}
&-\ln \Tr\bigl[(\hat{\rho}^{\text{eq}}_{0})^2\bigr]\notag\\
&=
S_{\text{vN}}[\hat{\rho}^{\text{can}}(2\beta_{0},f_{0})]
+2D(\hat{\rho}^{\text{can}}(2\beta_{0},f_{0})|\hat{\rho}^{\text{eq}}_{0})
\\
&\ge
S_{\text{vN}}[\hat{\rho}^{\text{can}}(2\beta_{0},f_{0})].
\label{eq:purity_KL}
\end{align}
Here
$S_{\text{vN}}[\hat{\rho}]=-\Tr\bigl[\hat{\rho}\ln\hat{\rho}\bigr]$ is the von Neumann entropy
and $D(\hat{\rho}|\hat{\sigma})=-\Tr\bigl[\hat{\rho}\ln\hat{\sigma}\bigr]-S_{\text{vN}}[\hat{\rho}]$ is the quantum relative entropy
for arbitrary density matrices $\hat{\rho}$ and $\hat{\sigma}$.
The above inequality is obtained from
$D(\hat{\rho}|\hat{\sigma})\ge 0$.
Since $S_{\text{vN}}[\hat{\rho}^{\text{can}}(2\beta_{0},f_{0})]$ gives the thermodynamic entropy of the equilibrium state $(2\beta_{0},f_{0})$ in the limit $N\to\infty$,
it satisfies
\begin{align}
S_{\text{vN}}[\hat{\rho}^{\text{can}}(2\beta_{0},f_{0})]
=\Theta(N).
\label{eq:S_vN=Theta(N)}
\end{align}
Equations~(\ref{eq:purity_KL}) and (\ref{eq:S_vN=Theta(N)}) yield Eq.~(\ref{eq:Exp_Fluc_3}).

Note that
Eq.~(\ref{eq:Exp_Fluc_2}) can be extended to
the case where $\mathcal{T}$ is finite,
by evaluating the term 
$\overline{e^{i(E_{\nu_{3}}-E_{\nu_{4}}-E_{\nu_{1}}+E_{\nu_{2}})t}}^{\mathcal{T}}$ 
in Eq.~(\ref{eq:SusFluc_FiniteTime})
more carefully,
as in Short and Farrelly~\cite{Short2012}.
However, 
such extension is meaningful
only when $1/\mathcal{T}$ is sufficiently small compared to the mean level spacing,
which means that $\mathcal{T}$ has to be exponentially 
large with respect to $N$~\cite{deOliveira2018}.
Hence we think that
the difference between such extension and the original result~(\ref{eq:Exp_Fluc_2}) 
is physically unimportant,
and for simplicity,
we omit the precise description of such a extension.
We also remark that a result similar to Eq.~(\ref{eq:Exp_Fluc_2}) was derived in Ref.~\cite{Alhambra2020}.

\subsection{\label{sec:Derivation_Var}Relation used in proof of \emph{Proposition}~2}

First we derive Eq.~(\ref{eq:VarChangeRate_can}).
Its left hand side 
consists of four terms,
\begin{align}
&\text{Var}_{0}[\hat{A}^{\Delta f}(t)]
-\text{Var}_{0}[\hat{A}]
\notag\\
&=\langle\bigl(\hat{A}^{\Delta f}(t)\bigr)^2\rangle^{\text{eq}}_{0}
-\bigl(\langle\hat{A}^{\Delta f}(t)\rangle^{\text{eq}}_{0}\bigr)^2
\notag\\
&\quad -\langle\hat{A}^2\rangle^{\text{eq}}_{0}
+\bigl(\langle\hat{A}\rangle^{\text{eq}}_{0}\bigr)^2.
\label{eq:Var_Change}
\end{align}
For the second and the fourth terms,
we have
\begin{align}
&\lim_{\Delta f\to 0}
\frac{
\bigl(\langle\hat{A}^{\Delta f}(t)\rangle^{\text{eq}}_{0}\bigr)^2
-\bigl(\langle\hat{A}\rangle^{\text{eq}}_{0}\bigr)^2
}{\Delta f}\notag\\
&=\lim_{\Delta f\to 0}
\frac{
\langle\hat{A}^{\Delta f}(t)\rangle^{\text{eq}}_{0}
\bigl(\langle\hat{A}^{\Delta f}(t)\rangle^{\text{eq}}_{0}
-\langle\hat{A}\rangle^{\text{eq}}_{0}\bigr)
}{\Delta f}\notag\\
&\hspace{10pt}
+\lim_{\Delta f\to 0}
\frac{\bigl(\langle\hat{A}^{\Delta f}(t)\rangle^{\text{eq}}_{0}
-\langle\hat{A}\rangle^{\text{eq}}_{0}\bigr)\langle\hat{A}\rangle^{\text{eq}}_{0}
}{\Delta f}
\notag\\
&=2\langle\hat{A}\rangle^{\text{eq}}_{0}
N\chi_{N}^{\text{QM}}(A|B;t),
\label{eq:Leibniz_chi^QM}
\end{align}
which corresponds to the Leibniz rule of calculus.
On the other hand,
because the first and the third terms of Eq.~(\ref{eq:Var_Change})
give the time-dependent susceptibility of $\hat{A}^2$
and
Eq.~(\ref{eq:chi^QM_t_BogoProd}) 
is also applicable to this susceptibility,
we have
\begin{align}
&\lim_{\Delta f\to 0}
\frac{
\langle\bigl(\hat{A}^{\Delta f}(t)\bigr)^2\rangle^{\text{eq}}_{0}
-\langle\hat{A}^2\rangle^{\text{eq}}_{0}
}{\Delta f}\notag\\
&=
\langle\beta_{0}\delta\hat{B};\hat{A}^2\rangle^{\text{eq}}_{0}
-\langle\beta_{0}\delta\hat{B};\bigl(\hat{A}^{0}(t)\bigr)^2\rangle^{\text{eq}}_{0}
\end{align}
Combining these with Eq.~(\ref{eq:chi^QM_t_BogoProd}),
we have
\begin{align}
&\lim_{\Delta f\to 0}
\frac{
\text{Var}_{0}[\hat{A}^{\Delta f}(t)]
-\text{Var}_{0}[\hat{A}]
}{\Delta f}\notag\\
&=
\langle\beta_{0}\delta\hat{B};\hat{A}^2\rangle^{\text{eq}}_{0}
-\langle\beta_{0}\delta\hat{B};\bigl(\hat{A}^{0}(t)\bigr)^2\rangle^{\text{eq}}_{0}
\notag\\
&\qquad-2\langle\hat{A}\rangle^{\text{eq}}_{0}
\bigl(\langle\beta_{0}\delta\hat{B};\hat{A}\rangle^{\text{eq}}_{0}
-\langle\beta_{0}\delta\hat{B};\hat{A}^{0}(t)\rangle^{\text{eq}}_{0}\bigr),
\label{eq:VarChangeRate_Tr}
\end{align}
which results in Eq.~(\ref{eq:VarChangeRate_can}).

Next,
inserting Eq.~(\ref{eq:sinh_cosh_can}) of Appendix~\ref{sec:Derivation_Exp_Fluc}
into Eq.~(\ref{eq:BogoProd}),
we can show 
the following inequality between
the canonical correlation 
and the symmetrized correlation:
\begin{align}
\langle\hat{X};\hat{X}\rangle^{\text{eq}}_{0}
\le
\frac{1}{2}\langle\hat{X}^{\dagger}\hat{X}+\hat{X}\hat{X}^{\dagger}\rangle^{\text{eq}}_{0}
\label{eq:BogoProd_Symmetrized}
\end{align}
for an arbitrary operator $\hat{X}$.
This was 
proved by Bogoliubov~\cite{Bogoliubov1962,Mermin1966}.

\subsection{\label{sec:Derivation_Timescale}Relations used in proof of \emph{Proposition}~3}

In this Appendix,
we show 
Eqs.~(\ref{eq:finite_Df_T_Exp_Ave})--(\ref{eq:finite_Df_T_Var}) 
of Sec.~\ref{sec:Proof_Timescale}.
We also show that
the time integration and the limit $\Delta f\to 0$
are interchangeable.

We introduce a function 
\begin{align}
\phi(f,t)&:=\langle e^{i\hat{H}(f)t}\hat{A}e^{-i\hat{H}(f)t}\rangle^{\text{eq}}_{0}/N.
\end{align}
This function satisfies
\begin{align}
\phi(f_{0}+\Delta f,t) - \phi(f_0,t)
&=
\frac{\langle \hat{A}^{\Delta f}(t)\rangle^{\text{eq}}_{0}
-\langle \hat{A}\rangle^{\text{eq}}_{0}}
{N},
\\
\frac{\partial\phi}{\partial f}(f_{0},t)
&=\chi_{N}^{\text{QM}}(A|B;t),
\end{align}
where $\chi_{N}^{\text{QM}}(A|B;t)$ 
is defined by Eq.~(\ref{eq:DEF_chi^QM_t}).
According to Taylor's theorem, for each $t$
there is a constant $\theta_{t}\in[0,1]$
such that
\begin{align}
&\frac{
\langle \hat{A}^{\Delta f}(t)\rangle^{\text{eq}}_{0}
-\langle \hat{A}\rangle^{\text{eq}}_{0}
}{\Delta f N}
-\chi_{N}^{\text{QM}}(A|B;t)
\notag\\
&=\frac{1}{2}
\frac{\partial^2\phi}{\partial f^2}(f_{0}+\theta_{t}\Delta f,t)\Delta f.
\label{eq:phi_Taylor}
\end{align}
(Here the differentiability of $\phi(f,t)$ with respect to $f$
follows from the concrete expression~(\ref{eq:phi''_Integral}) below.)

Let us derive
an upper bound of
the absolute value of the right hand side of Eq.~(\ref{eq:phi_Taylor}).
From the identity
\begin{align}
\pdv{f}e^{i\hat{H}(f)t}=\int_{0}^{t}\dd{t_{1}}e^{i\hat{H}(f)(t-t_{1})}
i\frac{\partial\hat{H}}{\partial f}(f)e^{i\hat{H}(f)t_{1}},
\end{align}
we have
\begin{widetext}
\begin{align}
&\frac{\partial^2\phi}{\partial f^2}(f,t)
=
\int_{0}^{t}\dd{t_{1}}\Bigl\langle e^{i\hat{H}(f)(t-t_{1})}
\Bigl[i\frac{\partial^2\hat{H}}{\partial f^2}(f),
e^{i\hat{H}(f)t_{1}}\frac{\hat{A}}{N}e^{-i\hat{H}(f)t_{1}}\Bigr]
e^{-i\hat{H}(f)(t-t_{1})}\Bigl\rangle^{\text{eq}}_{0}
\notag\\
&\hspace{12pt}+2\int_{0}^{t}\dd{t_{1}}\int_{0}^{t_{1}}\dd{t_{2}}
\Bigl\langle e^{i\hat{H}(f)(t-t_{1})}
\Bigl[i\frac{\partial\hat{H}}{\partial f}(f),
e^{i\hat{H}(f)(t_{1}-t_{2})}
\Bigl[i\frac{\partial\hat{H}}{\partial f}(f),
e^{i\hat{H}(f)t_{2}}\frac{\hat{A}}{N}e^{-i\hat{H}(f)t_{2}}\Bigr]
e^{-i\hat{H}(f)(t_{1}-t_{2})}\Bigr]
e^{-i\hat{H}(f)(t-t_{1})}\Bigl\rangle^{\text{eq}}_{0}.
\label{eq:phi''_Integral}
\end{align}
Since the integrands are bounded from above
by using the operator norms of $\hat{A}$ and the derivatives of $\hat{H}(f)$,
we have
\begin{align}
\Bigl|\frac{\partial^2\phi}{\partial f^2}(f_{0}+\theta_{t}\Delta f,t)\Bigr|
&
\le
\frac{\|\hat{A}\|_{\infty}}{N}
\Bigl(2\Bigl\|\frac{\mathrm{d}^2\hat{H}}{\mathrm{d} f^2}(f_{0}+\theta_{t}\Delta f)\Bigr\|_{\infty}\Bigr)
|t|
+
\frac{\|\hat{A}\|_{\infty}}{N}
\Bigl(2\Bigl\|\frac{\mathrm{d}\hat{H}}{\mathrm{d} f}(f_{0}+\theta_{t}\Delta f)\Bigr\|_{\infty}\Bigr)^2
|t|^2
\label{eq:phi''_bound_1}
\\
&
\le
\|\hat{A}\|_{\infty}(C_{2}|t|+C_{1}^2|t|^2)/N.
\label{eq:phi''_bound}
\end{align}
\end{widetext}
Here, we have introduced two constants
\begin{align}
C_{1}&:=2
\sup_{f \ \text{s.t.} \ |f-f_{0}|\le|\Delta f|}
\Bigl\|\frac{\mathrm{d}\hat{H}}{\mathrm{d} f}(f)\Bigr\|_{\infty},
\end{align}
\begin{align}
C_{2}&:=2
\sup_{f \ \text{s.t.} \ |f-f_{0}|\le|\Delta f|}
\Bigl\|\frac{\mathrm{d}^2\hat{H}}{\mathrm{d} f^2}(f)\Bigr\|_{\infty},
\end{align}
which are of $\mathcal{O}(|\Delta f|^0)$ 
since they decrease monotonically as $|\Delta f|\to 0$.
Using Eqs.~(\ref{eq:phi_Taylor}) and (\ref{eq:phi''_bound}),
we obtain Eq.~(\ref{eq:finite_Df_T_Exp_Ave}).

In addition,
using Eq.~(\ref{eq:phi_Taylor}) we have
\begin{align}
&\Bigl(\frac{
\langle \hat{A}^{\Delta f}(t)\rangle^{\text{eq}}_{0}
-\langle \hat{A}\rangle^{\text{eq}}_{0}
}{\Delta f N}\Bigr)^2
-\bigl(\chi_{N}^{\text{QM}}(A|B;t)\bigr)^2
\notag\\
&=
\chi_{N}^{\text{QM}}(A|B;t)
\frac{\partial^2\phi}{\partial f^2}(f_{0}+\theta_{t}\Delta f,t)
\Delta f
\notag\\
&\hspace{12pt}+\Bigl(\frac{1}{2}
\frac{\partial^2\phi}{\partial f^2}(f_{0}+\theta_{t}\Delta f,t)\Delta f\Bigr)^2.
\label{eq:Interchange_chi^2}
\end{align}
Furthermore, 
using Eqs.~(\ref{eq:VarChangeRate_CauchySchwarz}), (\ref{eq:BogoProd_Symmetrized}) and (\ref{eq:chi^QM_t_BogoProd}),
we find that
$|\chi_{N}^{\text{QM}}(A|B;t)|$ 
is bounded from above
by a quantity
independent of $t$,
\begin{align}
|\chi_{N}^{\text{QM}}(A|B;t)|\le
\frac{2\beta_{0}}{N}
\sqrt{\langle(\delta\hat{B})^2\rangle^{\text{eq}}_{0}
\langle(\delta\hat{A})^2\rangle^{\text{eq}}_{0}
}.
\label{eq:chi_t_bound}
\end{align}
Combining this with Eqs.~(\ref{eq:phi''_bound}) and (\ref{eq:Interchange_chi^2}),
we obtain Eq.~(\ref{eq:finite_Df_T_Exp_Fluc}).

Next we introduce a function 
\begin{align}
\psi(f,t)&:=\langle e^{i\hat{H}(f)t}\hat{A}^2e^{-i\hat{H}(f)t}\rangle^{\text{eq}}_{0}/N^2.
\end{align}
This function satisfies
\begin{align}
\frac{\partial\psi}{\partial f}(f_{0},t)
=
\tilde{\chi}_{N}(A^2;t),
\end{align}
where $\tilde{\chi}_{N}(A^2;t)$
is defined by Eq.~(\ref{eq:DEF_tilde_chi_t}).
According to Taylor's theorem, 
for each $t$
there is a constant $\theta_{t}\in [0,1]$ such that
\begin{align}
&\frac{
\langle \bigl(\hat{A}^{\Delta f}(t)\bigr)^2\rangle^{\text{eq}}_{0}
-\langle \hat{A}^2\rangle^{\text{eq}}_{0}
}{\Delta f N^2}
-
\tilde{\chi}_{N}(A^2;t)
\notag\\
&=\frac{1}{2}
\frac{\partial^2\psi}{\partial f^2}(f_{0}+\theta_{t}\Delta f,t)\Delta f.
\label{eq:psi_Taylor}
\end{align}
In a way similar to the derivation of
Eq.~(\ref{eq:phi''_bound}),
we have
\begin{align}
\Bigl|\frac{\partial^2\psi}{\partial f^2}(f_{0}+\theta_{t}\Delta f,t)\Bigr|
&
\le
\|\hat{A}\|_{\infty}^2(C_{2}|t|+C_{1}^2|t|^2)/N^2.
\label{eq:psi''_bound}
\end{align}
Combining Eqs.~(\ref{eq:psi_Taylor}) and (\ref{eq:psi''_bound}),
we obtain Eq.~(\ref{eq:finite_Df_T_Var}).

Finally, we notice 
that,
in the limit $\Delta f\to 0$
Eqs.~(\ref{eq:finite_Df_T_Exp_Ave})--(\ref{eq:finite_Df_T_Var}) yield
\begin{align}
&\lim_{\Delta f\to 0}
\frac{
\overline{\langle \hat{A}^{\Delta f}(t)\rangle^{\text{eq}}_{0}}^{\mathcal{T}}
-\langle \hat{A}\rangle^{\text{eq}}_{0}
}{\Delta f N}
=
\overline{\chi_{N}^{\text{QM}}(A|B;t)}^{\mathcal{T}},
\label{eq:FormulasInterchange}\\
&\lim_{\Delta f\to 0}
\overline{\biggl|
\frac{
\langle \hat{A}^{\Delta f}(t)\rangle^{\text{eq}}_{0}
-\langle \hat{A}\rangle^{\text{eq}}_{0}
}{\Delta f N}
\biggr|^2}^{\mathcal{T}}
=
\overline{
\bigl(\chi_{N}^{\text{QM}}(A|B;t)\bigr)^2
}^{\mathcal{T}},
\label{eq:FormulasInterchange_2}\\
&\lim_{\Delta f\to 0}
\frac{
\overline{\langle \bigl(\hat{A}^{\Delta f}(t)\bigr)^2\rangle^{\text{eq}}_{0}}^{\mathcal{T}}
-\langle \hat{A}^2\rangle^{\text{eq}}_{0}
}{\Delta f N^2}
=
\overline{
\tilde{\chi}_{N}(A^2;t)
}^{\mathcal{T}},
\label{eq:FormulasInterchange_3}
\end{align}
which show that 
the time integration and the limit $\Delta f\to 0$
can be interchanged.

\section{\label{sec:AdditionalNumerical}Additional numerical results}

\subsection{\label{sec:AdditionalNumerical_Suscep_I_II}Susceptibilities of Model~I and II}

In this Appendix,
we perform additional calculations of the susceptibilities in Model~I and II.
As explained in Sec.~\ref{sec:Models},
we choose the quench parameter as $f=J_{zz}$,
and hence
$\hat{B}$ is given by Eq.~(\ref{eq:Mzz}).
We take the initial state
as the canonical Gibbs state $\hat{\rho}^{\text{eq}}_{0}$ given by Eq.~(\ref{eq:rho_0})
with the inverse temperature $\beta_{0}=0.15$.
Using Eqs.~(\ref{eq:chi^QM_BogoProd}) and (\ref{eq:chi^TD_BogoProd}) of Appendix~\ref{sec:Derivation_Exp_Ave},
we calculate $\chi^{\text{QM}}(A|B)$ and $\chi^{\text{TD}}(A|B)$
by the exact diagonalization from $N=6$ to $19$.

The results are plotted in Figs.~\ref{fig:chi(xx)_XYZ,XY} and \ref{fig:chi(z1z)_XYZ,XY}.
See Sec.~\ref{sec:Numerical_chi_A} for discussions on these results.

\begin{figure}
\includegraphics[width=\linewidth]{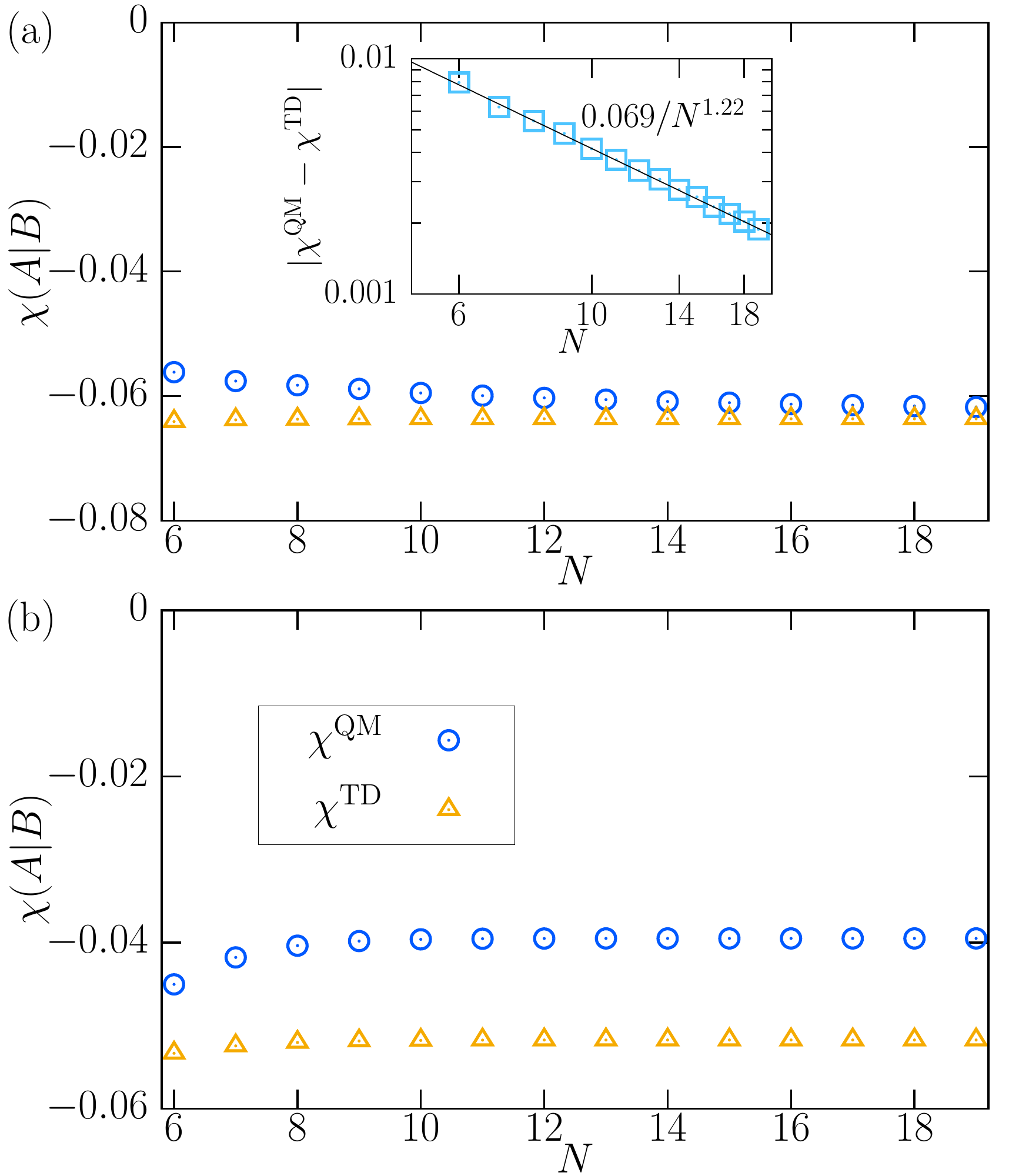}
\caption{\label{fig:chi(xx)_XYZ,XY}%
The same plots as in Fig.~\ref{fig:chi(x)_XYZ,XY} of Sec.~\ref{sec:Numerical_chi_A}
for $\hat{A}=\hat{M}^{xx}$.
In the inset of (a),
the points can be fitted by a function $a/N^b$ with constants $a=0.069(2)$, $b=1.22(1)$,
and the solid line shows the fitting function $0.069/N^{1.22}$.
}
\end{figure}%

\begin{figure}
\includegraphics[width=\linewidth]{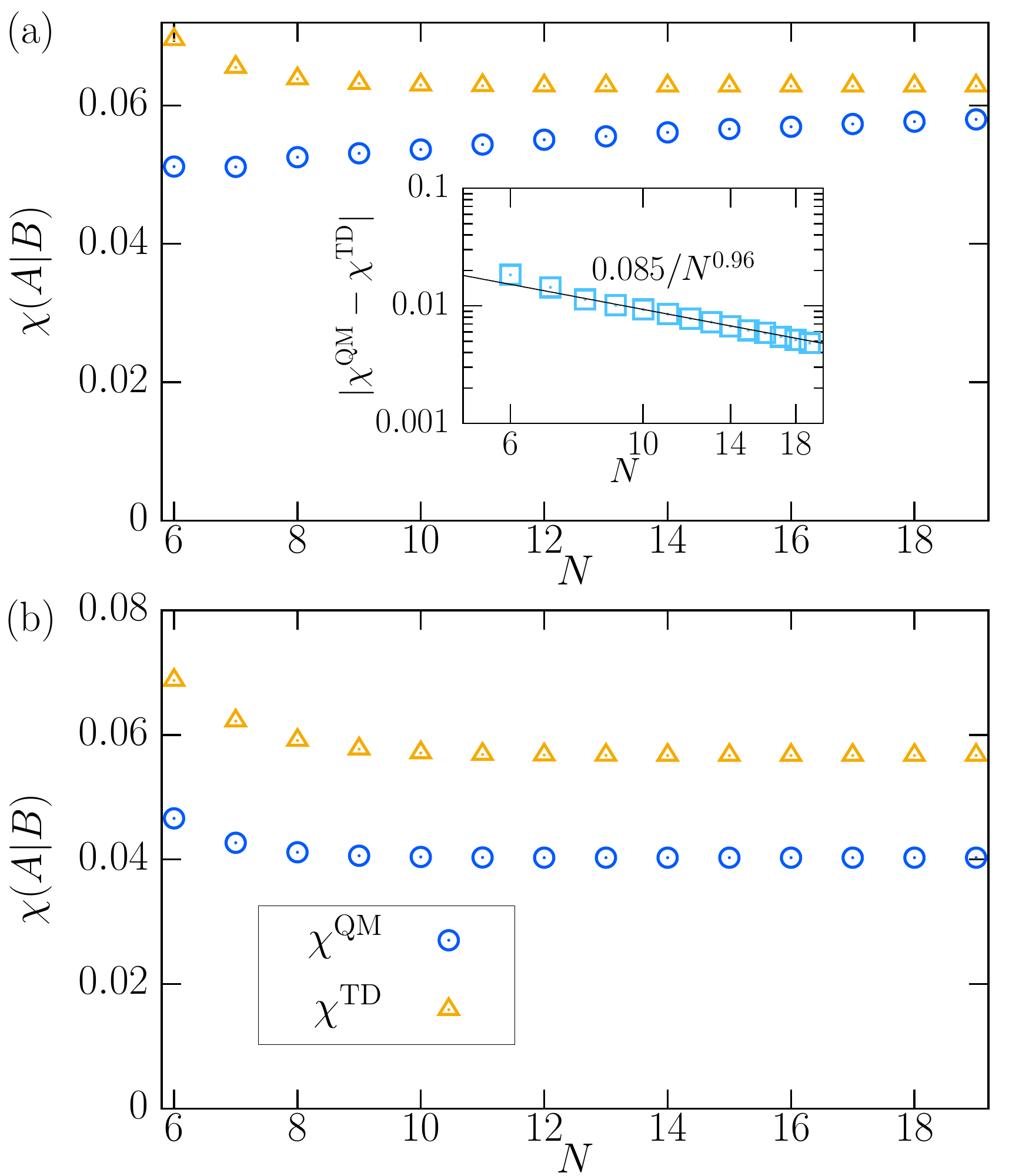}
\caption{\label{fig:chi(z1z)_XYZ,XY}%
The same plots as in Fig.~\ref{fig:chi(x)_XYZ,XY} of Sec.~\ref{sec:Numerical_chi_A}
for $\hat{A}=\hat{M}^{z1z}$.
In the inset of (a),
the points can be fitted by a function $a/N^b$ with constants $a=0.085(3)$, $b=0.96(1)$,
and the solid line shows the fitting function $0.085/N^{0.96}$.
}
\end{figure}%

\subsection{\label{sec:NumericalInteracting}Susceptibilities of Model~III}

In this Appendix, 
we calculate the susceptibilities of Model~III (see Table~\ref{tbl:Models})
in the same way as described in Sec.~\ref{sec:Models} or Appendix~\ref{sec:AdditionalNumerical_Suscep_I_II}.

\begin{figure}
\includegraphics[width=\linewidth]{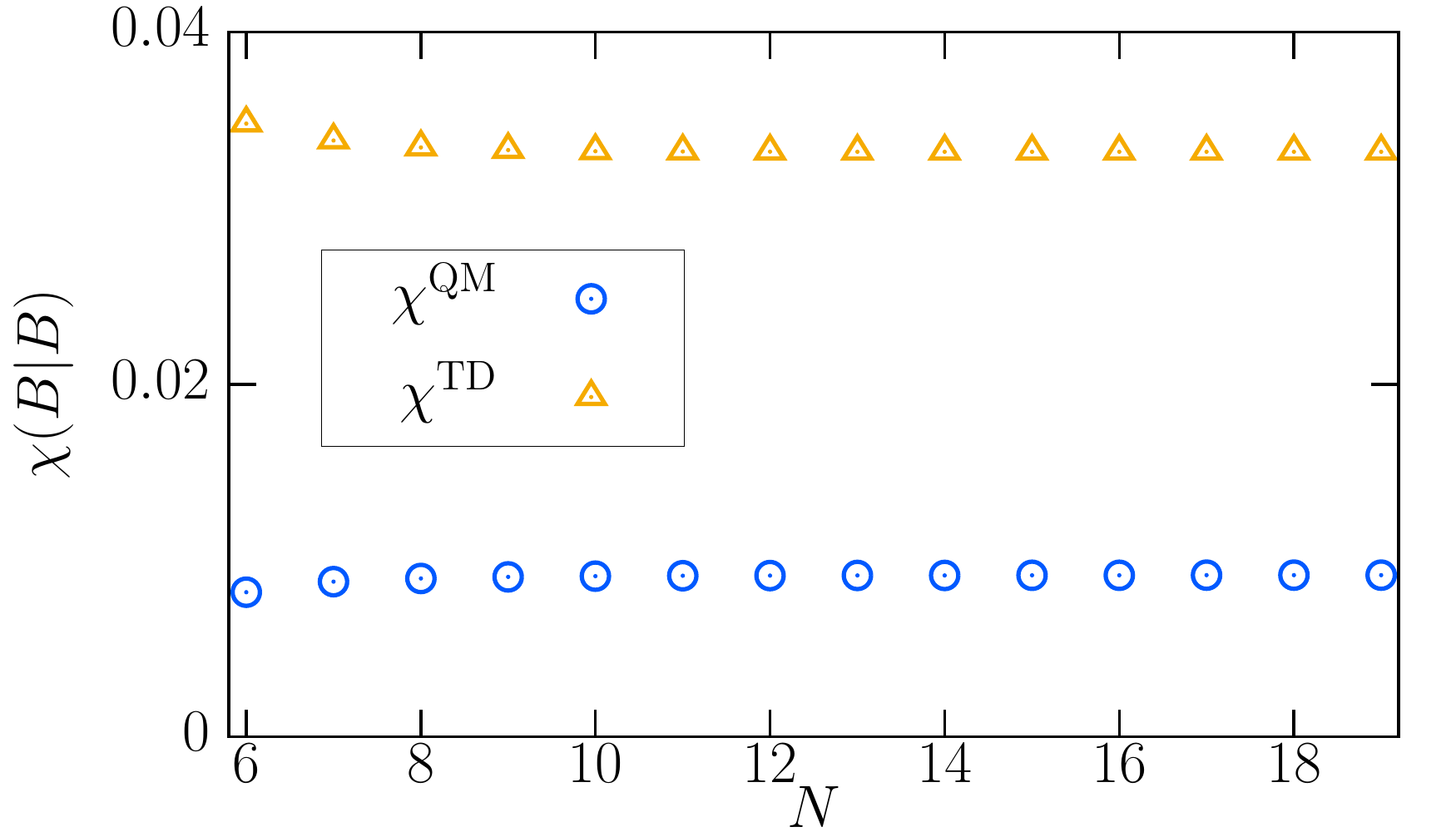}
\caption{\label{fig:chi(zz)_XXZ}%
$\chi^{\text{QM}}_{N}(B|B)$ (blue circle) and $\chi^{\text{TD}}_{N}(B|B)$ (orange triangle)
against the system size $N$
in an integrable system (Model~III).
The quench parameter is $f=J_{zz}$ and hence $\hat{B}=\hat{M}^{zz}$.
This plot reveals that
the condition~(\ref{eq:chi^QM(B|B)-chi^TD(B|B)}) is violated in Model~III.
}
\end{figure}

First we investigate whether Model~III satisfies 
condition~(\ref{eq:chi^QM(B|B)-chi^TD(B|B)}).
In Fig.~\ref{fig:chi(zz)_XXZ},
$\chi_{N}^{\text{QM}}(B|B)$ and $\chi_{N}^{\text{TD}}(B|B)$ of Model~III
are plotted
against the system size $N$.
This shows
Model~III violates condition~(\ref{eq:chi^QM(B|B)-chi^TD(B|B)}).
Hence, our \emph{Theorem} 
(in the form rephrased in Sec.~\ref{sec:GeneralizedSuscep})
indicates that \emph{some} of the additive observables $\hat{A}$
do not satisfy Eq.~(\ref{eq:chi^QM(A|B)-chi^TD(A|B)}).
To investigate this point, 
we calculate, 
as in Sec.~\ref{sec:Numerical_chi_A},
$\chi^{\text{QM}}(A|B)$ and $\chi^{\text{TD}}(A|B)$
for three additive observables $\hat{A}=\hat{M}^{x},\hat{M}^{xx},\hat{M}^{z1z}$.

\begin{figure}
\includegraphics[width=\linewidth]{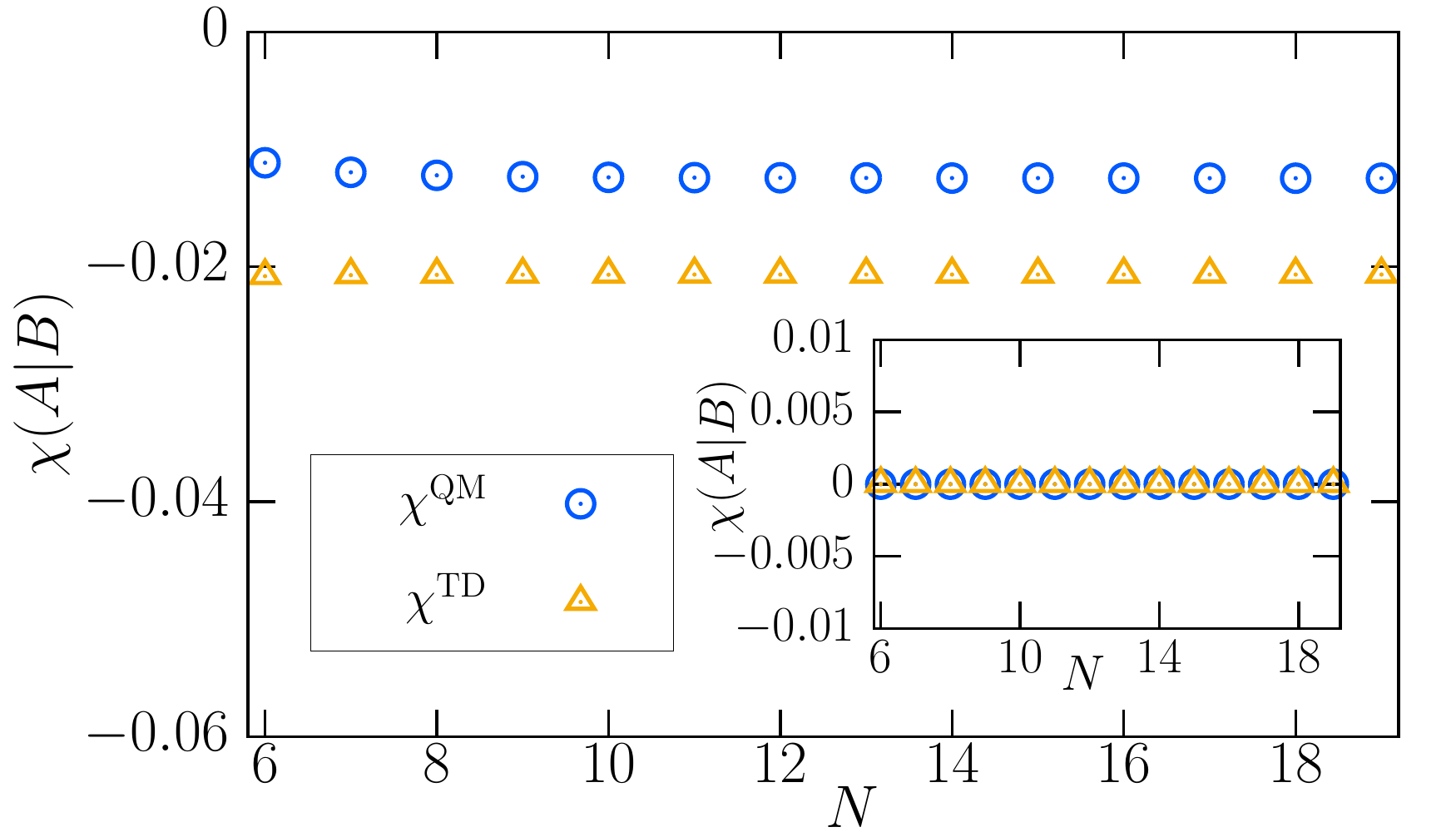}
\caption{\label{fig:chi(xx)_XXZ}%
$\chi^{\text{QM}}_{N}(A|B)$ (blue circle) and $\chi^{\text{TD}}_{N}(A|B)$ (orange triangle)
against the system size $N$
in an integrable system (Model~III).
The quench parameter is $f=J_{zz}$ and the observable of interest is $\hat{A}=\hat{M}^{xx}$.
This plot reveals Eq.~(\ref{eq:chi^QM(A|B)-chi^TD(A|B)}) is violated for $\hat{A}=\hat{M}^{xx}$ in Model~III.
Inset: The same plot as the main of this figure
for $\hat{A}=\hat{M}^{x}$.
The susceptibilities satisfy
$\chi^{\text{QM}}_{N}(A|B)=\chi^{\text{TD}}_{N}(A|B)=0$
because of the spin rotation symmetry around $z$ axis.
}
\end{figure}

In the main of Fig.~\ref{fig:chi(xx)_XXZ},
$\chi^{\text{QM}}_{N}(A|B)$ and $\chi^{\text{TD}}_{N}(A|B)$ are plotted
against the system size $N$ for $\hat{A}=\hat{M}^{xx}$.
It is seen that
these susceptibilities deviate from each other,
indicating the violation of Eq.~(\ref{eq:chi^QM(A|B)-chi^TD(A|B)}).

The inset of Fig.~\ref{fig:chi(xx)_XXZ} is the same plot
for $\hat{A}=\hat{M}^{x}$.
In this case, 
Eq.~(\ref{eq:chi^QM(A|B)-chi^TD(A|B)}) is satisfied in 
the trivial form 
$\chi^{\text{QM}}_{N}(A|B)=\chi^{\text{TD}}_{N}(A|B)=0$,
which follows from the spin rotation symmetry around $z$ axis.
This 
does not contradict our \emph{Theorem}, which does not state violation of
Eq.~(\ref{eq:chi^QM(A|B)-chi^TD(A|B)}) for all $\hat{A}$.

\begin{figure}
\includegraphics[width=\linewidth]{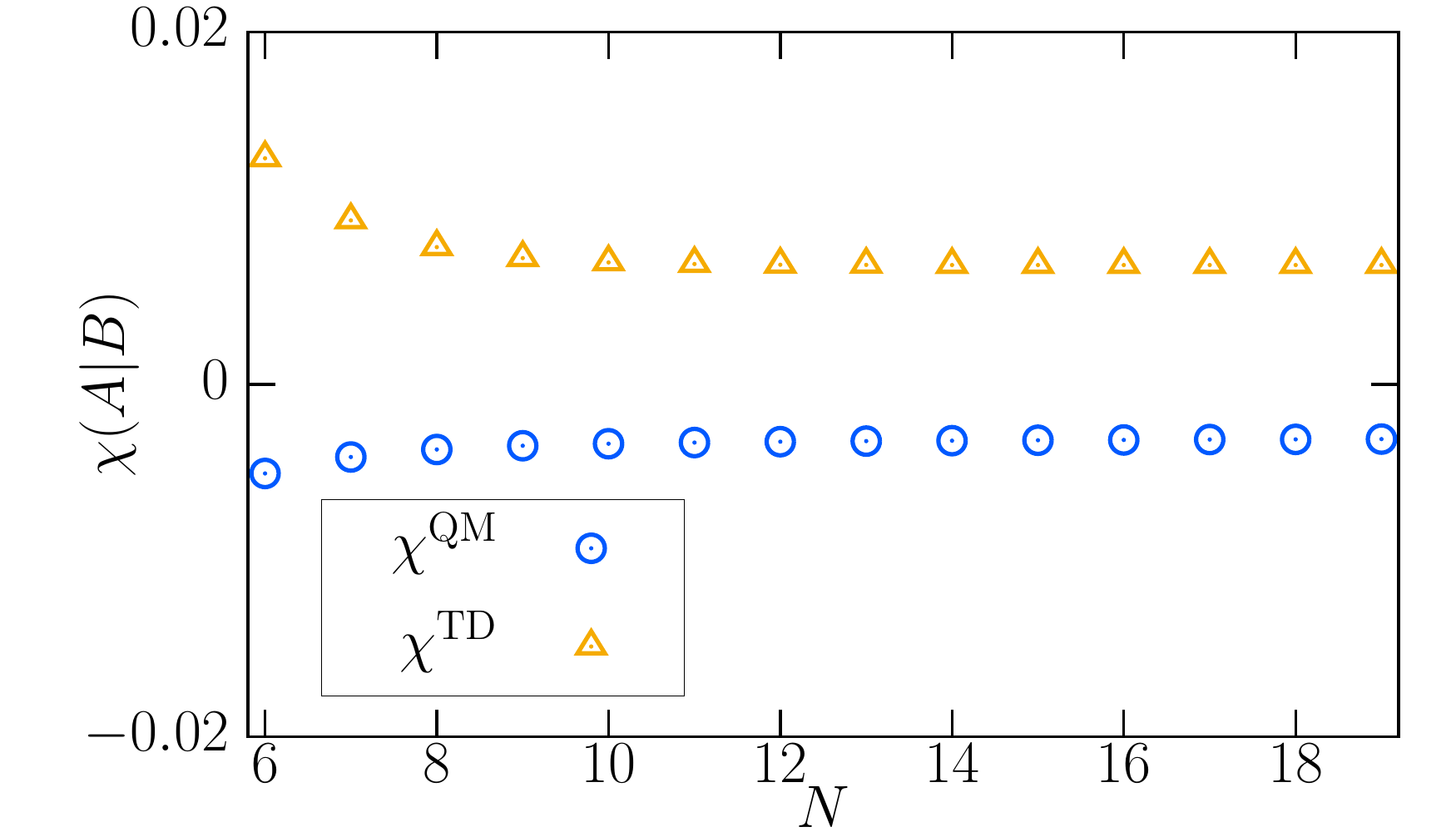}
\caption{\label{fig:chi(z1z)_XXZ}%
The same plots as in Fig.~\ref{fig:chi(xx)_XXZ}
for $\hat{A}=\hat{M}^{z1z}$.
This plot reveals Eq.~(\ref{eq:chi^QM(A|B)-chi^TD(A|B)}) is violated for $\hat{A}=\hat{M}^{z1z}$ in Model~III.
}
\end{figure}

Figure~\ref{fig:chi(z1z)_XXZ} shows
the same plot as in Fig.~\ref{fig:chi(xx)_XXZ}
for $\hat{A}=\hat{M}^{z1z}$.
Interestingly, 
$\chi^{\text{QM}}_{N}(A|B)$ and $\chi^{\text{TD}}_{N}(A|B)$ have
opposite signs,
which clearly indicates the violation of Eq.~(\ref{eq:chi^QM(A|B)-chi^TD(A|B)}).

\subsection{\label{sec:LevelSpacing}Energy level spacings}

In this Appendix, 
we study level spacing statistics 
of $\hat{H}_{0}$ for  Models~I, II and III,
which are defined in Table~\ref{tbl:Models}.

Let $D_{k}$ be the dimension of the eigenspace of the lattice translation
with an eigenvalue $e^{-ik}$ where $k=2\pi n_{k}/N$ ($n_{k}=0,1,...,N-1$) is wavenumber.
We here label eigenvalues of $\hat{H}_{0}$
in this subspace 
using an integer $j=0,...,D_{k}-1$
as $E^{k}_{j}$,
and sort them in ascending order,
$E^{k}_{j}\le E^{k}_{j+1}$ for all $j$.
We consider the ratio of two consecutive energy level spacings
$E^{k}_{j+1}-E^{k}_{j}$ and $E^{k}_{j+2}-E^{k}_{j+1}$
in the subspace 
\begin{align}
r^{k}_{j}:=\min\Bigl\{
\frac{E^{k}_{j+2}-E^{k}_{j+1}}{E^{k}_{j+1}-E^{k}_{j}},\ 
\frac{E^{k}_{j+1}-E^{k}_{j}}{E^{k}_{j+2}-E^{k}_{j+1}}
\Bigr\},
\label{eq:DEF_ratio}
\end{align}
which satisfies $0\le r^{k}_{j}\le 1$.
(When $E^{k}_{j+2}=E^{k}_{j+1}=E^{k}_{j}$, we define $r^{k}_{j}:=0$.)
In order to investigate the level statistics in the bulk of energy spectrum,
we use the ratios $r^{k}_{j}$ satisfying
$D_{k}/4\le j < 3D_{k}/4$.
By using all these ratios,
we define the histogram of the ratio, $P_{k}(r)$.
[We take the width of the interval of $r$ for the construction of $P_{k}(r)$ as $0.05$.]

Atas {\it et al}.~\cite{Atas2013} proposed the following functional forms 
of $P_{k}(r)$
using the random matrix theory.
When the energy levels obey the Poisson law,
$P_{k}(r)$ is given by 
\begin{align}
P_{\text{Poi}}(r):=2\frac{1}{(1+r)^2},
\end{align}
which is expected 
for typical integrable systems.
When the energy levels obey the GUE, 
its $P_{k}(r)$
is well approximated by
\begin{align}
&P_{\text{GUE}}(r):=
2\frac{81\sqrt{3}}{4\pi}
\frac{(r+r^2)^2}{(1+r+r^2)^{4}}\notag\\
&\hspace{36pt}+2\frac{0.578846}{(1+r)^2}\Bigl\{
\Bigl(r+\frac{1}{r}\Bigr)^{-2}
-4\frac{4-\pi}{3\pi-8}\Bigl(r+\frac{1}{r}\Bigr)^{-3}
\Bigr\}.
\end{align}
This is expected for typical nonintegrable systems
that have no symmetries other than the lattice translation.
Furthermore, when the levels obey 
the Gaussian orthogonal ensemble (GOE), 
its $P_{k}(r)$
is well approximated by
\begin{align}
&P_{\text{GOE}}(r):=
2\frac{27}{8}
\frac{(r+r^2)}{(1+r+r^2)^{5/2}}\notag\\
&\hspace{36pt}+2\frac{0.233378}{(1+r)^2}\Bigl\{
\Bigl(r+\frac{1}{r}\Bigr)^{-1}
-2\frac{\pi-2}{4-\pi}\Bigl(r+\frac{1}{r}\Bigr)^{-2}
\Bigr\},
\end{align}
which is expected 
for typical nonintegrable systems.

We have numerically calculated $P_{k}(r)$ of our Models
by the exact diagonalization,
and compared the results with the above forms.
Note that
the symmetry properties of the subspace of wavenumber $k$
sometimes depend on whether $N$ is even or odd.
In addition,
the properties at special wavenumbers $k=0,\pi$
are sometimes very different from those at other wavenumbers. 
Therefore we 
consider three values of wavenumber $k=0,2\pi/N, \pi$
and both even and odd $N$.
(Results for $k=\pi$ with odd $N$ are absent,
since $k=\pi$ is impossible when $N$ is odd.)

\begin{figure}
\includegraphics[width=\linewidth]{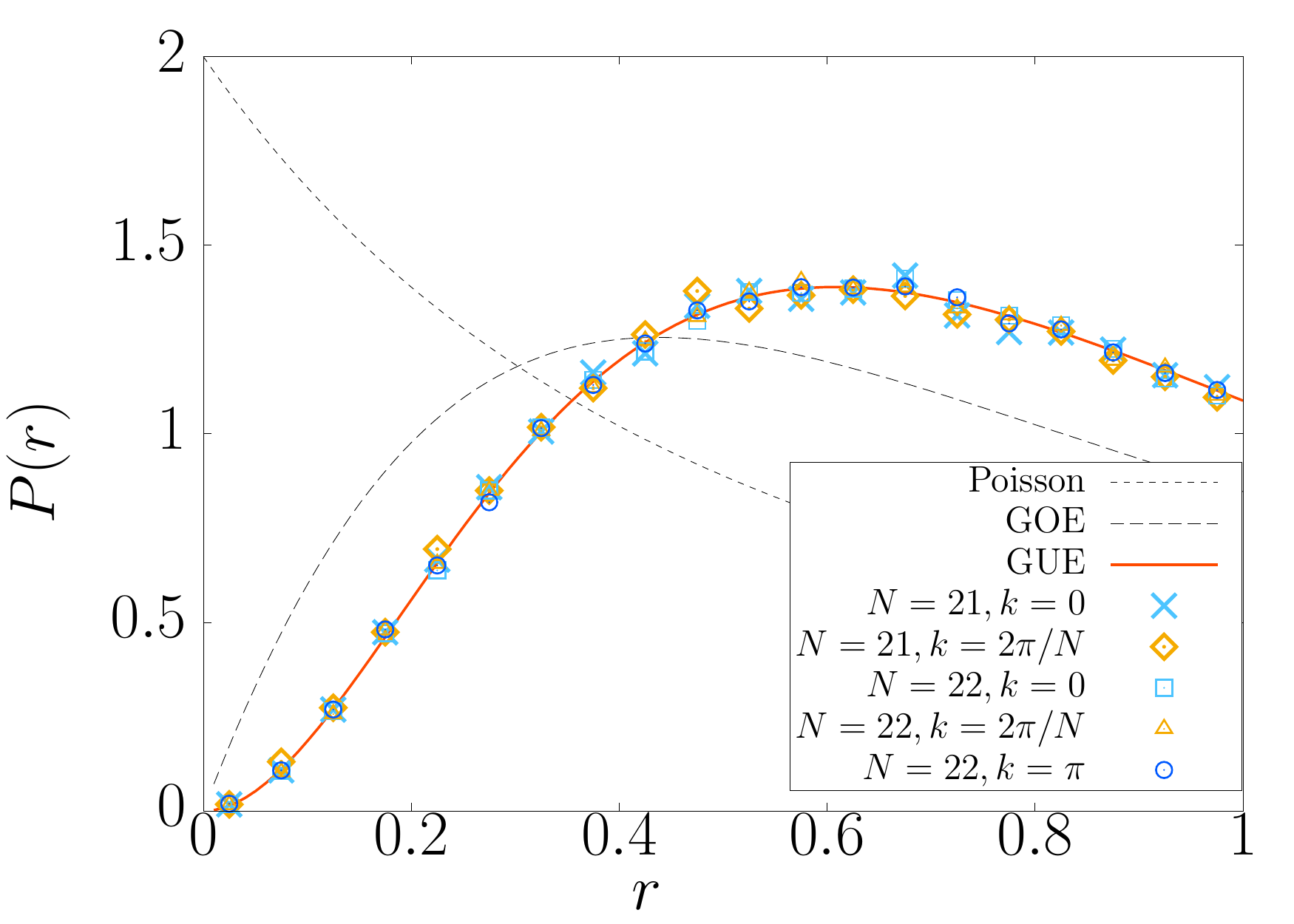}
\caption{\label{fig:Level_Nonintegrable}%
Histogram $P_{k}(r)$ of the ratio of consecutive energy level spacings of Model~I in a momentum sector of wavenumber $k$.
They are plotted for the system size $N=21,22$, and wavenumber $k=0,2\pi/N,\pi$.
The solid line shows the GUE prediction $P_{\text{GUE}}(r)$,
while two dashed lines show
the distribution for the Poisson case $P_{\text{Poi}}(r)$
and the GOE prediction $P_{\text{GOE}}(r)$.
}
\end{figure}%
We plot $P_{k}(r)$ of Model~I
for $N=21,22$
in Fig.~\ref{fig:Level_Nonintegrable}.
It is well fitted by 
the GUE form $P_{\text{GUE}}(r)$ (solid line).
For comparison, the Poisson form
$P_{\text{Poi}}(r)$
and the GOE form 
$P_{\text{GOE}}(r)$
are depicted by the dashed lines,
which clearly deviate from 
$P_{k}(r)$.
These results indicate 
that Model~I has no local conserved quantities
other than the Hamiltonian.

\begin{figure}
\includegraphics[width=\linewidth]{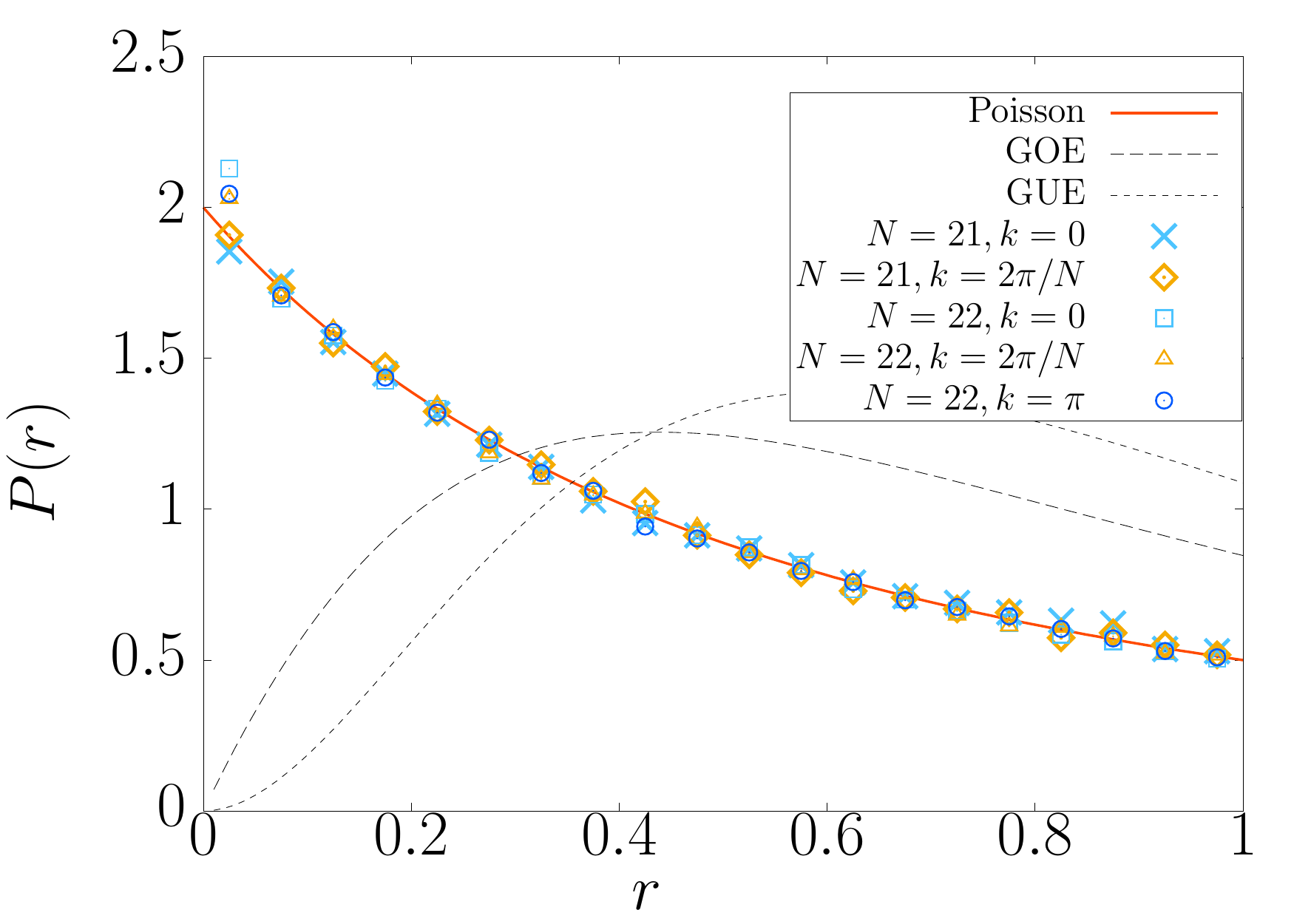}
\caption{\label{fig:Level_Noninteracting}%
Same plot as in Fig.~\ref{fig:Level_Nonintegrable} for Model~II.
The solid line shows the distribution for the Poisson case $P_{\text{Poi}}(r)$,
while two dashed lines show
the GUE prediction $P_{\text{GUE}}(r)$
and the GOE prediction $P_{\text{GOE}}(r)$.
}
\end{figure}%
In Fig.~\ref{fig:Level_Noninteracting},
$P_{k}(r)$ of Model~II is plotted
for $N=21,22$.
It is well fitted by
the Poisson form $P_{\text{Poi}}(r)$ (solid line),
while it clearly deviates from 
the GUE form $P_{\text{GUE}}(r)$ (dotted line)
and the GOE form $P_{\text{GOE}}(r)$ (dashed line).
These are consistent with the integrability of Model~II.

\begin{figure}
\includegraphics[width=\linewidth]{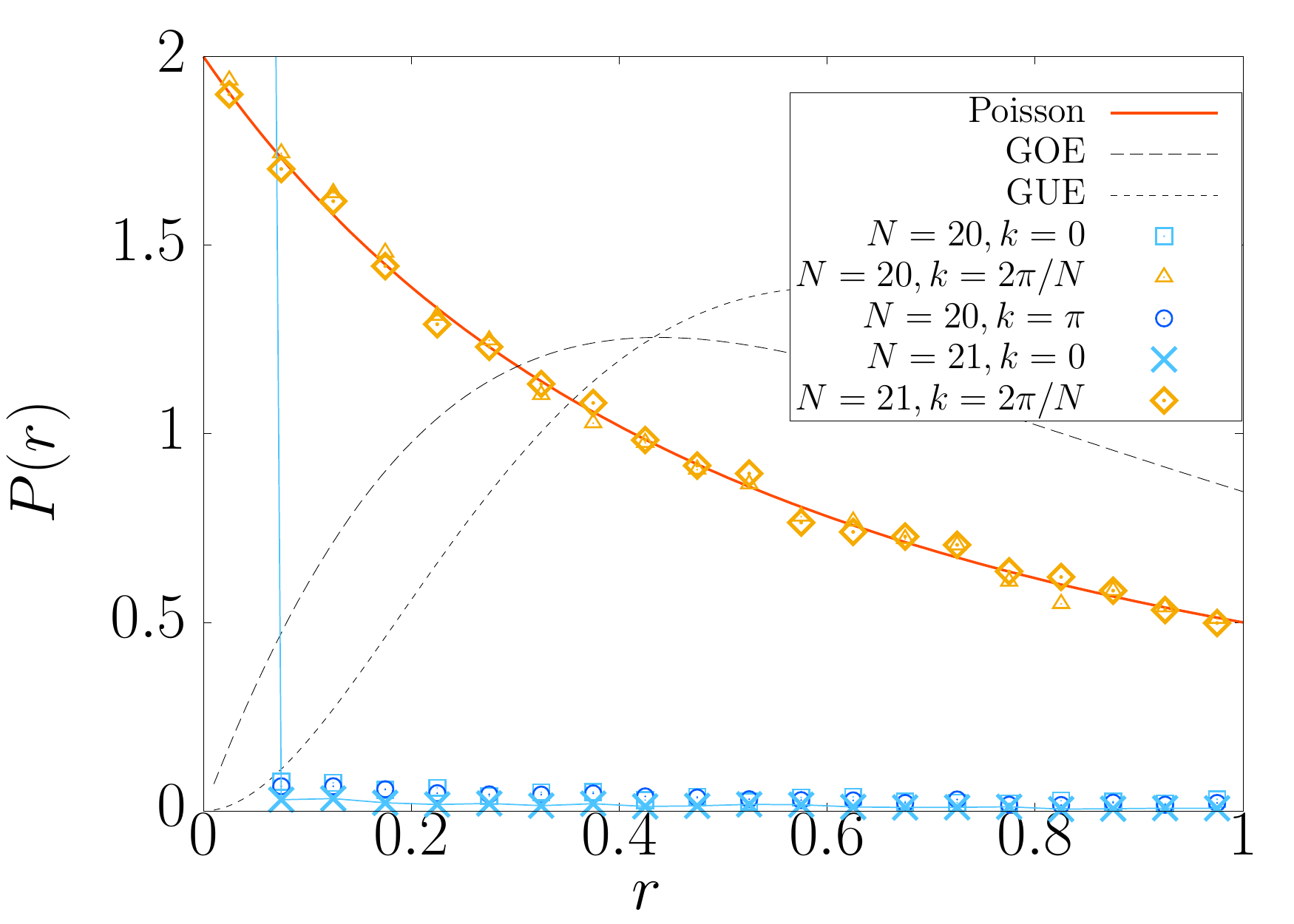}
\caption{\label{fig:Level_Interacting}%
Same plot as in Fig.~\ref{fig:Level_Nonintegrable} for Model~III.
They are plotted for the system size $N=20,21$ and wavenumber $k=0,2\pi/N,\pi$.
The orange solid line shows the distribution for the Poisson case $P_{\text{Poi}}(r)$,
while two dashed lines show
the GUE prediction $P_{\text{GUE}}(r)$
and the GOE prediction $P_{\text{GOE}}(r)$.
The light blue solid line depicts the $\delta$ function like behavior of $P_{k}(r)$.
}
\end{figure}%
In Fig.~\ref{fig:Level_Interacting},
$P_{k}(r)$ of Model~III is plotted
for $N=20,21$.
The $P_{k}(r)$ for $k=2\pi/N$ is well fitted by the Poisson form $P_{\text{Poi}}(r)$ (solid line).
This 
is consistent with the integrability of Model~III.
On the other hand,
$P_{k}(r)$ for $k=0,\pi$ show
$\delta$-function-like behaviors,
which are much different from 
the GUE form $P_{\text{GUE}}(r)$ (dotted line)
and the GOE form $P_{\text{GOE}}(r)$ (dashed line).
The $\delta$-function-like behaviors
come from the fact that
almost all 
energy eigenvalues
have certain degeneracies~\footnote{At least for $k=\pi$ with $N=8,10,12$ and for $k=0$ with $N=12$,
the degree of degeneracy larger than $2$ is observed.
However the majority of degeneracies are double degeneracy.
}.
These degeneracies indicate that
the subspaces of $k=0,\pi$ have additional non-Abelian symmetries,
which would be related to the integrability of Model~III~\footnote{
In fact,
the $\delta$-function-like behavior
is not observed
for $k=0,\pi$ sectors of 
the XXZ chain with next nearest neighbor interactions~\cite{Santos2010,Santos2012b,Santos2012a},
which possesses the lattice inversion symmetry,
the complex conjugate symmetry,
and the spin rotation symmetry around $z$ axis as in Model~III,
but is expected to be nonintegrable~\cite{Santos2010,Santos2012b,Santos2012a}
in contrast.
}.

\section{\label{sec:AnalyticNoninteracting}Analytic results for Model~II}

In this Appendix, 
we consider the following model
\begin{align}
\hat{H} &=-\sum_{r=1}^{N}
\Bigl(
 J_{yy}\hat{\sigma}_{r}^y\hat{\sigma}_{r+1}^y
 +J_{zz}\hat{\sigma}_{r}^z\hat{\sigma}_{r+1}^z
 \notag\\
 &\hspace{36pt}+J_{yz}\hat{\sigma}_{r}^y\hat{\sigma}_{r+1}^z
 +J_{zy}\hat{\sigma}_{r}^z\hat{\sigma}_{r+1}^y
 +h_{x}\hat{\sigma}_{r}^x
\Bigr),
\label{eq:H_XY}
\end{align}
which reduces to Model~II by setting $J_{zy}=0$.
This spin system can also be written as a fermionic system
\begin{align}
\hat{H} &=-\sum_{r=1}^{N-1}
\Bigl(
 (J_{+-}\hat{c}_{r}^{\dagger}\hat{c}_{r+1}+\text{H.c.})
 +(J_{++}\hat{c}_{r}^{\dagger}\hat{c}_{r+1}^{\dagger}+\text{H.c.})
 \notag\\
 &\hspace{36pt}
 +h_{x}(2\hat{c}_{r}^{\dagger}\hat{c}_{r}-1)
\Bigr)
\notag\\
&\hspace{12pt}+\hat{R}(J_{+-}\hat{c}_{N}^{\dagger}\hat{c}_{1}+\text{H.c.})
+\hat{R}(J_{++}\hat{c}_{N}^{\dagger}\hat{c}_{1}^{\dagger}+\text{H.c.})
\notag\\
 &\hspace{12pt}-h_{x}(2\hat{c}_{N}^{\dagger}\hat{c}_{N}-1),
 \label{eq:H_XY_JordanWigner}
\end{align}
by the Jordan-Wigner transformation
\begin{align}
\hat{c}_{r}^{\dagger}=\frac{\hat{\sigma}_{r}^y+i\hat{\sigma}_{r}^z}{2}\prod_{r'(<r)}(-\hat{\sigma}_{r'}^x).
\end{align}
Here
\begin{align}
\hat{R}:=\prod_{r=1}^{N}(-\hat{\sigma}_{r}^x)
=\prod_{r=1}^{N}(1-2\hat{c}_{r}^{\dagger}\hat{c}_{r})
\end{align}
and
\begin{align}
J_{+-}&:=J_{yy}+J_{zz}+iJ_{yz}-iJ_{zy},
\\
J_{++}&:=J_{yy}-J_{zz}-iJ_{yz}-iJ_{zy}.
\end{align}
The operator $\hat{R}$ takes $1$ for the states with even number of fermions,
whereas it takes $-1$ for the states with odd number of fermions.
Hence
depending on whether the number of fermions is even or odd,
terms containing $\hat{R}$ in Eq.~(\ref{eq:H_XY_JordanWigner})
become the antiperiodic or periodic boundary condition, respectively.

To resolve this boundary condition that depends on $\hat{R}$,
we consider two types of the Fourier transformation.
We introduce two sets of wavenumbers
\begin{align}
K^{\text{e}}&:=\Bigl\{\frac{2\pi n_{k}}{N}+\frac{\pi}{N}\Bigm|n_{k}=0,1,...,N-1\Bigr\},\\
K^{\text{o}}&:=\Bigl\{\frac{2\pi n_{k}}{N}\Bigm|n_{k}=0,1,...,N-1\Bigr\},
\end{align}
and perform the following Fourier transformation
\begin{align}
\tilde{c}_{k}^{a}:=\sum_{r=1}^{N}\frac{e^{-ikr}}{\sqrt{N}}\hat{c}_{r}\qquad\text{for }k\in K^{a} \ (a=\text{e},\text{o}).
\end{align}
Using new fermionic operators,
the Hamiltonian~(\ref{eq:H_XY_JordanWigner}) can be written as
\begin{align}
\hat{H}&=-\sum_{a=\text{e},\text{o}}\hat{P}_{R}^{a}\sum_{k\in K^{a}}\Bigl(
(J_{+-}e^{ik}\tilde{c}_{k}^{a\dagger}\tilde{c}_{k}^{a}+\text{H.c.})
\notag\\
&\hspace{36pt}
+(J_{++}e^{ik}\tilde{c}_{k}^{a\dagger}\tilde{c}_{-k}^{a\dagger}+\text{H.c.})
+h_{x}(2\tilde{c}_{k}^{a\dagger}\tilde{c}_{k}^{a}-1)
\Bigr),
\label{eq:H_XY_Fourier}
\end{align}
where 
\begin{align}
\hat{P}_{R}^{\text{e}}&:=\frac{1+\hat{R}}{2},\\
\hat{P}_{R}^{\text{o}}&:=\frac{1-\hat{R}}{2}
\end{align}
are the projection operators to the eigenspace of $\hat{R}$ with eigenvalues $1$ and $-1$,
respectively.

When $J_{++}\neq 0$,
Eq.~(\ref{eq:H_XY_Fourier}) includes the off-diagonal terms $\tilde{c}_{k}^{a\dagger}\tilde{c}_{-k}^{a\dagger}$.
To eliminate these,
we perform the following Bogoliubov transformation for $k\neq 0,\pi$,
\begin{align}
\hat{d}_{k}^{a}:=\cos\theta_{k}\tilde{c}_{k}^{a}+ie^{i\phi_{++}}\sin\theta_{k}\tilde{c}_{-k}^{a\dagger},
\end{align}
where $e^{i\phi_{++}}:=J_{++}/|J_{++}|$ and $\theta_{k}\in(-\pi/2,\pi/2)$ is determined from
\begin{align}
\cos 2\theta_{k}&=-\frac{2(\text{Re}[J_{+-}]\cos k + h_{x})}{r_{k}},\\
\sin 2\theta_{k}&=-\frac{2|J_{++}|\sin k}{r_{k}}
\end{align}
with
\begin{align}
r_{k}=2\sqrt{(\text{Re}[J_{+-}]\cos k + h_{x})^2+|J_{++}|^2\sin^2 k}.
\end{align}
As a result,
the Hamiltonian is given by 
\begin{align}
\hat{H}&=-\sum_{a=\text{e},\text{o}}\hat{P}_{R}^{a}\sum_{k\in K^{a}\cap\{0,\pi\}}
2(\text{Re}[J_{+-}]\cos k+h_{x})(\tilde{c}_{k}^{a\dagger}\tilde{c}_{k}^{a}-\frac{1}{2})
\notag\\
&\hspace{12pt}-\sum_{a=\text{e},\text{o}}\hat{P}_{R}^{a}\sum_{k\in K^{a}\setminus\{0,\pi\}}
\varepsilon_{k}(\hat{d}_{k}^{a\dagger}\hat{d}_{k}^{a}-\frac{1}{2}),
\label{eq:H_XY_Bogoliubov}
\end{align}
where
\begin{align}
\varepsilon_{k}:=r_{k}+2\text{Im}[J_{+-}]\sin k.
\end{align}
[Note that when $J_{++}=0$,
Eq.~(\ref{eq:H_XY_Fourier}) 
is already of the same form as the above.]

On the other hand,
$\hat{B}$ given by Eq.~(\ref{eq:Mzz})
can be written as
\begin{align}
\hat{B}
&=
\sum_{a=\text{e},\text{o}}\hat{P}_{R}^{a}
\sum_{k\in K^{a}}
(\cos k \cos 2\theta_{k}
+\sin k \sin 2\theta_{k} \cos\phi_{++})
\notag\\
&\hspace{12pt}\times(2\hat{d}_{k}^{a\dagger} \hat{d}_{k}^{a}-1)
\notag\\
&\hspace{12pt}
+\sum_{a=\text{e},\text{o}}\hat{P}_{R}^{a}\sum_{k\in K^{a}}
\Bigl(
\bigl(
-\cos k \, ie^{i\phi_{++}}\sin 2\theta_{k}
\notag\\
&\hspace{36pt}
+i\sin k (\cos^2\theta_{k} - e^{2i\phi_{++}}\sin^2\theta_{k})
\bigr) \hat{d}_{k}^{a\dagger} \hat{d}_{-k}^{a\dagger}
+\text{H.c.}\Bigr).
\end{align}

As a result,
the susceptibilities of $\hat{B}$ can be calculated by using Eqs.~(\ref{eq:chi^QM_BogoProd}) and (\ref{eq:chi^TD_BogoProd}) of Appendix~\ref{sec:Derivation_Exp_Ave},
whose thermodynamic limits are given by
\begin{align}
&\lim_{N\to\infty}
\chi^{\text{QM}}_{N}(\hat{B}|\hat{B})
\notag\\
&=\frac{1}{2\pi}\int_{0}^{2\pi}\dd{k}2
\frac{e^{2\beta_{0} r_{k}}-1}{r_{k}}
\frac{1}{e^{\beta_{0}\varepsilon_{k}}+1}
\frac{1}{e^{\beta_{0}\varepsilon_{-k}}+1}
\notag\\
&\hspace{5pt}
\times \bigl|
-\cos k \, ie^{i\phi_{++}}\sin 2\theta_{k}
+i\sin k (\cos^2\theta_{k} - e^{2i\phi_{++}}\sin^2\theta_{k})
\bigr|^2,
\label{eq:Analytic_chi^QM}
\end{align}
\begin{align}
&\lim_{N\to\infty}
\chi^{\text{TD}}_{N}(\hat{B}|\hat{B})
-\lim_{N\to\infty}
\chi^{\text{QM}}_{N}(\hat{B}|\hat{B})
\notag\\
&=
\frac{\beta_{0}}{2\pi}\int_{0}^{2\pi}\dd{k}
\frac{
(\cos k \cos 2\theta_{k}
+\sin k \sin 2\theta_{k} \cos\phi_{++}
-C \varepsilon_{k})^2
}{\cosh^2 (\beta_{0}\varepsilon_{k}/2)},
\label{eq:Analytic_chi^TD}
\end{align}
where
\begin{align}
C
&:=
\frac{1}{2\pi}
\int_{0}^{2\pi}\dd{k}
\frac{
\varepsilon_{k}
(\cos k \cos 2\theta_{k}
+\sin k \sin 2\theta_{k} \cos\phi_{++})
}{\cosh^2 (\beta_{0}\varepsilon_{k}/2)}
\notag\\
&\hspace{12pt}\Bigm/
\frac{1}{2\pi}
\int_{0}^{2\pi}\dd{k}
\frac{\varepsilon_{k}^2}{\cosh^2 (\beta_{0}\varepsilon_{k}/2)}.
\end{align}

\bibliography{2023_08_08.bbl}

\end{document}